\documentclass[
a4paper,
twocolumn,
10pt,
floatfix,
accepted=2025-06-29]{quantumarticle}
\pdfoutput=1

\usepackage{graphicx}
\usepackage[dvipsnames]{xcolor}
\usepackage{physics}
\usepackage{dcolumn}
\usepackage{bm}
\usepackage{comment}
\usepackage{siunitx}
\usepackage{mathtools}
\usepackage{cases}
\usepackage{xfrac}
\usepackage{pifont}

\usepackage{lipsum}
\usepackage{tikz}
\usepackage{float}
\usepackage{array}
\usepackage{amsbsy}
\usepackage[utf8x]{inputenc}
\usepackage{epstopdf}
\usepackage{amsmath,amssymb,amsfonts}
\usepackage{booktabs}
\usepackage{dsfont}
\usepackage{cprotect}
\usepackage{mathtools}
\usepackage{algpseudocode}
\usepackage[normalem]{ulem}
\usepackage{multirow}
\usepackage{verbatim}
\usepackage{tabularx}
\usepackage{enumitem}
\usepackage{url}

\usepackage[colorlinks]{hyperref}
\hypersetup{
plainpages=true,
breaklinks=true,
hypertexnames=false,
pageanchor=true,
colorlinks=true,
linkcolor={blue},
citecolor={red},
urlcolor={blue},
anchorcolor={black}
}
\usepackage[capitalise]{cleveref}
\usepackage{multirow}

\newcommand{\id}[0]{\mathds{1}}
\newcommand{\F}[0]{\mathcal{F}}
\newcommand{\E}[0]{\mathbb{E}}

\newcommand{\I}[0]{\mathds{1}}

\renewcommand{\H}[0]{\hat{H}}
\newcommand{\sz}[0]{\hat{\sigma}^z}

\newcommand{\LL}[0]{\mathcal{L}}

\newcommand{\psiv}[0]{\psi_{\theta}}
\newcommand{\kpsiv}[0]{\ket{\psi_{\theta}}}
\newcommand{\xmark}{\ding{55}}%

\newcommand{\argmin}[1]{\underset{#1}{\operatorname{argmin }}}

\usepackage{tikz}
\newcommand*\circled[1]{\tikz[baseline=(char.base)]{
            \node[shape=circle,draw,inner sep=2pt] (char) {\footnotesize #1};}}

\newsavebox{\mstrut}

\usepackage[normalem]{ulem}
\renewcommand\sout{\bgroup\markoverwith{\textcolor{red}{\rule[0.5ex]{2pt}{0.8pt}}}\ULon}

\begin{document}

\author{Luca Gravina}
\author{Vincenzo Savona}
\affiliation{Institute of Physics, Ecole Polytechnique Fédérale de Lausanne (EPFL), CH-1015 Lausanne, Switzerland}
\affiliation{Center for Quantum Science and Engineering, Ecole Polytechnique Fédérale de Lausanne (EPFL), CH-1015 Lausanne, Switzerland}
\author{Filippo Vicentini}
\affiliation{CPHT, CNRS, Ecole Polytechnique, Institut Polytechnique de Paris, 91120 Palaiseau, France.}
\affiliation{Coll\`ege de France, Universit\'e PSL, 11 place Marcelin Berthelot, 75005 Paris, France}

\title{Neural Projected Quantum Dynamics: a systematic study}
\begin{abstract}
We investigate the challenge of classical simulation of unitary quantum dynamics with variational Monte Carlo approaches, addressing the instabilities and high computational demands of existing methods. 
By systematically analyzing the convergence of stochastic infidelity optimizations, examining the variance properties of key stochastic estimators, and evaluating the error scaling of multiple dynamical discretization schemes, we provide a thorough formalization and significant improvements to the projected time-dependent Variational Monte Carlo (p-tVMC) method.
We benchmark our approach on a two-dimensional Ising quench, achieving state-of-the-art performance. This work establishes p-tVMC as a powerful framework for simulating the dynamics of large-scale two-dimensional quantum systems, surpassing alternative VMC strategies on the investigated benchmark problems.
\end{abstract}

\maketitle

\tableofcontents

\section{Introduction}
\label{sec:Intro}
Simulating the dynamics of a quantum system is essential for addressing various problems in material science, quantum chemistry, quantum optimal control, and for answering fundamental questions in quantum information \cite{Georgescu2014, PreskillQuantum18, Ezratty2021}. 
However, the exponential growth of the Hilbert space makes this one of the most significant challenges in computational quantum physics, with only few tools available to accurately simulate the dynamics of large, complex systems.

To manage the exponential growth of the Hilbert space, quantum states 
can be encoded using efficient compression schemes \cite{Yuan2019}.
While tensor network methods \cite{Bauls2023, Orus_Tensor_2019, schollwck2005, Feldman2022, eisert2013, werner2016}, particularly Matrix Product States  \cite{cirac2021, Paeckel2019, ors2014, schollwck2011}, excel in simulating large one-dimensional models with short-range interactions, extending them to higher dimensions is problematic.
Such extensions, either rely on uncontrolled approximations \cite{Lubasch2014, Verstraete2004}, 
or incur in an exponential costs when encoding area-law entangled states \cite{Tagliacozzo2009}, making them poorly suited for investigating strongly correlated, higher-dimensional systems or unstructured lattices, such as those encountered in chemistry or quantum algorithms \cite{Felser2021, Silvi2019, Jaschke2018, Jaschke2018a, Evenbly2011}.

Recently, Neural Quantum States (NQS) have garnered increasing attention as a non-linear variational encoding of the wave-function capable, in principle, of describing arbitrarily entangled states, both pure \cite{Carleo2017, Sharir2022, Deng2017, Urea2024, Passetti2023} and mixed \cite{Torlai2018_Latent,Vicentini2022_PositiveDefinite,Luo2022_AutoregOpen,Vicentini2019_Open,Eeltink2023_open,Reh2021}. 
This approach compresses the exponentially large wave-function into a polynomial set of parameters, with no restrictions on the geometry of the underlying system. 
The added flexibility, however, comes at a cost: unlike matrix product states whose bond dimension can be adaptively tuned via deterministic  algorithms, neural network optimizations are inherently stochastic, making it hard to establish precise error bounds.

Despite the limitations of neural networks not being fully understood \cite{Dash2024_qgt,Zhao2024_EmpiricalComplexity}, recent studies have demonstrated that NQS can be reliably optimized to represent the ground state of computationally challenging, non-stoquastic, fermionic, or frustrated Hamiltonians, arising across various domains of quantum physics \cite{Choo2019, Sharir2020, Vicentini2023, Viteritti2023a, Liang2018, Stokes2020a, Attila2020, Choo2020, Cassella2023, RobledoMoreno2022}. 
However, for the more complex task of simulating quantum dynamics, NQS have yet to show significant advantages over existing methods.

\paragraph{Neural Quantum Dynamics.}
There are two families of variational algorithms for approximating the direct integration of the Schrödinger equation using variational ansatze: time-dependent Variational Monte Carlo (tVMC) \cite{Carleo2017_tVMC} and projected tVMC (p-tVMC), formalized in Ref.~\cite{Sinibaldi2023}.
The former, tVMC, linearizes both the unitary evolution and the variational ansatz, casting the Schrödinger equation into an explicit algebraic-differential equation for the variational parameters \cite{Carleo2017_tVMC, Stokes2023}.
The latter, p-tVMC, relies on an implicit optimization problem to compute the parameters of the wave function at each timestep, using low-order truncations of the unitary evolution such as Taylor or Trotter expansions.

Of the two methods, tVMC is regarded as the least computationally expensive, as it avoids the need to solve a nonlinear optimization problem at every step. 
It has been successfully applied to simulate sudden quenches in large spin \cite{Schmitt_2020, Schmitt2023, Carleo2017,Joshi2024SkirmionDynamics,Czischek2018} and Rydberg \cite{Mauron2024} lattices, quantum dots \cite{Nys2024}, as well as finite temperature \cite{Wagner2024Temperature,Nys2024_thermofield} and out of equilibrium \cite{Eeltink2023_open, vicentini2022_dynamics,Lin2024Open} systems.
However, while stable for (log-)linear variational ansatze such as Jastrow \cite{Mauron2024} or Tensor Networks, the stiffness of the tVMC equation \cite{Shampine1979_stiff} appears to increase with the nonlinearity of the ansatz, making integration particularly hard for deep networks.
Contributing to this stiffness is the presence of a systematic statistical bias in the evaluation of the dynamical equation itself, which would be exponentially costly to correct \cite{Sinibaldi2023}. 
Although the effect of this noise can be partially regularised away \cite{Schmitt_2020}, this regularization procedure introduces additional bias that is difficult to quantify. 
As of today, the numerous numerical issues inherent to tVMC make its practical application to non-trivial problems difficult, with the estimation of the actual error committed by the method being unreliable at best.

\paragraph{Projected Neural Quantum Dynamics and open challenges.}

The projected time-dependent Variational Monte Carlo method offers a viable, albeit more computationally intensive, alternative by decoupling the discretization of the physical dynamics from the nonlinear optimization of the variational ansatz, thereby simplifying the analysis of each component.
So far, the discretization problem has been tackled using established schemes such as Runge-Kutta \cite{Donatella2023} or Trotter \cite{Sinibaldi2023, Poletti2024, Medvidovi2021, Carleo2018}.
These methods do not fully leverage the specific properties of VMC approaches and, as a result, the existing body of work \cite{Donatella2023, Sinibaldi2023, Gutirrez2022} has been limited to second-order accuracy in time, struggling to provide general, scalable solutions.
Similarly, the nonlinear optimization problem has mainly been addressed using first-order gradient descent techniques, neglecting the benefits offered by second-order optimization strategies.

In this manuscript, we investigate both aspects of p-tVMC — discretization and optimization — independently, addressing the shortcomings detailed above with the goal of enhancing accuracy, reducing computational costs, and improving stability and usability.

Specifically, in \cref{sec:integrators} we introduce a new family of discretization schemes tailored for p-tVMC, achieving higher accuracy for equivalent computational costs. 
In \cref{sec:fidelity} we conduct an in-depth analysis of the nonlinear optimization problem of infidelity minimization, identifying the most effective stochastic estimator and introducing a new adaptive optimization scheme that performs as well as manually tuned hyperparameters, eliminating the need for manual adjustment. 
Finally, in \cref{sec:results} we benchmark several of our methods against a challenging computational problem: a quench across the critical point of the two-dimensional transverse field Ising model.

\section{Integration schemes}
\label{sec:integrators}
Consider the generic evolution equation
\begin{equation}
\label{eqn:generic_evolution}
    \ket{\psi_{t+\dd t}} = e^{\hat \Lambda \dd t} \ket{\psi_t},
\end{equation}
where $\hat \Lambda = -i\hat H$ for some $K$-local time-independent Hamiltonian $\hat H$ constant in the the system size $N$. 
The fundamental challenge for the numerical integration of \cref{eqn:generic_evolution} lies in the dimensionality of the Hilbert space $\mathcal{H}$ scaling exponentially with system size, that is, $\operatorname{dim}(\mathcal{H}) \sim \operatorname{exp}(N)$. 
This makes it impossible to merely store in memory the state-vector $\ket{\psi}$, let alone numerically evaluate or apply the propagator.

Variational methods address the first problem by encoding an approximate representation of the state at time $t$ into the time-dependent parameter vector $\theta_t \in \mathbb{R}^{N_p}$ of a variational ansatz, while relying on Monte Carlo integration for computing expectation values \cite{Carleo2017, Medvidovi2024, Lange2024}.
Within this framework, the McLachlan variational principle is used to recast \cref{eqn:generic_evolution} 
as the optimization problem \cite{Yuan2019}
\begin{equation}
\label{eqn:minimization_problem}
    \theta_{t+\dd t} = \argmin{\theta}\,\, \LL\qty(\ket{\psi_\theta}, e^{\hat \Lambda \dd t} \ket{\psi_{\theta_t}}),
\end{equation}
where $\LL$ is a suitable loss function quantifying the discrepancy between two quantum states. 
Various choices for $\LL$ are possible, the one adopted throughout this work is presented and discussed in \cref{sec:fidelity}.

TVMC and p-tVMC confront \cref{eqn:minimization_problem} differently.
The former, tVMC, linearizes both the unitary evolution operator and the ansatz,  reducing \cref{eqn:minimization_problem} to an explicit first-order non-linear differential equation in the parameters \cite{Carleo2017, Stokes2023}.
In contrast, p-tVMC, relies on higher-order discretizations of the evolution operator to efficiently solve the optimization problem in \cref{eqn:minimization_problem} at each timestep.

In \cref{sec:p-tVMC_general}, we present a general formulation of p-tVMC and identify a set of fundamental requirements that discretization schemes for p-tVMC should satisfy. 
From this perspective, we revisit the well-established Trotter and Runge-Kutta methods in \cref{sec:trotter,sec:Taylor}. 
In \cref{sec:LPE,sec:PPE,sec:split-schemes}, we introduce a new family of discretization schemes tailored to the specific structure of p-tVMC, achieving higher-order accuracy in $\dd t$ with reduced computational complexity.

\subsection{Generic formulation of p-tVMC schemes}
\label{sec:p-tVMC_general}
The minimization problem in \cref{eqn:minimization_problem} aims at finding the parameters transformation that best approximates the infinitesimal evolution under $\hat\Lambda$, i.e.~
\begin{equation}
\label{eqn:non_general_problem}
    \ket{\psi_{\theta_{t+\dd t}}} = e^{\hat \Lambda \dd t}\ket{\psi_{\theta_t}}.
\end{equation}
The most generic problem of this form is that of finding the evolved parameters $\theta_{t+\dd t}$ such that
\begin{equation}
\label{eqn:general_problem}
    \hat V \ket{\psi_{\theta_{t+\dd t}}} = \hat U \ket{\psi_{\theta_t}},
\end{equation}
with arbitrary $\hat U$ and $\hat V$, not necessarily unitary.
Equation~\eqref{eqn:general_problem} reduces to \cref{eqn:non_general_problem} for $\hat V = \I$ and $\hat U = \operatorname{exp}(\hat \Lambda \dd t)$. Note that while the parameters $\theta_{t+\dd t}$ satisfying \cref{eqn:general_problem} also satisfy
\begin{equation}
     \ket{\psi_{\theta_{t+\dd t}}} = \hat V^{-1} \hat U \ket{\psi_{\theta_t}},
\end{equation}
the associated optimization problems
\begin{equation}
\begin{aligned}
    \theta_{t+\dd t}
    &= \argmin{\theta}\,\, \LL\qty(\hat V\ket{\psi_{\theta}}, \hat U \ket{\psi_{\theta_t}})\\
    &= \argmin{\theta}\,\, \LL\qty(\ket{\psi_{\theta}}, \hat V^{-1}\hat U \ket{\psi_{\theta_t}})
\end{aligned}
\end{equation}
share the same minimum but have different optimization landscapes. 
It is natural, therefore, to consider expansions of the infinitesimal time-independent propagator in the form of a product series
\begin{equation}
\label{eqn:optimization_UV_expanded}
\begin{aligned}
     e^{\hat \Lambda \dd t} 
     =\prod_{k=1}^{s} \hat V_{k}^{-1} \hat U_{k} + \order{\dd t^{o(s) + 1}},
\end{aligned}
\end{equation}
where the number of elements $s$ in the series is related to the order of the expansion $o = o(s)$.
This decomposition is convenient because:
\begin{itemize}
    \item There are no summations. Therefore, the terms $\hat V_{k}^{-1} \hat U_{k}$ in the series can be applied sequentially to a state, without the need to store intermediate states and recombine them.
    \item The single step of p-tVMC can efficiently embed an operator inverse at every sub-step.
\end{itemize}
By utilizing this discretization, the parameters after a single timestep $\dd t$ are found by solving a sequence of $s$ subsequent optimization problems, with the output of each substep serving as the input for the next.
Specifically, setting  $\theta_{t} \equiv \theta^{(0)}$, and $\theta_{t+\dd t} \equiv \theta^{(s)}$, we can decompose \cref{eqn:minimization_problem} as 
\begin{equation}
\label{eqn:optimization_UV}
    \theta^{(k)} = \argmin{\theta}\,\,\LL\qty(\hat V_k \ket{\psi_\theta}, \hat U_k \ket*{\psi_{\theta^{(k-1)}}}),
\end{equation}
with $0<k<s$.
This optimization does not directly compress the variational state $\ket{\psi_\theta}$ onto the target state $\ket{\phi}$. Instead, it matches two versions of these states transformed by the linear operators $\hat V_k$ and $\hat U_k$.
A careful tuning of $\hat V_k$ and $\hat U_k$ can be seen as a form of preconditioning, initializing the optimization closer to the solution and potentially accelerating convergence.

Equation \eqref{eqn:optimization_UV} can be solved efficiently with variational Monte Carlo methods provided all operators $\{\hat V_k, \hat U_k\}$ are log-sparse (or $K$-local). 
In what follows we explore proficient choices for the set $\{\hat V_k, \hat U_k\}$. 
Two conditions guide our search for an optimal expansion scheme:
\begin{enumerate}[label=(\roman*)]
    \item Equation~\eqref{eqn:optimization_UV_expanded} should match \cref{eqn:minimization_problem} to a specified order in $\dd t$, denoted as $o$, ensuring accurate time evolution up to this order.
    \item The computational complexity of solving \cref{eqn:optimization_UV}, which is proportional to $s N_c$ with $N_c$ the number of connected elements of $\{\hat V_k, \hat U_k\}$, must scale at most polynomially in the number of particles $N$ and in the order $o$ of the method. 
\end{enumerate}
Table~\ref{tab:UV} summarizes our analysis, including both established discretization schemes, and the four new ones introduced in this manuscript.

\setlength{\tabcolsep}{6pt}
\begin{table}[t!]
\centering
\begin{tabular}{c c c c c}
 Name & Sec.\! & substeps & $N_c$ & unitary \\[0.25em]
\hline \rule{0pt}{1.5em}

Trotter        & \ref{sec:trotter}      & $5^{\tfrac{o}{2}-1}\order{2N}$ & $\order{2}$   & \checkmark \\[0.2cm]

Taylor         & \ref{sec:Taylor}       & $1$                            & $\order{N^o}$ & \xmark     \\[0.2cm]

{LPE-}$o$      & \ref{sec:LPE}          & $o$                            & $\order{N}$   & \xmark     \\[0.2cm]

{PPE-}$o$      & \ref{sec:PPE}          & $\tfrac{o}{2}$                 & $\order{2N}$  & \checkmark \\[0.2cm]

{S-LPE-}$o$    & \ref{sec:split-schemes}& $\dagger$                      & $\order{N}$   & \xmark     \\[0.2cm]

{S-PPE-}$o$    & \ref{sec:split-schemes}& $\dagger$                      & $\order{2N}$  & $o=2$      \\

\end{tabular}
\caption{
Discretization schemes compatible with p-tVMC.
We denote by $s$ the number of substeps (optimizations), $o$ the order of the integration scheme, and $N_c$ the connected elements (complexity) of the operators entering the optimization problem.
We remark that the {PPE-$o$} scheme has only even orders $o$ and is the only scheme that is exactly unitary for any choice of the truncation order $o$. The S-PPE scheme is exactly unitary only for $o=2$. $\dagger$: We were not able to derive an analytic expression connecting the order of the diagonally-exact split schemes. Semi-analytically we could determine that for {S-LPE-$o$} the first few substeps and orders are $(s,o) = (1,1), (2,2), (4,3)$ and for {S-PPE-$o$} they are $(s,o) = (1,2), (2,3), (3,4)$.
}
\label{tab:UV}
\end{table}

\subsection{Trotter decomposition}
\label{sec:trotter}

A prototypical product series decomposition of a unitary operator is the Suzuki–Trotter decomposition \cite{Hatano2005}.
In this approach, $\hat\Lambda$ is expressed as a sum of local terms, and the exponential of the sum is approximated as a product of the exponentials of the individual terms. 
The decomposition of $\hat\Lambda$ is not unique and can be tailored to the specifics of the problem to maximize computational efficiency \cite{MllerHermes2012}.

While Suzuki–Trotter decompositions can be extended to arbitrary order, in practice, their use in NQS is typically limited to second order in $\dd t$, as seen in Refs.~\cite{Sinibaldi2023}~and~\cite{Poletti2024}. 
The key advantage of this approach is that it approximates the operator’s action in a manner where state changes are highly localized, which simplifies the individual optimization problems in \cref{eqn:optimization_UV} and tends to improve their convergence.

For all their benefits, Suzuki–Trotter decompositions face two main limitations: the truncation to second order in $\dd t$ and the scaling of the number of optimizations with the system size, both of which hinder computational efficiency in large-scale applications.

\begin{figure*}[t!]
\center
\hspace*{-1.em}
\includegraphics[width=1.02\linewidth]{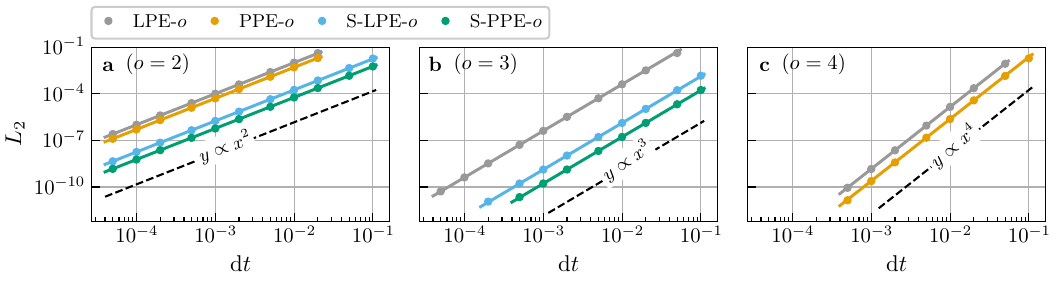}
\caption{
Global truncation error accumulated at time $t=1$ for LPE and PPE integrators, and their split counterparts, of orders $o=2$ (a), $o=3$ (b), and $o=4$ (c). 
The evolution is carried out under the Hamiltonian in \cref{eqn:TFIM} simulating on a $4 \times 4$ lattice the same quench dynamics investigated in \cref{sec:results_big}.
The accumulated error for an integrator of order $o$ scales as $\dd t^o$, with the local error scaling as $\dd t^{o+1}$. 
Both split and non-split integrators are shown, demonstrating the power of the splitting in reducing the prefactor of the error.
}
\label{fig:l2_integration_schemes}
\end{figure*}

\subsection{Taylor decomposition}
\label{sec:Taylor}

Another relevant decomposition to consider is the order-$o$ Taylor approximation of the propagator. Its expression
\begin{equation}
\label{eqn:taylor_expansion_exponential_map}
    e^{\hat \Lambda \dd t} = \sum_{k=0}^o \frac{(\hat\Lambda \dd t)^k }{k!} + \order{\dd t^{o+1}},
\end{equation} 
can be viewed within the framework of \cref{eqn:optimization_UV_expanded} as a single optimization problem ($s=1$) of the form in \cref{eqn:optimization_UV} with $\hat V_1 = \I $ and $\hat U_1 =\sum_{k=0}^o (\hat\Lambda \dd t)^k/k!$.
This approach satisfies condition (i) by matching the desired order of accuracy in $\dd t$, but it fails to meet condition (ii) related to computational efficiency.
Indeed, computing $\hat U_1 \ket{\psi_\theta}$ requires summing over all $N_c$ connected elements of $\hat U_1$. As the order $o$ increases, higher powers of $\hat \Lambda$ introduce a growing number of such connected elements, with $N_c \sim \order{N^o}$. 
This approach incurs a computational cost that scales exponentially in $o$, making it unviable at higher orders.
Furthermore, this approach cannot reasonably be  used in continuous space simulations, where the square of the Laplacian cannot be efficiently estimated.

\subsection{Linear Product Expansion (LPE)}
\label{sec:LPE}
We now introduce a new scheme circumventing the limitations of the previous approaches. 
We consider the linear operator $\hat T_{a} \equiv \I + a \hat \Lambda\dd t$ with $a\in\mathbb{C}$ and expand the evolution operator as a series of products of such terms,  
\begin{equation}
\label{eqn:LPE}
    e^{\hat\Lambda \dd t} = \prod_{i=1}^s \hat{T}_{a_i} + \mathcal{O}(\dd t^{s + 1}).
\end{equation}
The expansion is accurate to order $o(s) = s$. 
The complex-valued coefficients $a_i$ are determined semi-analitically by matching both sides of \cref{eqn:LPE} order by order up to $o(s)$.
For the second-order scheme ($s=2$), for example, we find $a_1=(1-i)/2$ and $a_2=(1+i)/2$.
Further details on the scheme and on the computation of the coefficients are provided in \cref{app:integration_scheme_details}.
Tabulated values for $a_i$ can be found in \cref{tab:LPE}.
We call this method LPE-$o$ for \emph{Linear product expansion}, where $o$ is the order of the method, related to the number of sub-steps $s$ by $o(s)=s$. 
Each substep corresponds to an optimization problem with $\hat V_k = \I$, and $\hat U_k = \hat T_{a_k}$.
The advantage of LPE over Taylor schemes (\cref{sec:Taylor}) is that the former defines an $s$-substep scheme of order $s$ with a step-complexity that is linear in $N$. This greatly outperforms Runge-Kutta style expansions for this particular application, enabling scaling to arbitrary order in $\dd t$, simultaneously satisfying conditions (i) and (ii).

It was noted in Ref.~\cite{Nys2024} that the coefficients $a_i$ of this expansion are the complex roots of the order $s$ Taylor polynomial. 
While this is a handy trick to compute them numerically, this approach is not general enough to represent the multi-operator expansions that we will analyze below in \cref{sec:PPE,sec:split-schemes}.

\subsection{Padé Product Expansion (PPE)}
\label{sec:PPE}

We now present schemes reaching order $2s$ with only $s$ sub-steps of marginally-increased complexity.
We consider the operator $\hat P_{b, a} \equiv \hat T_{b}^{-1} \hat T_{a}$ and expand the evolution operator as a series of products of such terms,
\begin{equation}
\label{eqn:PPE}
    e^{\hat \Lambda \dd t} 
    = \prod_{i=1}^s \hat P_{b_i,a_i} + \order{\dd t^{2s+1}}    .
\end{equation}
The expansion is accurate to order $o(s) = 2s$.
We call this method PPE-$s$ for \emph{Padé product expansion}, because the single term $\hat P_{b, a}$ corresponds to a (1,1) Padé approximation \cite{Moler2003}. 
The scheme is explicitly constructed to take advantage of the structure of the optimization problems in \cref{eqn:optimization_UV}, exploiting the presence of a matrix inverse in the expansion \eqref{eqn:optimization_UV_expanded}. 
While atypical for standard ODE integrators, as it would introduce an unjustified overhead, in our case this simply translates into optimizations where $\hat V_i = \hat T_{b_i}$, and $\hat U_i = \hat T_{a_i}$.
The coefficients $a_i$ and $b_i$ are again obtained by matching both sides of \cref{eqn:PPE} up to order $o$ (see \cref{app:integration_scheme_details}).
A remarkable property of PPE schemes is that they preserve the unitarity of the evolution map making them ideal candidates for the simulation of quantum circuits (proof in \cref{app:unitarity}). 
A comparison of PPE and LPE schemes of different orders is provided in \cref{fig:l2_integration_schemes} where we show the $L_2$-distance between the exact solution and the solution obtained from state-vector simulations of \cref{eqn:optimization_UV_expanded}.

\subsection{Diagonally-exact split schemes}
\label{sec:split-schemes}
Learning the parameter change connecting two states via state compression is challenging.
Restricting the problem to scenarios where state changes are highly localized has proven effective in mitigating this issue, easing optimization and generally improving convergence \cite{Sinibaldi2023}. 
This simplification, however, usually comes at the cost of an unfavourable scaling of the number of optimizations, typically scaling with $N$ (c.f.~\cref{sec:trotter}).

We propose to reduce the complexity of the nonlinear optimizations by splitting $\hat \Lambda$
as $\hat \Lambda = \hat X + \hat Z$, where $\hat Z$ acts diagonally in the computational basis\footnote{Acting diagonally in the computational basis means that $\bra{a}\hat{Z}\ket{b} \propto \delta_{ab}$.} while $\hat X$ is an off-diagonal matrix. 
The rationale will be to extract the diagonal operators which can be applied exactly to any variational parametrization.

We consider the decomposition
\begin{equation}
\label{eqn:LPE_S}
    e^{\hat\Lambda \dd t} =  \prod_{i=1}^s S^{(T)}_{\alpha_i,a_i} + \order{\dd t^{o(s) + 1}},
\end{equation}
where 
\begin{equation}
\label{eqn:LPE_S_unit}
    S^{(T)}_{\alpha,a} =  \qty(\I + a \hat X \dd t) \,e^{\alpha \hat Z \dd t} 
    \quad \text{with}\quad \alpha, a\in\mathbb{C}.
\end{equation}
The expansion is accurate to order $o(s)$ but the analytical dependence on $s$ is not straightforward to derive. 
For the lowest orders we find $o(1) = 1$, $o(2) = 2$, and $o(4) = 3$.
Each term in the product consists in principle of two optimizations: the first compressing the off-diagonal transformation, the second the diagonal one. The advantage of this decomposition is that the latter optimization can be performed exactly with negligible computational effort (see \cref{app:diag_ops}). 

The same approach can be extended to Padé-like schemes by substituting in \cref{eqn:LPE_S} the term
\begin{equation}
    S^{(P)}_{\alpha,b,a} =   
    \qty(\I + b \hat X \dd t)^{-1}  
    \qty(\I + a \hat X \dd t)
    e^{\alpha \hat Z \dd t},
\end{equation}
with $b,\alpha, a\in\mathbb{C}$. 
All coefficients are again obtained semi-analytically (see \cref{app:integration_scheme_details}).
Though we do not have an explicit expression for the order $o(s)$ resulting from an $s$-substep expansion of this form, we find for the shallower schemes $o(1) = 2$, $o(2) = 3$, and $o(3) = 4$.

We will refer to these schemes as \emph{split LPE} (S-LPE) and \emph{split PPE} (S-PPE), respectively. 
They have the two advantages. 
First, they reduce the complexity of the optimizations in \cref{eqn:optimization_UV}, and second, in many cases they reduce the prefactor of the error of their reciprocal non-split counterpart of the same order, as evidenced in \cref{fig:l2_integration_schemes}.

\section{State compression Optimizations}
\label{sec:fidelity}

In \cref{sec:integrators}, we introduced various schemes for efficiently decomposing unitary dynamics into a sequence of minimization problems, intentionally omitting the specific form of the loss function. The sole requirement was that the loss function quantify the dissimilarity between quantum states, reaching its minimum when the states perfectly match.

This section is organized as follows:
In \cref{sec:fidelity-specific}, we discuss a specific choice for the loss function — the infidelity — and explore some of its general properties.
In \cref{sec:fidelity-estimators}, we examine the properties of different stochastic estimators for the fidelity and its gradient, identifying those that are most stable and perform best in practice.
Finally, in \cref{sec:ngd} we review results on natural gradient optimization and introduce, in \cref{sec:auto_damping}, an automatic regularization strategy that simplifies hyperparameter tuning for these simulations.

\subsection{The generic fidelity optimization problem}
\label{sec:fidelity-specific}

A common measure of similarity between two pure quantum states is the fidelity, defined as  \cite{Sinibaldi2023, Havlicek2023, Medvidovi2021}
\begin{equation}
    \label{eqn:fidelity}
    \F\qty(\hat V \ket{\psi}, \hat U \ket{\phi}) 
    = \frac{\bra{\psi}\hat V^\dagger \hat U\ket{\phi}\bra{\phi}\hat U^\dagger \hat V\ket{\psi}}{\bra{\psi}\hat V^\dagger \hat V\ket{\psi}\bra{\phi}\hat U^\dagger \hat U\ket{\phi}},
\end{equation} 
where the operators $\hat U$ and $\hat V$ are included as in \cref{eqn:optimization_UV}.
In this work, we adopt the infidelity $\LL \equiv \mathcal{I} = 1 - \F$ as the loss function for each substep of \cref{eqn:optimization_UV}, though alternative, less physically motivated metrics are also possible \cite{Gutirrez2022, Ledinauskas2023}.

The choice of operators $\hat U$ and $\hat V$ in \cref{eqn:optimization_UV} significantly influences the numerical complexity of the optimization.
In Trotter-like decompositions, $\hat U$ and $\hat V$ act locally on a few particles, inducing minor changes to the wavefunctions. 
These localized transformations yield smoother optimization landscapes, which are effectively navigated using standard stochastic gradient methods, such as Adam \cite{Adam}.
This likely explains the findings of Sinibaldi et al.~\cite{Sinibaldi2023}, where natural gradient optimization provided no significant improvements in performance.

In contrast, Taylor, LPE, and PPE schemes encode global transformations in $\hat U$ and $\hat V$, causing the target and variational wavefunctions to diverge substantially.
These global transformations result in more complex optimization landscapes, where we find standard stochastic gradient descent methods inadequate (not shown).
This issue is further exacerbated when training deep neural network architectures.
To address these challenges, we employ parameterization-invariant optimization strategies, specifically natural gradient descent (NGD) which we discuss in \cref{sec:ngd}.
NGD adjusts the optimization path based on the geometry of the output space, enabling more efficient convergence in complex, high-dimensional problems.
We find that NGD plays a critical role in improving both convergence and the overall efficiency of our proposed schemes (c.f.~\cref{app:numerically_exact}).

\subsection{Stochastic estimators}
\label{sec:fidelity-estimators}
Estimators of the fidelity [\cref{eqn:fidelity}] take the general form
\begin{equation}
\label{eqn:generic_mc}
    \F(\ket\psi, \ket\phi) = \E_{(x,y)\sim\chi}[f(x,y)],
\end{equation}
for a suitable choice of random variable $\sigma$, sampling distribution $\chi(\sigma)$, and local estimator $f(\sigma)$. 
While this expression is exact, its direct evaluation incurs exponential scaling with system size $N$, rendering it computationally infeasible.
Equation~\eqref{eqn:generic_mc} is thus generally approximated by its sample mean 
\begin{equation}
    \bar f_{N_s} = \frac{1}{N_s} \sum_{i=1}^{N_s} f(x_i, y_i) \quad \text{with} \quad (x_i, y_i) \sim \chi,
\end{equation} 
evaluated over a number of samples $N_s$ scaling polynomially in system size.

Importantly, different stochastic estimators, while sharing identical expectation values in the limit of infinite samples,  can exhibit significantly different behavior over finite sample sizes.
Though some attention has been given to characterizing the properties of different fidelity estimators \cite{Sinibaldi2023},
a systematic study of the variance properties of the estimators of the gradient is still lacking. 
As an accurate estimation of the gradient is crucial for driving fidelity optimization to convergence, this represents a central issue for quantum many-body dynamics.

In \cref{sec:fidelity-fidelity}  we extend the analysis of Sinibaldi et al.~\cite{Sinibaldi2023}, further improving on the variance properties of the fidelity estimator.
In \cref{sec:gradient} we examine the properties of various gradient estimators, decoupling their analysis from that of the fidelity itself. 
Our findings are summarized in \cref{tab:fidelity_estimators,tab:gradient_estimators}, respectively.

\begin{table*}[htb]
    \centering
        \begin{minipage}[t]{0.45\textwidth} 
        \centering
        \setlength{\tabcolsep}{4pt}
        \begin{tabular}{c c c c c }
            Name & Ref. & Eq. & +CV & +RW \\[0.25em]
            \hline \rule{0pt}{1.25em}
            smc$\,$ & \cite{Sinibaldi2023}, \ref{sec:fidelity-fidelity} & \eqref{eqn:smc} & \eqref{eqn:smc_cv} & \eqref{eqn:smc_rw} \\[0.2cm]
            cmc & \cite{Poletti2024, Medvidovi2021, Carleo2018}, \ref{sec:fidelity-fidelity} & \eqref{eqn:cmc} & \eqref{eqn:cmc_cv}  & \eqref{eqn:cmc_rw}
            \\[0.2cm]
            $\quad$
        \end{tabular}
        \caption{List of stochastic fidelity estimators, their definition, and their expression with control variables (CV) and reweighting (RW).
        +CV~links to the expressions of the estimators with control variables.
        +RW~links to the expressions allowing sampling from the Born distribution of $\ket{\phi}$ rather than $\hat U\ket{\phi}$ when linear transformations are applied to the target or variational state. 
        Figure~\ref{fig:fidelity_estimator} shows that both estimators perform similarly, and need CV to give accurate values.
        }
        \label{tab:fidelity_estimators}
    \end{minipage}
    \hfill 
    \begin{minipage}[t]{0.52\textwidth} 
        \centering
        \setlength{\tabcolsep}{3.pt}
        \begin{tabular}{c c c c c c }
            Name & Ref. & Eq. & +RW & NTK & Stability \\[0.25em]
            \hline \rule{0pt}{1.25em}
            $\nabla$cmc\phantom{-}$\,\,$ & \cite{Poletti2024, Medvidovi2021, Carleo2018}, \ref{sec:gradient} & \eqref{eqn:grad_cmc}  & \eqref{eqn:cmc_rw} & \checkmark & High\\[0.2cm]
            $\nabla$smc\phantom{-} & \ref{sec:gradient} & \eqref{eqn:grad_smc} & \eqref{eqn:smc_rw} & \checkmark & Medium \\[0.2cm]
            $\nabla$cv-smc & \cite{Sinibaldi2023, Nys2024} & \eqref{eqn:grad_cv} & \eqref{eqn:smc_rw} & \xmark & Low
        \end{tabular}
        \caption{List of estimators for the gradient of the infidelity. The NTK column lists the equation to implement the natural gradient estimator in the limit of large number of parameters. Not all estimators can be expressed that way. 
        +RW~links to the expressions allowing sampling from the Born distribution of $\ket{\psi}$ rather than $\hat V\ket{\psi}$ when linear transformations are applied to the target or variational state. 
        The stability score is determined by the empirical results discussed in \cref{fig:state_matching}.
        \label{tab:gradient_estimators}
        }
    \end{minipage}%
\end{table*}

\subsubsection{Fidelity}
\label{sec:fidelity-fidelity}
\begin{figure}[t]
\center
\includegraphics[width=\columnwidth]{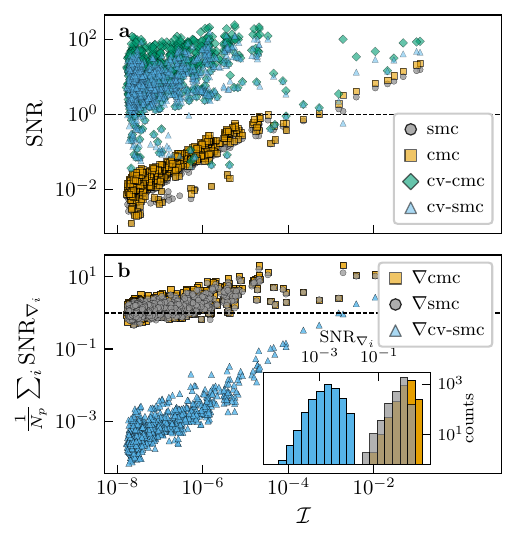}
\caption{
(a) Signal-to-noise ratio [\cref{eqn:SNR}] of the infidelity as a function of its exact value for the estimators given in Eqs.~\eqref{eqn:smc} (smc), \eqref{eqn:cmc} (cmc), \eqref{eqn:smc_cv} (cv-smc), and \eqref{eqn:cmc_cv} (cv-cmc).
For all control-variate-enhanced estimators, the control coefficient is set to the asymptotically optimal value $c=-1/2$. 
(b) Component-averaged signal-to-noise ratio of the infidelity gradient estimators in Eqs.~\eqref{eqn:grad_cmc} ($\nabla$cmc), \eqref{eqn:grad_smc} ($\nabla$smc), and \eqref{eqn:grad_cv} ($\nabla$cv-smc).
The inset shows the distribution of the component-wise signal-to-noise ratio at the 300-th iterate.
All estimators are evaluated on the same states, specifically those visited along the curve with lowest final infidelity among those reported in \cref{fig:state_matching}(a). 
Parameters: 
$N=16$, $J=1$, $h=h_c/10$,
$N_s = 2^{12}$, $n_{\rm iter}=10^{3}$, $\alpha = 0.27$, $\lambda = 3.17\times10^{-7}$, and $\bm\Theta_{\rm CNN} = (10,10,10,10;3)$.
}
\label{fig:fidelity_estimator}
\end{figure}

\begin{figure*}[!t]
\center
\hspace*{-1.3em}
\includegraphics[width=1.04\linewidth]{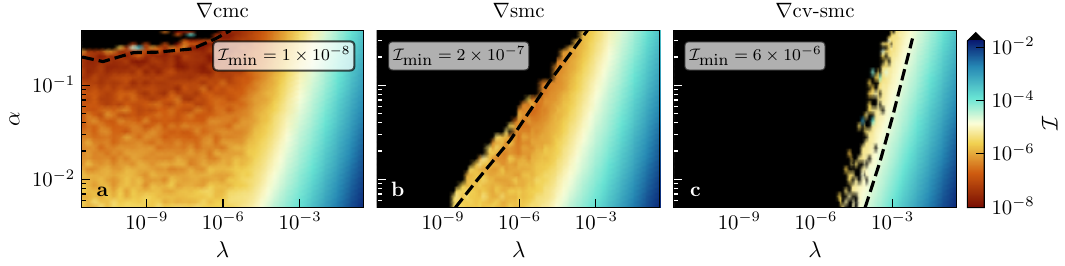}
\caption{
Stability diagram for the (a) $\nabla$cmc [\cref{eqn:grad_cmc}], (b) $\nabla$smc [\cref{eqn:grad_smc}], and (c) $\nabla$cv-smc [\cref{eqn:grad_cv}] gradient estimators as a function of the regularization coefficient $\lambda$ and learning rate $\alpha$.
Stability is quantified by the lowest infidelity (color scale) achieved across an ensemble of 10 optimizations with identical hyperparameters, identical initial parameters, but different random seeds for the Monte Carlo sampling.
Comparisons are made for a fixed computational cost as all optimizations consist of $n_{\rm iter}=10^3$ iterations.
Analogous conclusions can be drawn from the stability diagrams where $n_{\rm iter}/\alpha = \rm{const.}$ (not shown).
Regions in black correspond to $(\alpha, \lambda)$ combinations where all simulations diverged, indicating instability. Dashed black lines outline areas where more than $50\%$ of simulations diverged, providing a second visual stability threshold supporting the stability summary in \cref{tab:gradient_estimators}.
The inset in each panel shows the minimal infidelity $\mathcal{I}_{\rm min}$ achieved
within the stable region (fewer than $50\%$ of simulations diverged). 
For reliability and consistency, regardless of the chosen gradient estimator, we display the exact fidelity evaluated in full-summation.
While the reference state wave function is exactly encoded into the parameters of a log-state-vector ansatz, the variational state is approximated with a complex CNN architecture.
Parameters: $N=16$, $J=1$, $h=h_c/10$, $N_s=2^{12}$, $n_{\rm{iter}}=10^3$, and $\bm\Theta_{\rm{CNN}} = (10,10,10,10;3)$.
}
\label{fig:state_matching}
\end{figure*}

As explained in Ref.~\cite{Havlicek2023}, the natural choice for the fidelity estimator in \cref{eqn:generic_mc} is the \emph{simple Monte Carlo} (smc) estimator defined by
\begin{equation}
\label{eqn:smc}
\begin{aligned}
&\chi(x,y) =\pi_\psi(x) \pi_\phi(y) \equiv \pi(x,y),\\
&f(x,y) = \frac{\phi(x)}{\psi(x)} \frac{\psi(y)}{\phi(y)} \equiv A(x,y),
\end{aligned}
\end{equation}
where the samples $\sigma = (x,y)$ are drawn from the joint Born distribution $\pi(x,y)$ of the two states. 
The estimator of the infidelity is $1-f(x,y)$.  
Since $\pi(x,y)$ is separable, sampling can be carried out independently over the Born distributions $\pi_\psi(x) = |\psi(x)|^2/\braket{\psi}$ and $\pi_\phi(y) = |\phi(y)|^2/\braket{\phi}$.
While Ref.~\cite{Havlicek2023} demonstrated that the variance of this estimator vanishes as $\mathcal{I}\rightarrow 0$, this does not guarantee we will be able to resolve arbitrarily small values of $\mathcal{I}$ from MC noise. Our ability to do so is measured by the signal to noise ratio (SNR) of the infidelity estimator, defined as
\begin{equation}
    \label{eqn:SNR} \frac{\operatorname{SNR}}{\sqrt{N_s}}
   =
   \frac{\lvert 1-\bar f_{N_s}\rvert}
        {\sqrt{\dfrac{1}{N_s-1}\displaystyle\sum_{i=1}^{N_s}
               \bigl\lvert f(\sigma_i)-\bar f_{N_s}\bigr\rvert^{2}}}.
\end{equation}
Reference~\cite{Sinibaldi2023} shows that the SNR of the smc estimator vanishes for $\mathcal{I}\to0$ making it an unreliable indicator of progress in state compression problems \cite{Sinibaldi2023}.
In \cref{fig:fidelity_estimator}(a), we provide evidence of this by displaying the infidelity estimator's SNR as a function of the exact infidelity recorded along the best optimization amongst those presented in \cref{fig:state_matching}(a).

To address this issue, Sinibaldi et al.~\cite{Sinibaldi2023} leveraged the identity $\E_\pi[|A(x,y)|^2] = 1$ to construct the \emph{control-variate-enhanced smc} (cv-smc) estimator
\begin{equation}
\label{eqn:smc_cv}
\begin{aligned}
    &\chi(x,y) = \pi(x,y)\\
    &f(x,y) = \Re{A(x,y)} + c \qty( \abs{A(x,y)}^2 - 1) \equiv F(x,y).
\end{aligned}
\end{equation}
As shown in \cref{fig:fidelity_estimator}(a), the analytical control variable $|A(x,y)|^2$ allows for drastic variance reduction upon appropriate tuning of the control parameter $c$\footnote{As $\ket{\psi} \to \ket{\phi}$, the optimal choice for $c$ has been shown to converge to $c = -1/2$.}.

An alternative variance reduction technique, readily applicable to \cref{eqn:smc}, is Rao-Blackwellization.
In its simplest form, Rao-Blackwellization reduces the variance of an estimator by replacing it with the conditional expectation with respect to a subset of its variables \cite{Rubinstein2016, ranganath2014black}.
For the particular case of a factorizable distribution such as $\chi(x,y)=\pi_\psi(x)\pi_\phi(y)$, Rao-Blackwellization amounts to marginalizing over the target state's distribution. 
In the following, we will refer to the Rao-Blackwellized smc estimator as the \emph{conditional Monte Carlo} (cmc) estimator, expressed as (details in \cref{app:grad_variance})
\begin{equation}
\label{eqn:cmc}
\begin{aligned}
&\chi(x,y) = \chi(x) =\pi_\psi(x), \\[0.1cm]
&f(x) = \E[A(x,y)|x] =A_x(x)\,\,\E_{y\sim\pi_\phi}\qty[A_y(y)] \equiv H_{\rm loc}(x),\\[-1.2em]
&\ 
\end{aligned}
\vspace*{0.5em}
\end{equation}
where $A(x,y) = A_x(x) A_y(y)$, $A_x(x) = \phi(x)/\psi(x)$, and $A_y(y) = \psi(y)/\phi(y)$.

We remark that the computational cost of \cref{eqn:smc,eqn:cmc} is equivalent.
Indeed, they use the same set of samples $\{(x_i, y_i)\}_{i=1}^{N_s}$ and evaluate $\psi$ and $\phi$ over all of them. 
The difference is that while \cref{eqn:smc} sums over the diagonal pairs $(x_i, y_i)$ alone, the sample average in \cref{eqn:cmc} includes all cross terms $(x_i, y_j)$.

Though the connection to conditional Monte Carlo was not drawn, the cmc estimator was used in Refs.~\cite{Poletti2024, Medvidovi2021, Carleo2018}. We demonstrate in \cref{fig:fidelity_estimator}(a) that just like the smc estimator, the cmc estimator also suffers from a vanishing SNR in the limit of $\mathcal{I}\to 0$.

In the following, we combine the two variance reduction techniques detailed above, control variates and Rao-Blackwellization, to produce the \emph{control-variate-enhanced cmc} (cv-cmc) estimator (derivation in \cref{app:cv_cmc})
\begin{equation}
\label{eqn:cmc_cv}
\begin{aligned}
    &\chi(x) = \pi_\psi(x) \\[0.2cm]
    &f(x) 
    = \E[F(x,y)|x]\\[-0.1cm]
    &= \Re{H_{\rm loc}(x)} + c\qty(\qty|A_x(x)|^2\, \E_{y\sim\pi_\phi}\bigl[\qty|A_y(y)|^2\bigr]-1).
\end{aligned}
\end{equation}
As a direct consequence of the Rao-Blackwell theorem, the cv-cmc estimator displays an SNR performance superior to all others, as illustrated in \cref{fig:fidelity_estimator}(a). 
We adopt this estimator in all simulations presented below.

We finally remark that p-tVMC might require evaluating $\F(\hat V\ket{\psi}, \hat U\ket{\phi})$ thereby implicating the necessity of sampling from the Born distributions of the transformed states $\hat V\ket{\psi}$ and $\hat U\ket{\phi}$. This introduces an additional overhead which can however be sidestepped by means of importance sampling. We show this in \cref{app:reweighting}.

\subsubsection{Gradient}
\label{sec:gradient}
While the preceding section provides compelling evidence that control variates are essential for accurate fidelity estimation, it does not address how this impacts the estimation of the gradient.
In gradient-based optimization problems, such as \cref{eqn:optimization_UV}, managing the variance of gradient estimates is crucial for achieving stable and efficient convergence.
Historically, however, infidelity-optimization strategies have overlooked this aspect, simply adopting the gradient estimator obtained from automatic differentiation of the chosen fidelity estimator according to:
\begin{equation}
\begin{aligned}
\label{eqn:grad_general}
    &\grad \F = \grad \E_\chi[f(\sigma)] = \E_\chi[g(\sigma)],\\[0.1cm]
    &g(\sigma) = 2\Re{\Delta J(\sigma)}f(\sigma) + \grad f(\sigma).
\end{aligned}
\end{equation}
While for an infinite number of samples, this expression is of course valid, when approximating the average with the sample mean $\bar g_{N_s}$, there is no guarantee that the selected gradient estimator will perform well.
Indeed, as we show below, good variance properties of the fidelity estimator do not necessarily translate, under \cref{eqn:grad_general}, into good variance properties of the gradient estimator.
As a result, stable low-variance estimators for both fidelity and gradient have, until now, never been simultaneously employed. 
Specifically, while Refs.~\cite{Sinibaldi2023, Nys2024} employ a control-variate-enhanced estimator for the fidelity, their choice of gradient estimator suffers from a vanishing SNR, rendering it ineffective for optimization. 
Conversely, the gradient estimator adopted in Refs.~\cite{Medvidovi2021, Carleo2018, Poletti2024} has good variance properties, while the fidelity estimator exhibits a vanishing SNR. In the latter case, we observe this often leads to an underestimation of the quality of the results. 

In support to this point, we explicitly evaluate the gradient estimators obtained from the smc, cmc, and cv-smc fidelity estimators, and compare their performance.
We refer to such estimators as $\nabla$smc, $\nabla$cmc, and $\nabla$cv-smc, respectively.
For the smc estimator [\cref{eqn:smc}], automatic differentiation of \cref{eqn:generic_mc} yields: 
\begin{equation}
\label{eqn:grad_smc}
\begin{aligned}
    &\chi(x,y) = \pi(x,y), \\
    &g(x,y) = 2\Re{\Delta J(x) A(x,y)^*},
\end{aligned}
\end{equation}
with $\Delta J(x)$ defined in \cref{eqn:QGT}.
Applying the same procedure to the cmc estimator [\cref{eqn:cmc}] results in: 
\begin{equation}
\label{eqn:grad_cmc}
\begin{aligned}
    &\chi(x,y) = \chi(x) =\pi_\psi(x), \\
    &g(x,y) = 2\Re{\Delta J(x) H_{\rm loc}(x)^*}.
\end{aligned}
\end{equation}
This choice underlies the investigations in Refs.~\cite{Poletti2024, Medvidovi2021, Carleo2018}. 
Notably, the expression in \cref{eqn:grad_cmc} can be derived directly via Rao-Blackwellization of \cref{eqn:grad_smc}, guaranteeing a reduction in variance.
Finally, the gradient obtained by differentiation of the cv-smc estimator [\cref{eqn:smc_cv}] reads: 
\begin{equation}
\label{eqn:grad_cv}
\begin{aligned}
&\chi(x,y) = \pi(x,y), \\
&g(x,y) = \Re\Big\{2 \Delta F(x,y) J(x) \,+\\[-0.2em]
&\qquad\qquad+ \qty(A(x,y) + 2c \abs{A(x,y)}^2) [J(y) - J(x)]\Big\},
\end{aligned}    
\end{equation}
with $F(x,y)$ as defined in \cref{eqn:smc_cv}.
This expression underlies the studies in Refs.~\cite{Sinibaldi2023, Nys2024}.

To evaluate the performance of these gradient estimators, we first compare their SNRs.
Since $\grad\F\in\mathbb{R}^{N_p}$, we compute the SNR of each component of the gradient vector.
In \cref{fig:fidelity_estimator}(b), we present the component-averaged SNR as a function of the true infidelity recorded during the best optimization run of \cref{fig:state_matching}(a).
The inset shows a distribution of per-component SNR values.
Interestingly, while the cv-smc estimator achieves superior SNR for fidelity estimation, this advantage is reversed when estimating the gradient. Specifically, the SNR of the cv-smc gradient estimator declines rapidly as the simulation progresses toward convergence. In contrast, gradient estimators derived from the bare smc and cmc fidelity estimators maintain an SNR of $\order{1}$ as convergence is approached.
We attribute this disparity to the covariance structure present in \cref{eqn:grad_smc,eqn:grad_cmc}, but absent in \cref{eqn:grad_cv}.
This covariance structure introduces sample-specific control variates that dynamically adapt to the non-stationary nature of the variables, significantly reducing variance and enhancing estimator stability (c.f.~\cref{app:grad_variance}).

To further characterize the performance of the gradient estimators, we examine their ability to drive infidelity optimizations to convergence. 
As a benchmark, we consider a state-matching problem in which we optimize the variational state to match a given reference state.
As a physical reference we take $\ket{\psi(t)} = \operatorname{exp}(-i\hat H t)\ket{\rightarrow\ldots\rightarrow}$ with $\H$ the transverse field Ising Hamiltonian in \cref{eqn:TFIM}.
In general, we observe that states at shorter times are easier to learn, with smaller differences in the results produced by the various gradient estimators. Conversely, states at longer times are significantly more challenging to accurately match. 
It is important to emphasize that this state-matching setup is inherently more demanding than the dynamics itself. 
In the simulation of dynamics, the state at each timestep is initialized from the preceding timestep, offering a more informed starting point closer to the target state. 
By contrast, the state-matching problem begins in general from a random state, making the optimization considerably more difficult.

In \cref{fig:state_matching} we present the accuracies with which we were able to recover the state at $Jt=0.5$, a time point that we consider \textit{relatively challenging} and representative of the general trends observed.
Convergence is quantified by the final optimization infidelity (colormap) which we display as a function of the NGD regularization parameter $\lambda$ (see \cref{sec:ngd}) and learning rate $\alpha$ for the different gradient estimators detailed above. 
For each pair $(\alpha, \lambda)$ we consider $10$ different replicas with different sampling histories. Black regions in the figure represent instability zones, where all replicas diverged, indicating poor stability of the method for the selected hyperparameter combination.

Overall, our results indicate that the $\nabla$cmc gradient estimator consistently delivers the best convergence and exhibits the highest stability among all the gradient estimators examined in this study and reported in the available literature. 
We rely on this estimator for all optimizations presented in \cref{sec:results}.

\subsection{Natural Gradient Descent}
\label{sec:ngd}
Throughout this work we use NGD to solve the optimizations in \cref{eqn:optimization_UV}.
In its simplest implementation, given the current iterate $\theta_k \in \mathbb{R}^{N_p}$, NGD proposes a new iterate $\theta_{k+1} = \theta_k - \alpha_k\, \delta_0$ , where $\alpha_k$ is a schedule of learning rates, and $\delta_0$  the natural gradient at the current iterate. 
$\delta_0$ is determined by minimizing a local quadratic model $M(\delta)$ of the objective, formed using gradient and curvature information at $\theta_k$ \cite{Martens2012, Martens2020, Wright2006}. Formally,
\begin{equation}
\begin{aligned}
\label{eqn:quadratic_model}
    M(\delta) &= \LL(\theta_k) + \grad \LL(\theta_k)^\top \delta + \frac{1}{2} \delta^\top \bm B\, \delta,\\
    \delta_0 &= \argmin{\delta}\,\, M(\delta) = \bm B^{-1}\grad \LL,
\end{aligned}
\end{equation}
where $\bm B$ is a symmetric positive definite \emph{curvature matrix}. 
This matrix is taken to be the Fisher information matrix when modeling probability distributions, or the Fubini-Study (FS) metric tensor for quantum states \cite{Stokes2023, Cheng2010, Stokes2020}. 
The latter is a Gram matrix estimated as\footnote{
The expression of the FS tensor in \cref{eqn:QGT} applies to a generic variational state $\ket{\psi}$ and is used to precondition the gradient of the objective as $\bm{S}^{-1} \grad \LL(\ket{\psi}, \ket{\phi})$, where $\ket{\phi}$ is an arbitrary target state. This formulation remains valid if the state $\ket{\psi}$ is transformed by an operator $\hat{V}$, with the corresponding replacement $\psi \to \tilde{\psi}$.
}
\begin{equation}
    \label{eqn:QGT}
    \bm S = \operatorname{Re}\qty{\E_{\pi_\psi}\!\bigl[\Delta J^*(x) \Delta J(x)^\top\bigl]} = \bm X^{\!\top}\! \bm X \in \mathbb{R}^{N_p\times N_p},
\end{equation}
where $J(x) = \grad\log\psi(x) \in \mathbb{C}^{N_p}$ is the quantum score function, $\Delta J(x) = J(x) - \E_{\pi_\psi}[J(x)]$, and 
\begin{equation}
\label{eqn:X}
    \bm X = 
    \begin{pmatrix}
    \Re{\bm O} \\[0.5ex]
    \Im{\bm O}
    \end{pmatrix} \in \mathbb{R}^{2N_s\times N_p},
\end{equation}
with 
\begin{equation}
    \bm O = \frac{1}{\sqrt{N_s}}
    \begin{pmatrix}
    \Delta J(x_1)^\top \\
    \vdots\\
    \Delta J(x_{N_s})^\top
    \end{pmatrix}\in \mathbb C^{N_s\times N_p}.
\end{equation}
Here, the samples $\{x_1, \ldots, x_{N_s}\}$ are taken from $\pi_\psi$.
In essence, NGD is a way of implementing steepest descent in state space rather than parameter space. 
In practice, however, NGD still operates in parameter space, computing directions in the space of distributions and translating them back to parameter space before implementing the step \cite{Martens2012, Martens2020}.
As \cref{eqn:quadratic_model} stems from a Taylor approximation of the loss, it is important that the step in parameter space be small, for the expansion (and the update direction) to be reliable. 
As the FS tensor is often ill-conditioned or rank-deficient, this requirement is enforced by hand by adding to the objective a damping term with a regularization coefficient $\lambda$ that penalizes large moves in parameter space. 
This yields the update
\begin{equation}
\label{eqn:damping}
\begin{aligned}
    \delta_\lambda 
     &= \argmin{\delta}\,M(\delta) + \frac{\lambda}{2} \norm{\delta}^2 
     = (\bm X^\top \!\bm X + \lambda \I_{\!N_p})^{-1} \bm X^\top\! \bm \varepsilon,
\end{aligned}
\end{equation}
where we used that $\bm B = \bm S = \bm X^\top\! \bm X\in\mathbb{C}^{N_p\times N_p}$ and $\grad \LL = \bm X^\top \bm \varepsilon$ with $\bm \varepsilon\in\mathbb{R}^{2N_s}$\footnote{
Note that the identity $\grad \LL = \bm X^\top\! \bm \varepsilon$ does not hold universally for all loss functions, although it is verified for many prototypical choices, such as the mean squared error or the variational energy.
In \cref{app:gradient_derivations}, we demonstrate that the fidelity can exhibit this structure, although this is not guaranteed for all gradient estimators.}.
Alternative ways of formulating this constraint exist, such as trust-region methods \cite{Wright2006}, proximal optimization, or Tikhonov damping \cite{Martens2020}, all eventually leading to \cref{eqn:damping}.

It is important to note that evaluating the curvature matrix over a subset of configurations stochastically sampled from the full Hilbert space ensures at most a decrease of the loss over the sampled configurations. 
This does not guarantee a reduction in the overall loss or in the loss evaluated on a different set of configurations.
A desirable property of any stochastic optimization method is its ability to approach optimal (or near-optimal) behavior in the limit of infinite samples.
The expectation is that, with a sufficiently large sample size, a reduction in the loss on the sampled data will correspond to a global decrease in the loss.

The main challenge with NGD is the high computational cost of inverting $(\bm X^\top \!\bm X + \lambda \I_{\!N_p})$ in large-scale models with many parameters ($N_p\gg N_s$). 
Various approximate approaches have been proposed to address this, such as layer-wise block diagonal approximations \cite{Heskes2000, Grosse2015}, Kronecker-factored approximate curvature  \cite{Martens2015, Grosse2016}, and unit-wise approximations \cite{Ollivier2015, Amari2019}.
At the moment, the only method enabling the use of NGD in deep architectures without approximating the curvature matrix is the tangent kernel method \cite{Karakida2021, Wei2022}, recently rediscovered in the NQS community as minSR \cite{Chen2024}. 
This approach leverages a simple linear algebra identity \cite{Karakida2021, Petersen2012, Rende2024} to rewrite \cref{eqn:damping} as
\begin{equation}
    \label{eqn:damping_ntk}
    \delta_\lambda = \bm X^\top(\bm X \bm X^{\top} + \lambda \I_{2N_s})^{-1} \bm\varepsilon,    
\end{equation}
where the matrix $\bm T = \bm X \bm X^\top \in \mathbb{R}^{2N_s\times 2N_s}$ is known as the \emph{neural tangent kernel} (NTK) \cite{NTK2018, Novak2022}. 
In the limit where $N_p \gg N_s$, inverting the NTK becomes much more tractable than inverting the FS tensor, shifting the computational bottleneck on the number of parameters to the number of samples.
Importantly, the NTK formulation relies on the gradient of the objective being of the form $\grad \F = \bm X^\top\!\bm\varepsilon$.
As shown in \cref{app:gradient_derivations}, this is not a prerogative of all estimators investigated in \cref{sec:gradient}.
Specifically, while the $\nabla$smc and $\nabla$cmc estimators admit such decomposition with 
\begin{equation}
\begin{aligned}
    \bm\epsilon &=\frac{2}{\sqrt{N_s}}\bigl(\Re{A(x_1,y_1)},\ldots,\Im{A(x_1,y_1)},\ldots\bigr)^\top,\\
    \bm\epsilon &= \frac{2}{\sqrt{N_s}}\bigl(\Re{H_{\rm loc}(x_1)},\ldots,\Im{H_{\rm loc}(x_1)},\ldots\bigr)^\top,
\end{aligned}
\end{equation}
respectively, the $\nabla$cv-smc estimator does not.
This restriction prevents these estimators from being efficiently computed in the NTK framework, making them a poor choice for scaling to the deep network limit.

In addition to not supporting an NTK formulation, the absence of a form like $\grad \F = \bm X^\top \bm\varepsilon$ also precludes the use of L-curve and generalized cross-validation methods for adaptively selecting the regularization coefficient $\lambda$.
While these automatic damping strategies have been used in the literature \cite{Gacon2024}, they are less suited to the NGD problem compared to those discussed in \cref{sec:auto_damping}.

\subsubsection{Autonomous damping}
\label{sec:auto_damping}
Choosing an appropriate value for $\lambda$ is crucial for successful optimization. 
If $\lambda$ is too large, the update resembles standard gradient descent with a very small learning rate.
Conversely, if $\lambda$ is too small, updates can become excessively large, particularly in low-curvature directions, possibly leading to increasing the loss rather than decreasing it \cite{Martens2012}. 
Identifying an optimal value for $\lambda$ is nontrivial, and it is rare for a single value of $\lambda$ to be suitable across all optimization iterations.
This challenge is exacerbated in p-tVMC calculations, where numerous successive infidelity optimizations, with very different target and initial states, must be performed. 
Performing an hyperparameter search analogous to that in \cref{fig:state_matching} for every state compression is of course impractical, and an automated approach is required.
Hereunder, we introduce a custom adaptive control strategy inspired by heuristics commonly employed in the numerical optimization literature \cite{Martens2012, Martens2020, Wright2006, Martens2015, martens2010}. 

\begin{figure}[htb]
\center
\includegraphics[width=\columnwidth]{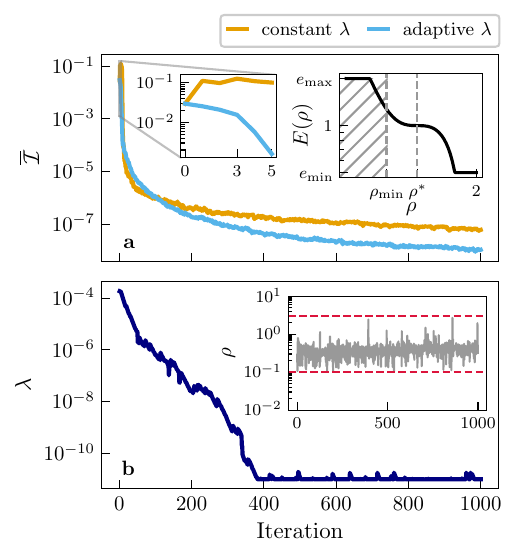}
\caption{
Performance of the autonomous damping algorithm described in \cref{eqn:ad0}.
(a) Comparison of the infidelity across the optimization. 
The infidelity is computed exactly and averaged over 10 Monte Carlo runs, with parameter updates derived from noisy estimates of the $\nabla$cmc gradient and curvature matrix.
Results are compared to the best configuration ($\alpha=0.27$, $\lambda = 3\times 10^{-7}$) from \cref{fig:state_matching}(a). 
The first inset highlights an initial increase in the loss due to a suboptimal initial choice of $\lambda$.
The second inset characterizes the behavior of the error function $E(\rho)$ in \cref{eqn:ad1}. 
Hatching is used to highlight the region $\rho<\rho_{\rm min}$ where updates are discarded and recomputed. 
(b) Evolution of the diagonal shift. The data corresponds to the replica achieving the lowest infidelity.
The inset displays the value of reduction ratio $\rho$ at each substep.
Parameters: $q=1$, $\beta_1=0.9$, $\beta_2=0.1$, $\alpha=0.05$, $\rho_{\rm max}=3$, $\rho_{\rm min}=0.1$, $\rho^*=0.35$, $\gamma_s=0.9$, $e_{\rm max}=2$, and $e_{\rm min}=0.05$.}
\label{fig:autodamping}
\end{figure}

Let $\theta_k$ be the parameters at the $k$-th iterate.
The parameters at the following iterate are $\theta_{k+1} = \theta_k + \delta\theta_k$ with 
$\delta\theta_k = -\alpha_k \delta_{\lambda_k}$.
Here, $\delta_{\lambda_k}$ is the natural gradient update, and $\alpha_k$ is the learning rate.
Under ideal conditions (accurate curvature matrix, infinite sample size, and proper regularization), the optimal learning rate $\alpha_k = 1$ maximizes the decrease of the quadratic model. However, these conditions are rarely satisfied in practice and mitigating the overfitting caused by finite-sample effects typically requires lowering $\alpha_k$ \cite{Martens2012}. We find that  $\alpha_k \gtrsim 5 \times 10^{-2}$ is a reasonable choice in most cases.

To dynamically adapt $\lambda_k$, we build on the Levenberg-Marquardt heuristic \cite{martens2010}, which relies on the reduction ratio 
\begin{equation} 
    \rho_k = \frac{\LL(\theta_k + \delta\theta_k) - \LL(\theta_k)}{M(\delta\theta_k) - \LL(\theta_k)}.
\end{equation} 
This ratio measures the accuracy of the quadratic model $M(\delta\theta_k)$ [c.f.~\cref{eqn:quadratic_model}] in predicting $\LL(\theta_k + \delta\theta_k)$.
Small $\rho_k$ ($\rho_k \ll 1$) suggests poor agreement between the model and the actual loss, indicating the need to increase $\lambda_k$.
Larger $\rho_k$ ($\rho_k \gtrsim 1$) indicates good agreement, allowing $\lambda_k$ to be reduced \footnote{
In a trust-region approach to second-order optimization problems, an increase in $\lambda_k$ corresponds to tightening the trust region. Conversely, a reduction in $\lambda_k$ corresponds to a relaxation of the trust region \cite{Wright2006}.
}.
The standard Levenberg-Marquardt algorithm aims at keeping $1/4 \leq \rho_k \leq 3/4$ by taking $\lambda_{k+1} = m_k \lambda_k$ with 
\begin{equation}
    m_{k} = 
    \begin{cases}
        3/2 & \text{if } \phantom{\,0.< }\rho_k \leq \frac{1}{4}\\[0.2em]
        2/3 & \text{if } \phantom{\,0.< }\rho_k > \frac{3}{4}\\[0.2em]
         1  & \text{if } \frac{1}{4}< \rho_k \leq \frac{3}{4}
    \end{cases}.
\end{equation}
Steps where $\rho_k < 0$ would actually lead to an increase of the loss are thus discarded and recomputed with the adjusted $\lambda_k$.

While effective in many cases, the standard heuristic can result in rapid and uncontrolled variations in $\rho_k$ and $\lambda_k$, leading to suboptimal performance. 
Drawing inspiration from step-size control in differential equations, we enhance this heuristic by incorporating proportional-integral controllers, which enhance stability by combining proportional responses to current errors with integral corrections from accumulated past errors \cite{wanner1996solving}.
As a measure of the error we take the deviation of $\rho_k$ from a desired target value $\rho^*$. The multiplier $m_k$ for adjusting $\lambda_k$ is then taken to be
\begin{equation}
\label{eqn:ad0}
    m_k = \min\!\Big[m_{\rm max} \,, \max\!\Big[ m_{\rm min}\,,E^{\beta_1}(\rho_k) \, E^{-\beta_2}(\rho_{k-1})\Big]\Big],
\end{equation}
where $e_{\rm min}$ and $e_{\rm max}$ constrain the scaling factor within a safe range, and
$\beta_1$ and $\beta_2$ control the influence of current and past errors, respectively.
Contrary to standard controllers where the magnitude of the (positive) error alone modulates $m_k$, here the sign of the error plays an important role as well. Specifically, the error functional $E(\rho)$ is chosen to ensure that $m_k > 1$ for $\rho_k < \rho^*$ and $m_k < 1$ for $\rho_k > \rho^*$. 
While various choices are possible, we choose 
\begin{equation}
\label{eqn:ad1}
    E(\rho) = \operatorname{sgn}(\rho^* - \rho) |\rho^* - \rho|^{q} + 1,
\end{equation}
whose functional form we display as an inset to \cref{fig:autodamping}(a).
To further enhance robustness, we reject steps where $\rho_k > \rho_{\rm max}$ or $\rho_k < \rho_{\rm min}$ and recompute the update with the newly adjusted $\lambda_k$.

We find the proposed approach to work well in practice and to be insensitive to minor changes in the selected thresholds. In general, we find $q=1$, $\beta_1=0.9$, $\beta_2=0.1$, $\alpha=0.05$, $\rho_{\rm max} = 3$, $\rho_{\rm min}=0.1$, $\rho^*=0.35$, $e_{\rm max}=2$, and $e_{\rm min}=0.05$ to be good choices.

\section{Numerical Experiments}
\label{sec:results}
In the following sections, we benchmark our methods on the paradigmatic example of the two-dimensional transverse-field Ising model (TFIM), a widely used testbed for NQS dynamics \cite{Carleo2017, Sinibaldi2023, Schmitt_2020, Schmitt2023, Czischek2018, Donatella2023, Gutirrez2022}. The Hamiltonian is given by 
\begin{equation} 
\label{eqn:TFIM}
\H = -J\sum_{\langle i,j\rangle} \hat\sigma_i^z\sigma_j^z - h\sum_i \hat\sigma_i^x, 
\end{equation} 
where $J$ is the nearest-neighbor coupling strength and $h$ represents the transverse field strength. Throughout this work, we set $J=1$, we adopt periodic boundary conditions, and we fix the $z$-axis as the quantization axis.

At zero temperature, this model undergoes a quantum phase transition at the critical point $h_c = 3.044$. 
This separates the ferromagnetic phase ($h < h_c$), from the paramagnetic phase ($h > h_c$). 
Deep in the ferromagnetic phase ($h\to0$) the ground state is degenerate and lies in the subspace spanned by $\ket{\uparrow\ldots\uparrow}$ and $\ket{\downarrow\ldots\downarrow}$. 
In the opposite limit ($h \to \infty$), the ground state is $\ket{\rightarrow\ldots\rightarrow}$, with spins aligned along the transverse field direction.

We demonstrate that the far-from-equilibrium dynamics characteristic of the quantum quenches of this model can be efficiently captured using p-tVMC. 
Our results are consistent across architectures, integration schemes, and quench parameters.

\subsection{Small-scale experiments}
\label{results_small}
Before presenting results for large system sizes, we validate our method against exact diagonalization results on a $4 \times 4$ lattice: small enough to allow simulating the exact dynamics but large enough for MC sampling to be non-trivial.
We consider the quench dynamics in which the system is initialized in the paramagnetic ground state at $h = \infty$, and evolved under a finite transverse field of strength $h = 2h_c$. 
For these simulations we adopt the Convolutional Neural Network (CNN) ansatz described in \cref{app:CNN}.
 
We compare several integration schemes: S-LPE-3, S-PPE-3 (third-order in $\dd t$), and S-LPE-2, S-PPE-2 (second-order in $\dd t$). 
We intentionally choose a fixed step size of $h\dd t = 3 \times 10^{-2}$, too big for product schemes of second order to accurately approximate the dynamics at hand, to underscore the advantages of our higher-order schemes.
Optimizations are performed using NGD and the autonomous damping strategies detailed in \cref{sec:auto_damping}.

\begin{figure}[htb]
\center
\includegraphics[width=\columnwidth]{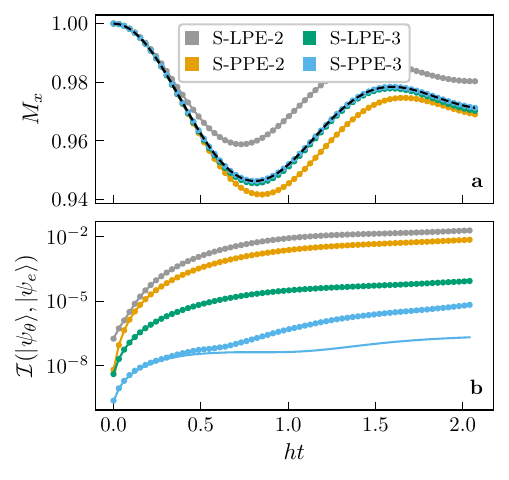}
\caption{
Quench dynamics ($h=\infty \to 2h_c$) of the transverse-field Ising model on a $4\times 4$ lattice obtained using different integration schemes: S-LPE-2, S-LPE-3, S-PPE-2, S-PPE-3. We show the evolution of the average magnetization $M_x$ (a), and of the infidelity between the exact solution $\ket{\psi_e}$ and its variational approximation $\ket{\psi_\theta}$ (b).
Full dots are used to mark variational data, obtained by solving the optimization problems in \cref{eqn:optimization_UV}. 
Solid lines detail the ideal behavior of each integrator scheme, estimated from a state-vector simulation of the dynamics resulting from the product expansion of the evolution operator. The dashed black line in (a) corresponds to the exact evolution.
Parameters: $N=16$, $J=1$, $h=2h_c$, 
$N_s=2^{14}$, $\bm\Theta_{\rm{CNN}} = (5,4,3;3)$, and $\dd t=0.025$.
}
\label{fig:4x4_TFIM}
\end{figure}

\begin{figure}[htb]
\center
\includegraphics[width=\columnwidth]{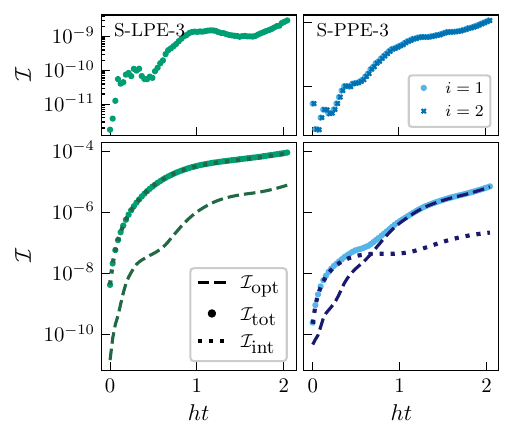}
\caption{Evolution of the total infidelity ($\mathcal{I}_{\rm tot}$), integration infidelity ($\mathcal{I}_{\rm int}$), and optimization infidelity ($\mathcal{I}_{\rm opt}$) as a function of time. The exact dynamics $\ket{\psi_e(t)}$, the dynamics obtained from state-vector simulations of product schemes $\ket{\psi_a(t)}$, and the variational dynamics $\ket{\psi_\theta(t)}$ are compared.
The infidelities are defined as follows: $\mathcal{I}_{\rm{tot}} = \mathcal{I}(\ket{\psi_e(t)}, \ket{\psi_\theta(t)})$, $\mathcal{I}_{\rm int} = \mathcal{I}(\ket{\psi_e(t)}, \ket{\psi_a(t)})$, and $\mathcal{I}_{\rm opt} = \mathcal{I}(\ket{\psi_a(t)}, \ket{\psi_\theta(t)})$. 
The final step infidelities obtained at the end of the optimization problems in \cref{eqn:optimization_UV} are reported in the top panels. The different substeps of the S-PPE-3 scheme are indexed by $i$. Parameters: same as in \cref{fig:4x4_TFIM}.
}
\label{fig:4x4_TFIM_fidelity}
\end{figure}

The variational evolution closely follows the dynamics predicted by the integration scheme, achieving infidelities with respect to the exact solution below $10^{-5}$ for the best-performing S-PPE-3 scheme. The ideal behavior of each integrator is estimated by applying the product expansion directly to the full state vector [equivalent to solving exactly the optimization problem in \cref{eqn:optimization_UV} on a log-state vector ansatz]. This analysis reveals that the variational dynamics is influenced by two primary sources of error: optimization error and integration error. In the absence of optimization error, the variational dynamics would follow the approximate integrator's dynamics which, in the absence of integration error, would in turn match the exact evolution.
For most schemes shown in \cref{fig:4x4_TFIM}, the dynamics is limited by integration error. An exception is the S-PPE-3 scheme, where the discretization is sufficiently accurate that optimization error becomes the dominant source at $ht \gtrsim 0.6$. This crossover between integration and optimization errors is more clearly illustrated in \cref{fig:4x4_TFIM_fidelity}, where we analyze the error contributions for S-LPE-3 and S-PPE-2, the two best-performing schemes. While S-LPE-3 remains dominated by integration error throughout, S-PPE-3 exhibits a crossover point where optimization error overtakes, as indicated by the intersection of the dashed and dotted lines.

\subsection{Large-scale experiments}
\label{sec:results_big}

We now demonstrate the effectiveness of our methods for simulating large-scale quantum dynamics. 
We again focus on the quench dynamics of the TFIM, this time on a $10\times 10$ lattice and for the more challenging quench from $h = \infty$ to $h = h_c/10$.
While no exact solution is available at this scale, this problem is widely used as a benchmark in quantum dynamics, allowing for comparison with established methods. 

\begin{figure}[!t]
\center
\includegraphics[width=\columnwidth]{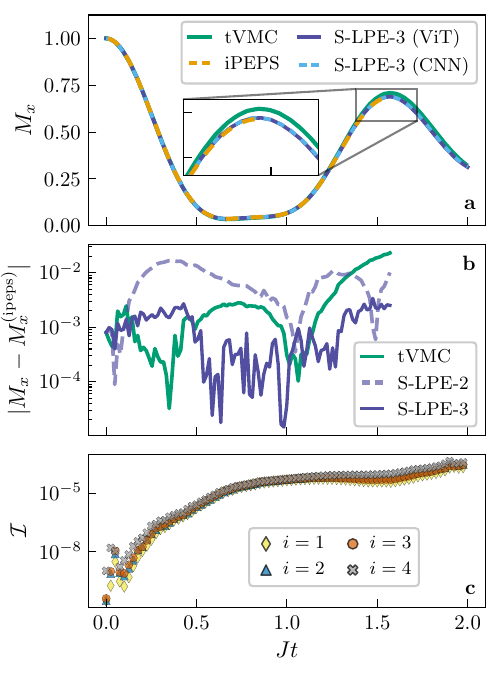}
\caption{
Quench dynamics across the critical point of the TFIM ($h=\infty \to h_c/10$) on a $10 \times 10$ lattice.
(a) Average magnetization as a function of time. 
Results are compared with existing methods: tVMC \cite{Schmitt_2020} and iPEPS \cite{Czarnik2019}. 
P-tVMC simulations are obtained using the S-LPE-3 scheme on two architectures: CNN and ViT. 
(b) Absolute difference in average magnetization between the iPEPS baseline and the variational simulations. P-tVMC results obtained using the SLPE2 and SLPE3 schemes are displayed. 
(c) Final optimization infidelity (step infidelity) achieved at each substep $i = 1, \ldots, 4$ of the S-LPE-3 scheme.
For clarity, in panels (b) and (c) only the results for the ViT architecture are displayed. 
Similar curves were observed with the CNN (not shown). 
Parameters: $N=100$, $J=1$, $h=h_c/10$, $N_s=2^{17}$ (CNN), $\bm\Theta_{\rm{CNN}} = (5,4,3;6)$, 
$N_s=2^{15}$ (ViT), $\bm\Theta_{\rm{ViT}} = (2,12,6,4;6)$, and $\dd t=0.025$.
}
\label{fig:Mx_10x10}
\end{figure}

In \cref{fig:Mx_10x10}, we present results obtained using the S-LPE-3 integration scheme (c.f.~\cref{sec:split-schemes}) for times up to $Jt = 2$ where comparison with existing results is meaningful.
To rigorously evaluate our approach, we utilize two architectures: a CNN with configuration $\bm \Theta_{\rm CNN} = (5,4,3;6)$ to explore the regime $N_s \gg N_p$, and a Vision Transformer (ViT) with $\bm \Theta_{\rm ViT} = (2,12,6,4;6)$ for the opposite case where $N_s \ll N_p$. 
In the latter regime, direct inversion of the FS tensor becomes prohibitively expensive and the gradient estimator introduced in \cref{eqn:grad_cmc} proves essential for the evaluation of the natural gradient. 
As shown in \cref{fig:Mx_10x10}(a), both models provide consistent predictions, underscoring the effectiveness of p-tVMC in yielding consistent results across different architectures, provided they are sufficiently expressive. 
These findings establish the ViT as a promising ansatz not only for ground-state simulations, as previously shown in Refs.~\cite{Viteritti2023, Viteritti2023a}, but also for large-scale simulations of quantum dynamics.

While the p-tVMC results, the iPEPS data from Ref.~\cite{Czarnik2019}, and the largest tVMC simulations from Ref.~\cite{Schmitt_2020}, generally show strong agreement, subtle differences emerge around the more challenging region $Jt \gtrsim 1.4$. 
Here, the tVMC curves deviate from the iPEPS predictions, while p-tVMC maintains closer alignment, suggesting superior stability at longer times.
This is further quantified in \cref{fig:Mx_10x10}(b), where we report the absolute deviation from the iPEPS baseline. 
Notably, p-tVMC captures long-time dynamics more accurately, highlighting its potential for enhanced long-term stability.
We further observe, at the early stages of the simulation ($0.1 \lesssim Jt \lesssim 0.4$), the existence of a small region where tVMC data are marginally better aligned with the iPEPS baseline compared to p-tVMC.
This region coincides with the high-activity region of the adaptive timestepping algorithm used in the tVMC simulations suggesting the origin of this discrepancy to be found in the larger, fixed, timestep employed in our calculations.
While our current results are competitive with state-of-the-art methods, implementing adaptive timestepping in future iterations will further enhance the overall accuracy of the results.
Finally, in \cref{fig:Mx_10x10}(c) we display the final step-infidelities at each time step.
We attribute the degradation of the quality of the fidelity optimizations to finite-sample effects that become more pronounced at later times.
To validate this hypothesis we ran the same optimizations in a smaller $4\times 4$ lattice evaluating expectation values exactly by summing over the entire Hilbert space. In the absence of MC noise, we observe convergence to numerical precision (see \cref{app:numerically_exact}). 
We remark that to stabilize our simulations, we employed the autonomous damping strategy outlined in \cref{sec:auto_damping}. 
All systems were initialized using a two-step process where we first use traditional VMC to approximate the ground state of $\hat{H}x = \sum_i \hat{\sigma}^x_i$, and then perform infidelity minimization to align the variational state with the constant-amplitude target state $\phi_{\rm gs}(y) = \text{const}$ for all $y \in [-1, +1]^N$. .

\section{Conclusions and outlooks}
\label{Sec:Conclusions}
In this work, we provide a rigorous formalization of the p-tVMC method, decoupling the discretization of time evolution from the state compression task, performed by infidelity minimization.

In our analysis of the discretization scheme, we identify key criteria for constructing schemes that are simultaneously accurate and computationally efficient. 
Building on these principles, we address the limitations in prior approaches to p-tVMC through two novel families of integration schemes capable of achieving arbitrary-order accuracy in time while scaling favorably with the number of degrees of freedom in the system. 
This is made possible by making efficient use of the specific structure of the p-tVMC problem.

In the study of the fidelity optimization, we demonstrate the critical role of natural gradient descent in compressing non-local transformations of the wavefunction into parameter updates. 
Additionally, we introduce an automatic damping mechanism for NGD which provides robust performance without the need for extensive hyperparameter tuning at each timestep of the dynamics.

We further clarify which among the available stochastic estimators are most reliable for computing both the infidelity and its gradient, addressing open questions regarding the role of control variates in the estimation of infidelity and their necessity in gradient-based optimization.

By integrating these advances into the p-tVMC framework, we demonstrate the potential to achieve machine precision in small-scale state compression and time evolution tasks. Applying these methods to larger systems, we show that p-tVMC not only reproduces state-of-the-art tVMC results with higher accuracy and improved control, but also surpasses previous methods in terms of generality and stability.

While fully numerically exact simulations on large systems remain beyond reach — likely due to limitations in Monte Carlo sampling — this work establishes p-tVMC as a highly promising approach, capable of overcoming several intrinsic challenges faced by other methods, and bringing us closer to achieving precise, large-scale classical quantum simulations.

\tocless{\section*{Software}}
Simulations were performed with NetKet \cite{netket3:2022,netket2:2019}, and at times parallelized with mpi4JAX \cite{mpi4jax:2021}.
This software is built on top of JAX \cite{jax2018github}, equinox \cite{kidger2021equinox} and Flax \cite{flax2020github}.
We used QuTiP \cite{qutip1,qutip2} for exact benchmarks.
The accompanying code and Mathematica scripts for the manuscript are available at \href{https://github.com/NeuralQXLab/ptvmc-systematic-study}{github.com/NeuralQXLab/ptvmc-systematic-study}.

\tocless{\section*{Acknowledgments}}
We acknowledge the useful contributions of R. Chen towards the derivation of PPE and S-PPE schemes.
We acknowledge insightful discussions with F. Becca, D. Poletti, G. Carleo, F. Ferrari and F. Minganti. 
We thank F. Caleca, M. Schmitt, and J. Dziarmaga for sharing their data with us.
We are grateful to L. L. Viteritti, C. Giuliani, A. Kahn, A. Shokry, L. Fioroni, and A. Sinibaldi for assisting in our fight against non-converging simulations, Jax bugs and complicated equations.
F.V. acknowledges support by the French Agence Nationale de la Recherche through the NDQM project, grant ANR-23-CE30-0018.
V.S. acknowledges support by the Swiss National Science Foundation through Projects No. 200020\_185015, 200020\_215172, and support from the EPFL Science Seed Fund 2021.
This project was provided with computing HPC and storage resources by GENCI at IDRIS thanks to the grant 2023-AD010514908 on the supercomputer Jean Zay's V100/A100 partition.\\
During the redaction of this manuscript we were made aware of a late revision of Ref.~\cite{Nys2024} applying the LPE scheme to fermionic dynamics, referring to it as \emph{Taylor Root Expansion}.

\bibliographystyle{quantum}
\bibliography{main}

\begin{thebibliography}{100}

\bibitem{Georgescu2014}
I.M. Georgescu, S.~Ashhab, and Franco Nori.
\newblock ``Quantum simulation''.
\newblock \href{https://dx.doi.org/10.1103/revmodphys.86.153}{Reviews of Modern Physics {\bf 86}, 153–185}~(2014).

\bibitem{PreskillQuantum18}
John Preskill.
\newblock ``Quantum {C}omputing in the {NISQ} era and beyond''.
\newblock {Quantum} {\bf 2}, 79~(2018).
\newblock  url:~\url{https://doi.org/10.22331/q-2018-08-06-79}.

\bibitem{Ezratty2021}
Olivier Ezratty.
\newblock ``Understanding quantum technologies 2023''~(2021) \href{http://arxiv.org/abs/2111.15352}{arXiv:arxiv:2111.15352}.

\bibitem{Yuan2019}
Xiao Yuan, Suguru Endo, Qi~Zhao, Ying Li, and Simon~C. Benjamin.
\newblock ``Theory of variational quantum simulation''.
\newblock \href{https://dx.doi.org/10.22331/q-2019-10-07-191}{Quantum {\bf 3}, 191}~(2019).

\bibitem{Bauls2023}
Mari~Carmen Bañuls.
\newblock ``Tensor network algorithms: A route map''.
\newblock \href{https://dx.doi.org/10.1146/annurev-conmatphys-040721-022705}{Annual Review of Condensed Matter Physics {\bf 14}, 173–191}~(2023).

\bibitem{Orus_Tensor_2019}
Rom{\'a}n Or{\'u}s.
\newblock ``Tensor networks for complex quantum systems''.
\newblock \href{https://dx.doi.org/10.1038/s42254-019-0086-7}{Nature Reviews Physics {\bf 1}, 538--550}~(2019).

\bibitem{schollwck2005}
U.~Schollw\"{o}ck.
\newblock ``The density-matrix renormalization group''.
\newblock \href{https://dx.doi.org/10.1103/revmodphys.77.259}{Reviews of Modern Physics {\bf 77}, 259–315}~(2005).

\bibitem{Feldman2022}
Noa Feldman, Augustine Kshetrimayum, Jens Eisert, and Moshe Goldstein.
\newblock ``Entanglement estimation in tensor network states via sampling''.
\newblock \href{https://dx.doi.org/10.1103/prxquantum.3.030312}{PRX Quantum{\bf 3}}~(2022).

\bibitem{eisert2013}
J.~Eisert.
\newblock ``Entanglement and tensor network states''~(2013) \href{http://arxiv.org/abs/1308.3318}{arXiv:1308.3318}.

\bibitem{werner2016}
A.~H. Werner, D.~Jaschke, P.~Silvi, M.~Kliesch, T.~Calarco, J.~Eisert, and S.~Montangero.
\newblock ``Positive tensor network approach for simulating open quantum many-body systems''.
\newblock \href{https://dx.doi.org/10.1103/PhysRevLett.116.237201}{Phys. Rev. Lett. {\bf 116}, 237201}~(2016).

\bibitem{cirac2021}
J.~Ignacio Cirac, David Pérez-García, Norbert Schuch, and Frank Verstraete.
\newblock ``Matrix product states and projected entangled pair states: Concepts, symmetries, theorems''.
\newblock \href{https://dx.doi.org/10.1103/revmodphys.93.045003}{Reviews of Modern Physics{\bf 93}}~(2021).

\bibitem{Paeckel2019}
Sebastian Paeckel, Thomas K\"{o}hler, Andreas Swoboda, Salvatore~R. Manmana, Ulrich Schollw\"{o}ck, and Claudius Hubig.
\newblock ``Time-evolution methods for matrix-product states''.
\newblock \href{https://dx.doi.org/10.1016/j.aop.2019.167998}{Annals of Physics {\bf 411}, 167998}~(2019).

\bibitem{ors2014}
Román Orús.
\newblock ``A practical introduction to tensor networks: Matrix product states and projected entangled pair states''.
\newblock \href{https://dx.doi.org/10.1016/j.aop.2014.06.013}{Annals of Physics {\bf 349}, 117–158}~(2014).

\bibitem{schollwck2011}
Ulrich Schollw\"{o}ck.
\newblock ``The density-matrix renormalization group in the age of matrix product states''.
\newblock \href{https://dx.doi.org/10.1016/j.aop.2010.09.012}{Annals of Physics {\bf 326}, 96–192}~(2011).

\bibitem{Lubasch2014}
Michael Lubasch, J~Ignacio Cirac, and Mari-Carmen Bañuls.
\newblock ``Unifying projected entangled pair state contractions''.
\newblock \href{https://dx.doi.org/10.1088/1367-2630/16/3/033014}{New Journal of Physics {\bf 16}, 033014}~(2014).

\bibitem{Verstraete2004}
F.~Verstraete and J.~I. Cirac.
\newblock ``Renormalization algorithms for quantum-many body systems in two and higher dimensions''~(2004) \href{http://arxiv.org/abs/cond-mat/0407066}{arXiv:cond-mat/0407066}.

\bibitem{Tagliacozzo2009}
L.~Tagliacozzo, G.~Evenbly, and G.~Vidal.
\newblock ``Simulation of two-dimensional quantum systems using a tree tensor network that exploits the entropic area law''.
\newblock \href{https://dx.doi.org/10.1103/PhysRevB.80.235127}{Phys. Rev. B {\bf 80}, 235127}~(2009).

\bibitem{Felser2021}
Timo Felser, Simone Notarnicola, and Simone Montangero.
\newblock ``Efficient tensor network ansatz for high-dimensional quantum many-body problems''.
\newblock \href{https://dx.doi.org/10.1103/physrevlett.126.170603}{Physical Review Letters{\bf 126}}~(2021).

\bibitem{Silvi2019}
Pietro Silvi, Ferdinand Tschirsich, Matthias Gerster, Johannes J\"{u}nemann, Daniel Jaschke, Matteo Rizzi, and Simone Montangero.
\newblock ``The tensor networks anthology: Simulation techniques for many-body quantum lattice systems''.
\newblock \href{https://dx.doi.org/10.21468/scipostphyslectnotes.8}{SciPost Physics Lecture Notes}~(2019).

\bibitem{Jaschke2018}
Daniel Jaschke, Simone Montangero, and Lincoln~D Carr.
\newblock ``One-dimensional many-body entangled open quantum systems with tensor network methods''.
\newblock \href{https://dx.doi.org/10.1088/2058-9565/aae724}{Quantum Science and Technology {\bf 4}, 013001}~(2018).

\bibitem{Jaschke2018a}
Daniel Jaschke, Michael~L. Wall, and Lincoln~D. Carr.
\newblock ``Open source matrix product states: Opening ways to simulate entangled many-body quantum systems in one dimension''.
\newblock \href{https://dx.doi.org/10.1016/j.cpc.2017.12.015}{Computer Physics Communications {\bf 225}, 59–91}~(2018).

\bibitem{Evenbly2011}
G.~Evenbly and G.~Vidal.
\newblock ``Tensor network states and geometry''.
\newblock \href{https://dx.doi.org/10.1007/s10955-011-0237-4}{Journal of Statistical Physics {\bf 145}, 891–918}~(2011).

\bibitem{Carleo2017}
Giuseppe Carleo and Matthias Troyer.
\newblock ``Solving the quantum many-body problem with artificial neural networks''.
\newblock \href{https://dx.doi.org/10.1126/science.aag2302}{Science {\bf 355}, 602–606}~(2017).

\bibitem{Sharir2022}
Or~Sharir, Amnon Shashua, and Giuseppe Carleo.
\newblock ``Neural tensor contractions and the expressive power of deep neural quantum states''.
\newblock \href{https://dx.doi.org/10.1103/PhysRevB.106.205136}{Phys. Rev. B {\bf 106}, 205136}~(2022).

\bibitem{Deng2017}
Dong-Ling Deng, Xiaopeng Li, and S.~Das~Sarma.
\newblock ``Quantum entanglement in neural network states''.
\newblock \href{https://dx.doi.org/10.1103/PhysRevX.7.021021}{Phys. Rev. X {\bf 7}, 021021}~(2017).

\bibitem{Urea2024}
Julio Ureña, Antonio Sojo, Juani Bermejo-Vega, and Daniel Manzano.
\newblock ``Entanglement detection with classical deep neural networks''.
\newblock \href{https://dx.doi.org/10.1038/s41598-024-68213-0}{Scientific Reports{\bf 14}}~(2024).

\bibitem{Passetti2023}
Giacomo Passetti and Dante~M. Kennes.
\newblock ``Entanglement transition in deep neural quantum states''~(2023) \href{http://arxiv.org/abs/2312.11941}{arXiv:2312.11941}.

\bibitem{Torlai2018_Latent}
Giacomo Torlai and Roger~G. Melko.
\newblock ``Latent space purification via neural density operators''.
\newblock \href{https://dx.doi.org/10.1103/physrevlett.120.240503}{Physical Review Letters{\bf 120}}~(2018).

\bibitem{Vicentini2022_PositiveDefinite}
Filippo Vicentini, Riccardo Rossi, and Giuseppe Carleo.
\newblock ``Positive-definite parametrization of mixed quantum states with deep neural networks''~(2022).

\bibitem{Luo2022_AutoregOpen}
Di~Luo, Zhuo Chen, Juan Carrasquilla, and Bryan~K. Clark.
\newblock ``Autoregressive neural network for simulating open quantum systems via a probabilistic formulation''.
\newblock \href{https://dx.doi.org/10.1103/physrevlett.128.090501}{Physical Review Letters{\bf 128}}~(2022).

\bibitem{Vicentini2019_Open}
Filippo Vicentini, Alberto Biella, Nicolas Regnault, and Cristiano Ciuti.
\newblock ``Variational neural-network ansatz for steady states in open quantum systems''.
\newblock \href{https://dx.doi.org/10.1103/physrevlett.122.250503}{Physical Review Letters{\bf 122}}~(2019).

\bibitem{Eeltink2023_open}
Debbie Eeltink, Filippo Vicentini, and Vincenzo Savona.
\newblock ``Variational dynamics of open quantum systems in phase space''~(2023).

\bibitem{Reh2021}
Moritz Reh, Markus Schmitt, and Martin G\"arttner.
\newblock ``Time-dependent variational principle for open quantum systems with artificial neural networks''.
\newblock \href{https://dx.doi.org/10.1103/PhysRevLett.127.230501}{Phys. Rev. Lett. {\bf 127}, 230501}~(2021).

\bibitem{Dash2024_qgt}
Sidhartha Dash, Luca Gravina, Filippo Vicentini, Michel Ferrero, and Antoine Georges.
\newblock ``Efficiency of neural quantum states in light of the quantum geometric tensor''.
\newblock \href{https://dx.doi.org/10.1038/s42005-025-02005-4}{Communications Physics{\bf 8}}~(2025).

\bibitem{Zhao2024_EmpiricalComplexity}
Haimeng Zhao, Giuseppe Carleo, and Filippo Vicentini.
\newblock ``Empirical sample complexity of neural network mixed state reconstruction''.
\newblock \href{https://dx.doi.org/10.22331/q-2024-05-23-1358}{Quantum {\bf 8}, 1358}~(2024).

\bibitem{Choo2019}
Kenny Choo, Titus Neupert, and Giuseppe Carleo.
\newblock ``Two-dimensional frustrated ${J}_{1}\text{\ensuremath{-}}{J}_{2}$ model studied with neural network quantum states''.
\newblock \href{https://dx.doi.org/10.1103/PhysRevB.100.125124}{Phys. Rev. B {\bf 100}, 125124}~(2019).

\bibitem{Sharir2020}
Or~Sharir, Yoav Levine, Noam Wies, Giuseppe Carleo, and Amnon Shashua.
\newblock ``Deep autoregressive models for the efficient variational simulation of many-body quantum systems''.
\newblock \href{https://dx.doi.org/10.1103/PhysRevLett.124.020503}{Phys. Rev. Lett. {\bf 124}, 020503}~(2020).

\bibitem{Vicentini2023}
Dian Wu, Riccardo Rossi, Filippo Vicentini, Nikita Astrakhantsev, Federico Becca, Xiaodong Cao, Juan Carrasquilla, Francesco Ferrari, Antoine Georges, Mohamed Hibat-Allah, Masatoshi Imada, Andreas~M. L\"{a}uchli, Guglielmo Mazzola, Antonio Mezzacapo, Andrew Millis, Javier Robledo~Moreno, Titus Neupert, Yusuke Nomura, Jannes Nys, Olivier Parcollet, Rico Pohle, Imelda Romero, Michael Schmid, J.~Maxwell Silvester, Sandro Sorella, Luca~F. Tocchio, Lei Wang, Steven~R. White, Alexander Wietek, Qi~Yang, Yiqi Yang, Shiwei Zhang, and Giuseppe Carleo.
\newblock ``Variational benchmarks for quantum many-body problems''.
\newblock \href{https://dx.doi.org/10.1126/science.adg9774}{Science {\bf 386}, 296–301}~(2024).

\bibitem{Viteritti2023a}
Luciano~Loris Viteritti, Riccardo Rende, and Federico Becca.
\newblock ``Transformer variational wave functions for frustrated quantum spin systems''.
\newblock \href{https://dx.doi.org/10.1103/PhysRevLett.130.236401}{Phys. Rev. Lett. {\bf 130}, 236401}~(2023).

\bibitem{Liang2018}
Xiao Liang, Wen-Yuan Liu, Pei-Ze Lin, Guang-Can Guo, Yong-Sheng Zhang, and Lixin He.
\newblock ``Solving frustrated quantum many-particle models with convolutional neural networks''.
\newblock \href{https://dx.doi.org/10.1103/PhysRevB.98.104426}{Phys. Rev. B {\bf 98}, 104426}~(2018).

\bibitem{Stokes2020a}
James Stokes, Javier~Robledo Moreno, Eftychios~A. Pnevmatikakis, and Giuseppe Carleo.
\newblock ``Phases of two-dimensional spinless lattice fermions with first-quantized deep neural-network quantum states''.
\newblock \href{https://dx.doi.org/10.1103/PhysRevB.102.205122}{Phys. Rev. B {\bf 102}, 205122}~(2020).

\bibitem{Attila2020}
Attila Szab\'o and Claudio Castelnovo.
\newblock ``Neural network wave functions and the sign problem''.
\newblock \href{https://dx.doi.org/10.1103/PhysRevResearch.2.033075}{Phys. Rev. Res. {\bf 2}, 033075}~(2020).

\bibitem{Choo2020}
Kenny Choo, Antonio Mezzacapo, and Giuseppe Carleo.
\newblock ``Fermionic neural-network states for ab-initio electronic structure''.
\newblock \href{https://dx.doi.org/10.1038/s41467-020-15724-9}{Nature Communications{\bf 11}}~(2020).

\bibitem{Cassella2023}
Gino Cassella, Halvard Sutterud, Sam Azadi, N.~D. Drummond, David Pfau, James~S. Spencer, and W.~M.~C. Foulkes.
\newblock ``Discovering quantum phase transitions with fermionic neural networks''.
\newblock \href{https://dx.doi.org/10.1103/PhysRevLett.130.036401}{Phys. Rev. Lett. {\bf 130}, 036401}~(2023).

\bibitem{RobledoMoreno2022}
Javier Robledo~Moreno, Giuseppe Carleo, Antoine Georges, and James Stokes.
\newblock ``Fermionic wave functions from neural-network constrained hidden states''.
\newblock \href{https://dx.doi.org/10.1073/pnas.2122059119}{Proceedings of the National Academy of Sciences{\bf 119}}~(2022).

\bibitem{Carleo2017_tVMC}
Giuseppe Carleo, Lorenzo Cevolani, Laurent Sanchez-Palencia, and Markus Holzmann.
\newblock ``Unitary dynamics of strongly interacting bose gases with the time-dependent variational monte carlo method in continuous space''.
\newblock \href{https://dx.doi.org/10.1103/physrevx.7.031026}{Physical Review X{\bf 7}}~(2017).

\bibitem{Sinibaldi2023}
Alessandro Sinibaldi, Clemens Giuliani, Giuseppe Carleo, and Filippo Vicentini.
\newblock ``Unbiasing time-dependent variational monte carlo by projected quantum evolution''.
\newblock \href{https://dx.doi.org/10.22331/q-2023-10-10-1131}{Quantum {\bf 7}, 1131}~(2023).

\bibitem{Stokes2023}
James Stokes, Brian Chen, and Shravan Veerapaneni.
\newblock ``Numerical and geometrical aspects of flow-based variational quantum monte carlo''.
\newblock \href{https://dx.doi.org/10.1088/2632-2153/acc8b9}{Machine Learning: Science and Technology {\bf 4}, 021001}~(2023).

\bibitem{Schmitt_2020}
Markus Schmitt and Markus Heyl.
\newblock ``Quantum many-body dynamics in two dimensions with artificial neural networks''.
\newblock \href{https://dx.doi.org/10.1103/physrevlett.125.100503}{Physical Review Letters{\bf 125}}~(2020).

\bibitem{Schmitt2023}
Tiago Mendes-Santos, Markus Schmitt, and Markus Heyl.
\newblock ``Highly resolved spectral functions of two-dimensional systems with neural quantum states''.
\newblock \href{https://dx.doi.org/10.1103/PhysRevLett.131.046501}{Phys. Rev. Lett. {\bf 131}, 046501}~(2023).

\bibitem{Joshi2024SkirmionDynamics}
Ashish Joshi, Robert Peters, and Thore Posske.
\newblock ``Quantum skyrmion dynamics studied by neural network quantum states''.
\newblock \href{https://dx.doi.org/10.1103/physrevb.110.104411}{Physical Review B{\bf 110}}~(2024).

\bibitem{Czischek2018}
Stefanie Czischek, Martin G\"arttner, and Thomas Gasenzer.
\newblock ``Quenches near ising quantum criticality as a challenge for artificial neural networks''.
\newblock \href{https://dx.doi.org/10.1103/PhysRevB.98.024311}{Phys. Rev. B {\bf 98}, 024311}~(2018).

\bibitem{Mauron2024}
Linda Mauron, Zakari Denis, Jannes Nys, and Giuseppe Carleo.
\newblock ``Predicting topological entanglement entropy in a rydberg analog simulator''~(2024) \href{http://arxiv.org/abs/2406.19872}{arXiv:2406.19872}.

\bibitem{Nys2024}
Jannes Nys, Gabriel Pescia, Alessandro Sinibaldi, and Giuseppe Carleo.
\newblock ``Ab-initio variational wave functions for the time-dependent many-electron schr\"{o}dinger equation''.
\newblock \href{https://dx.doi.org/10.1038/s41467-024-53672-w}{Nature Communications{\bf 15}}~(2024).

\bibitem{Wagner2024Temperature}
Dennis Wagner, Andreas Kl\"{u}mper, and Jesko Sirker.
\newblock ``Thermodynamics based on neural networks''.
\newblock \href{https://dx.doi.org/10.1103/physrevb.109.155128}{Physical Review B{\bf 109}}~(2024).

\bibitem{Nys2024_thermofield}
Jannes Nys, Zakari Denis, and Giuseppe Carleo.
\newblock ``Real-time quantum dynamics of thermal states with neural thermofields''.
\newblock \href{https://dx.doi.org/10.1103/PhysRevB.109.235120}{Phys. Rev. B {\bf 109}, 235120}~(2024).

\bibitem{vicentini2022_dynamics}
Filippo Vicentini, Riccardo Rossi, and Giuseppe Carleo.
\newblock ``Positive-definite parametrization of mixed quantum states with deep neural networks''~(2022) \href{http://arxiv.org/abs/2206.13488}{arXiv:2206.13488}.

\bibitem{Lin2024Open}
Joshua Lin, Di~Luo, Xiaojun Yao, and Phiala~E. Shanahan.
\newblock ``Real-time dynamics of the schwinger model as an open quantum system with neural density operators''.
\newblock \href{https://dx.doi.org/10.1007/jhep06(2024)211}{Journal of High Energy Physics{\bf 2024}}~(2024).

\bibitem{Shampine1979_stiff}
L.~F. Shampine and C.~W. Gear.
\newblock ``A user’s view of solving stiff ordinary differential equations''.
\newblock \href{https://dx.doi.org/10.1137/1021001}{SIAM Review {\bf 21}, 1–17}~(1979).

\bibitem{Donatella2023}
Kaelan Donatella, Zakari Denis, Alexandre Le~Boit\'e, and Cristiano Ciuti.
\newblock ``Dynamics with autoregressive neural quantum states: Application to critical quench dynamics''.
\newblock \href{https://dx.doi.org/10.1103/PhysRevA.108.022210}{Phys. Rev. A {\bf 108}, 022210}~(2023).

\bibitem{Poletti2024}
Wenxuan Zhang, Bo~Xing, Xiansong Xu, and Dario Poletti.
\newblock ``Paths towards time evolution with larger neural-network quantum states''~(2024) \href{http://arxiv.org/abs/2406.03381}{arXiv:2406.03381}.

\bibitem{Medvidovi2021}
Matija Medvidović and Giuseppe Carleo.
\newblock ``Classical variational simulation of the quantum approximate optimization algorithm''.
\newblock \href{https://dx.doi.org/10.1038/s41534-021-00440-z}{npj Quantum Information{\bf 7}}~(2021).

\bibitem{Carleo2018}
Bjarni Jónsson, Bela Bauer, and Giuseppe Carleo.
\newblock ``Neural-network states for the classical simulation of quantum computing''~(2018).

\bibitem{Gutirrez2022}
Irene~López Gutiérrez and Christian~B. Mendl.
\newblock ``Real time evolution with neural-network quantum states''.
\newblock \href{https://dx.doi.org/10.22331/q-2022-01-20-627}{Quantum {\bf 6}, 627}~(2022).

\bibitem{Medvidovi2024}
Matija Medvidović and Javier~Robledo Moreno.
\newblock ``Neural-network quantum states for many-body physics''.
\newblock \href{https://dx.doi.org/10.1140/epjp/s13360-024-05311-y}{The European Physical Journal Plus{\bf 139}}~(2024).

\bibitem{Lange2024}
Hannah Lange, Anka Van~de Walle, Atiye Abedinnia, and Annabelle Bohrdt.
\newblock ``From architectures to applications: A review of neural quantum states''~(2024) \href{http://arxiv.org/abs/2402.09402}{arXiv:2402.09402}.

\bibitem{Hatano2005}
Naomichi Hatano and Masuo Suzuki.
\newblock ``Finding exponential product formulas of higher orders''.
\newblock \href{https://dx.doi.org/10.1007/11526216_2}{Page 37–68}.
\newblock Springer Berlin Heidelberg. ~(2005).

\bibitem{MllerHermes2012}
Alexander M\"{u}ller-Hermes, J~Ignacio~Cirac, and Mari~Carmen Bañuls.
\newblock ``Tensor network techniques for the computation of dynamical observables in one-dimensional quantum spin systems''.
\newblock \href{https://dx.doi.org/10.1088/1367-2630/14/7/075003}{New Journal of Physics {\bf 14}, 075003}~(2012).

\bibitem{Moler2003}
Cleve Moler and Charles Van~Loan.
\newblock ``Nineteen dubious ways to compute the exponential of a matrix, twenty-five years later''.
\newblock \href{https://dx.doi.org/10.1137/s00361445024180}{SIAM Review {\bf 45}, 3–49}~(2003).

\bibitem{Havlicek2023}
Vojtech Havlicek.
\newblock ``Amplitude ratios and neural network quantum states''.
\newblock \href{https://dx.doi.org/10.22331/q-2023-03-02-938}{Quantum {\bf 7}, 938}~(2023).

\bibitem{Ledinauskas2023}
Eimantas Ledinauskas and Egidijus Anisimovas.
\newblock ``Scalable imaginary time evolution with neural network quantum states''.
\newblock \href{https://dx.doi.org/10.21468/scipostphys.15.6.229}{SciPost Physics{\bf 15}}~(2023).

\bibitem{Adam}
Diederik~P. Kingma and Jimmy Ba.
\newblock ``Adam: A method for stochastic optimization''~(2014) \href{http://arxiv.org/abs/1412.6980}{arXiv:1412.6980}.

\bibitem{Rubinstein2016}
Reuven~Y. Rubinstein and Dirk~P. Kroese.
\newblock ``Simulation and the monte carlo method''.
\newblock \href{https://dx.doi.org/10.1002/9781118631980}{Chapter~5}.
\newblock Wiley. ~(2016).

\bibitem{ranganath2014black}
Rajesh Ranganath, Sean Gerrish, and David Blei.
\newblock ``Black box variational inference''.
\newblock In Artificial intelligence and statistics.
\newblock \href{https://dx.doi.org/10.48550/arXiv.1401.0118}{Pages 814--822}.
\newblock PMLR~(2014).

\bibitem{Martens2012}
James Martens and Ilya Sutskever.
\newblock ``Training deep and recurrent networks with hessian-free optimization''.
\newblock \href{https://dx.doi.org/10.1007/978-3-642-35289-8_27}{Page 479–535}.
\newblock Springer Berlin Heidelberg. ~(2012).

\bibitem{Martens2020}
James Martens.
\newblock ``New insights and perspectives on the natural gradient method''.
\newblock Journal of Machine Learning Research {\bf 21}, 1--76~(2020).
\newblock  url:~\url{http://jmlr.org/papers/v21/17-678.html}.

\bibitem{Wright2006}
Jorge Nocedal and Stephen~J. Wright.
\newblock ``Numerical optimization''.
\newblock \href{https://dx.doi.org/10.1007/978-0-387-40065-5}{Volume~2 of Springer Series in Operations Research and Financial Engineering}.
\newblock Springer New York. ~(2006).

\bibitem{Cheng2010}
Ran Cheng.
\newblock ``Quantum geometric tensor (fubini-study metric) in simple quantum system: A pedagogical introduction''~(2010) \href{http://arxiv.org/abs/1012.1337}{arXiv:1012.1337}.

\bibitem{Stokes2020}
James Stokes, Josh Izaac, Nathan Killoran, and Giuseppe Carleo.
\newblock ``Quantum natural gradient''.
\newblock \href{https://dx.doi.org/10.22331/q-2020-05-25-269}{Quantum {\bf 4}, 269}~(2020).

\bibitem{Heskes2000}
Tom Heskes.
\newblock ``On “natural” learning and pruning in multilayered perceptrons''.
\newblock \href{https://dx.doi.org/10.1162/089976600300015637}{Neural Computation {\bf 12}, 881–901}~(2000).

\bibitem{Grosse2015}
Roger Grosse and Ruslan Salakhudinov.
\newblock ``Scaling up natural gradient by sparsely factorizing the inverse fisher matrix''.
\newblock In Francis Bach and David Blei, editors, Proceedings of the 32nd International Conference on Machine Learning.
\newblock Volume~37 of Proceedings of Machine Learning Research, pages 2304--2313.
\newblock Lille, France~(2015). PMLR.
\newblock  url:~\url{https://proceedings.mlr.press/v37/grosse15.html}.

\bibitem{Martens2015}
James Martens and Roger Grosse.
\newblock ``Optimizing neural networks with kronecker-factored approximate curvature''.
\newblock In Francis Bach and David Blei, editors, Proceedings of the 32nd International Conference on Machine Learning.
\newblock Volume~37 of Proceedings of Machine Learning Research, pages 2408--2417.
\newblock Lille, France~(2015). PMLR.
\newblock  url:~\url{https://proceedings.mlr.press/v37/martens15.html}.

\bibitem{Grosse2016}
Roger Grosse and James Martens.
\newblock ``A kronecker-factored approximate fisher matrix for convolution layers''.
\newblock In Maria~Florina Balcan and Kilian~Q. Weinberger, editors, Proceedings of The 33rd International Conference on Machine Learning.
\newblock Volume~48 of Proceedings of Machine Learning Research, pages 573--582.
\newblock New York, New York, USA~(2016). PMLR.
\newblock  url:~\url{https://proceedings.mlr.press/v48/grosse16.html}.

\bibitem{Ollivier2015}
Y.~Ollivier.
\newblock ``Riemannian metrics for neural networks i: feedforward networks''.
\newblock \href{https://dx.doi.org/10.1093/imaiai/iav006}{Information and Inference {\bf 4}, 108–153}~(2015).

\bibitem{Amari2019}
Shun-ichi Amari, Ryo Karakida, and Masafumi Oizumi.
\newblock ``Fisher information and natural gradient learning in random deep networks''.
\newblock In Kamalika Chaudhuri and Masashi Sugiyama, editors, Proceedings of the Twenty-Second International Conference on Artificial Intelligence and Statistics.
\newblock Volume~89 of Proceedings of Machine Learning Research, pages 694--702.
\newblock PMLR~(2019).
\newblock  url:~\url{https://proceedings.mlr.press/v89/amari19a.html}.

\bibitem{Karakida2021}
Ryo Karakida and Kazuki Osawa.
\newblock ``Understanding approximate fisher information for fast convergence of natural gradient descent in wide neural networks*''.
\newblock \href{https://dx.doi.org/10.1088/1742-5468/ac3ae3}{Journal of Statistical Mechanics: Theory and Experiment {\bf 2021}, 124010}~(2021).

\bibitem{Wei2022}
Qinxun Bai, Steven Rosenberg, and Wei Xu.
\newblock ``A geometric understanding of natural gradient''~(2022) \href{http://arxiv.org/abs/2202.06232}{arXiv:2202.06232}.

\bibitem{Chen2024}
Ao~Chen and Markus Heyl.
\newblock ``Empowering deep neural quantum states through efficient optimization''.
\newblock \href{https://dx.doi.org/10.1038/s41567-024-02566-1}{Nature Physics {\bf 20}, 1476–1481}~(2024).

\bibitem{Petersen2012}
K.~B. Petersen and M.~S. Pedersen.
\newblock ``The matrix cookbook''~(2012).

\bibitem{Rende2024}
Riccardo Rende, Luciano~Loris Viteritti, Lorenzo Bardone, Federico Becca, and Sebastian Goldt.
\newblock ``A simple linear algebra identity to optimize large-scale neural network quantum states''.
\newblock \href{https://dx.doi.org/10.1038/s42005-024-01732-4}{Communications Physics{\bf 7}}~(2024).

\bibitem{NTK2018}
Arthur Jacot, Franck Gabriel, and Cl\'{e}ment Hongler.
\newblock ``Neural tangent kernel: convergence and generalization in neural networks''.
\newblock In Proceedings of the 32nd International Conference on Neural Information Processing Systems.
\newblock \href{https://dx.doi.org/10.48550/arXiv.1806.07572}{Page 8580–8589}.
\newblock NIPS'18Red Hook, NY, USA~(2018). Curran Associates Inc.

\bibitem{Novak2022}
Roman Novak, Jascha Sohl-Dickstein, and Samuel~S Schoenholz.
\newblock ``Fast finite width neural tangent kernel''.
\newblock In Kamalika Chaudhuri, Stefanie Jegelka, Le~Song, Csaba Szepesvari, Gang Niu, and Sivan Sabato, editors, Proceedings of the 39th International Conference on Machine Learning.
\newblock Volume 162 of Proceedings of Machine Learning Research, pages 17018--17044.
\newblock PMLR~(2022).
\newblock  url:~\url{https://proceedings.mlr.press/v162/novak22a.html}.

\bibitem{Gacon2024}
Julien Gacon, Jannes Nys, Riccardo Rossi, Stefan Woerner, and Giuseppe Carleo.
\newblock ``Variational quantum time evolution without the quantum geometric tensor''.
\newblock \href{https://dx.doi.org/10.1103/PhysRevResearch.6.013143}{Phys. Rev. Res. {\bf 6}, 013143}~(2024).

\bibitem{martens2010}
James Martens.
\newblock ``Deep learning via hessian-free optimization''.
\newblock In Proceedings of the 27th International Conference on International Conference on Machine Learning.
\newblock Page 735–742.
\newblock ICML'10Madison, WI, USA~(2010). Omnipress.
\newblock  url:~\url{https://dl.acm.org/doi/10.5555/3104322.3104416}.

\bibitem{wanner1996solving}
Gerhard Wanner and Ernst Hairer.
\newblock ``Solving ordinary differential equations ii''.
\newblock \href{https://dx.doi.org/https://doi.org/10.1007/978-3-642-05221-7}{Volume 375}.
\newblock Springer Berlin Heidelberg New York. ~(1996).

\bibitem{Viteritti2023}
Luciano~Loris Viteritti, Riccardo Rende, Alberto Parola, Sebastian Goldt, and Federico Becca.
\newblock ``Transformer wave function for the shastry-sutherland model: emergence of a spin-liquid phase''~(2023) \href{http://arxiv.org/abs/2311.16889}{arXiv:2311.16889}.

\bibitem{Czarnik2019}
Piotr Czarnik, Jacek Dziarmaga, and Philippe Corboz.
\newblock ``Time evolution of an infinite projected entangled pair state: An efficient algorithm''.
\newblock \href{https://dx.doi.org/10.1103/PhysRevB.99.035115}{Phys. Rev. B {\bf 99}, 035115}~(2019).

\bibitem{netket3:2022}
Filippo Vicentini, Damian Hofmann, Attila Szabó, Dian Wu, Christopher Roth, Clemens Giuliani, Gabriel Pescia, Jannes Nys, Vladimir Vargas-Calderón, Nikita Astrakhantsev, and Giuseppe Carleo.
\newblock ``{NetKet 3: Machine Learning Toolbox for Many-Body Quantum Systems}''.
\newblock \href{https://dx.doi.org/10.21468/SciPostPhysCodeb.7}{SciPost Phys. CodebasesPage~7}~(2022).

\bibitem{netket2:2019}
Giuseppe Carleo, Kenny Choo, Damian Hofmann, James E.~T. Smith, Tom Westerhout, Fabien Alet, Emily~J. Davis, Stavros Efthymiou, Ivan Glasser, Sheng-Hsuan Lin, Marta Mauri, Guglielmo Mazzola, Christian~B. Mendl, Evert van Nieuwenburg, Ossian O'Reilly, Hugo Th{\'e}veniaut, Giacomo Torlai, Filippo Vicentini, and Alexander Wietek.
\newblock ``Netket: A machine learning toolkit for many-body quantum systems''.
\newblock \href{https://dx.doi.org/10.1016/j.softx.2019.100311}{SoftwareXPage 100311}~(2019).

\bibitem{mpi4jax:2021}
Dion Häfner and Filippo Vicentini.
\newblock ``mpi4jax: Zero-copy mpi communication of jax arrays''.
\newblock \href{https://dx.doi.org/10.21105/joss.03419}{Journal of Open Source Software {\bf 6}, 3419}~(2021).

\bibitem{jax2018github}
James Bradbury, Roy Frostig, Peter Hawkins, Matthew~James Johnson, Chris Leary, Dougal Maclaurin, George Necula, Adam Paszke, Jake Vander{P}las, Skye Wanderman-{M}ilne, and Qiao Zhang.
\newblock ``{JAX}: composable transformations of {P}ython+{N}um{P}y programs''.
\newblock \url{https://github.com/google/jax}~(2018).
\newblock Version 0.3.13.

\bibitem{kidger2021equinox}
Patrick Kidger and Cristian Garcia.
\newblock ``Equinox: neural networks in jax via callable pytrees and filtered transformations''~(2021) \href{http://arxiv.org/abs/2111.00254}{arXiv:2111.00254}.

\bibitem{flax2020github}
Jonathan Heek, Anselm Levskaya, Avital Oliver, Marvin Ritter, Bertrand Rondepierre, Andreas Steiner, and Marc van {Z}ee.
\newblock ``{F}lax: A neural network library and ecosystem for {JAX}''.
\newblock \url{https://github.com/google/flax}~(2024).
\newblock Version 0.10.6.

\bibitem{qutip1}
J.R. Johansson, P.D. Nation, and F.~Nori.
\newblock ``{QuTiP}: An open-source {P}ython framework for the dynamics of open quantum systems''.
\newblock \href{https://dx.doi.org/10.1016/j.cpc.2012.02.021}{Computer Physics Communications {\bf 183}, 1760--1772}~(2012).

\bibitem{qutip2}
J.R. Johansson, P.D. Nation, and F.~Nori.
\newblock ``{QuTiP} 2: A {P}ython framework for the dynamics of open quantum systems''.
\newblock \href{https://dx.doi.org/10.1016/j.cpc.2012.11.019}{Computer Physics Communications {\bf 184}, 1234--1240}~(2013).

\bibitem{stanley1999}
Richard~P. Stanley.
\newblock ``Enumerative combinatorics, volume 2''.
\newblock \href{https://dx.doi.org/http://dx.doi.org/10.1017/CBO9780511609589}{Volume~2 of Cambridge Studies in Advanced Mathematics}.
\newblock Cambridge University Press. ~(1999).

\bibitem{Ascher1995}
Uri~M. Ascher, Steven~J. Ruuth, and Brian T.~R. Wetton.
\newblock ``Implicit-explicit methods for time-dependent partial differential equations''.
\newblock \href{https://dx.doi.org/10.1137/0732037}{SIAM Journal on Numerical Analysis {\bf 32}, 797–823}~(1995).

\bibitem{Ascher1997}
Uri~M. Ascher, Steven~J. Ruuth, and Raymond~J. Spiteri.
\newblock ``Implicit-explicit runge-kutta methods for time-dependent partial differential equations''.
\newblock \href{https://dx.doi.org/10.1016/s0168-9274(97)00056-1}{Applied Numerical Mathematics {\bf 25}, 151–167}~(1997).

\bibitem{Li2024}
Ruichen Li, Haotian Ye, Du~Jiang, Xuelan Wen, Chuwei Wang, Zhe Li, Xiang Li, Di~He, Ji~Chen, Weiluo Ren, and Liwei Wang.
\newblock ``A computational framework for neural network-based variational monte carlo with forward laplacian''.
\newblock \href{https://dx.doi.org/10.1038/s42256-024-00794-x}{Nature Machine Intelligence}~(2024).

\bibitem{Minganti2019}
Fabrizio Minganti, Adam Miranowicz, Ravindra~W. Chhajlany, and Franco Nori.
\newblock ``Quantum exceptional points of non-hermitian hamiltonians and liouvillians: The effects of quantum jumps''.
\newblock \href{https://dx.doi.org/10.1103/physreva.100.062131}{Physical Review A{\bf 100}}~(2019).

\end{thebibliography}

\clearpage
\appendix

\onecolumngrid

\section{Details on LPE and PPE schemes}
\label{app:integration_scheme_details}

The two tables below report the coefficients for the first few orders of the LPE and PPE schemes. 
The details on the derivation can be found in the subsections following the tables.

\begin{table}[!h]
    \centering
    \begin{minipage}[t]{0.53\textwidth}
        \centering
        \setlength{\tabcolsep}{5pt}
        \begin{tabular}{c  cccc}
          $o$ & $1$ & $2$ & $3$ & $4$ \\[0.25em]
          \hline \rule{0pt}{1.25em}
         $a_1\,\,$ & $\,\,1\,\,$ & $(1 - i)/2$ & $0.6265$ & $0.0426 - 0.3946 i$ \\
         $a_2\,\,$ &  & $(1+ i)/2$ & $0.1867 - 0.4808 i$ & $0.0426 + 0.3946 i$ \\
         $a_3\,\,$ &  &  & $0.1867 + 0.4808 i$ & $0.4573 - 0.2351 i$ \\
         $a_4\,\,$ &  &  &  & $0.4573 + 0.2351 i$ \\[0.5em]
        \,\, &  &  &  & \\[0.25em]
        \,\, &  &  &  & 
        \end{tabular}
        \caption{Coefficients for the LPE schemes of lowest order. The sets of coefficients presented for each order are not unique: all $s!$ permutations are also solutions. Irrational numbers are reported with a precision of four decimal points. 
        We remark that the first order scheme with a single timestep is equivalent to a standard Euler scheme.
        }
        \label{tab:LPE}
    \end{minipage}\hfill
    \begin{minipage}[t]{0.43\textwidth}
        \centering
        \setlength{\tabcolsep}{3pt}
        \begin{tabular}{c ccc}
          $o$ & $2$ & $4$ & $6$ \\[0.25em]
          \hline
          \rule{0pt}{1.25em} 
         $a_1\,\,$ & $\,\phantom{+}1/2\,$ & $(3 - \sqrt{3} i)/12$ & $0.2153$  \\
         $a_2\,\,$ &  & $(3 + \sqrt{3} i)/12$ & $0.1423 - 0.1358 i$ \\
         $a_3\,\,$ &  &  & $0.1423 + 0.1358 i$ \\[0.5em]
         $b_1\,\,$ & $\,-1/2\,$ & $(-3 - \sqrt{3} i)/12$ & $-0.2153$ \\
         $b_2\,\,$ &  & $(-3 + \sqrt{3} i)/12$ & $-0.1423 - 0.1358 i$\\
         $b_3\,\,$ &  &  & $-0.1423 + 0.1358 i$ \\
        \end{tabular}
        \caption{Coefficients for the PPE schemes of lowest order. The sets of coefficients presented for each order are not unique. All the permutations of the $a_j$s and of $b_j$s are also solutions leading to a total of $[(s/2)!]^2$ combinations.
        We remark that the second order scheme with a single substep corresponds to a simple midpoint scheme.
        }
        \label{tab:PPE}
    \end{minipage}
\end{table}
\noindent The two tables below report the coefficients for the first few orders of the S-LPE and S-PPE schemes. 
\begin{table}[!ht]
    \centering
    \begin{minipage}[t]{0.48\textwidth}
        \centering
        \setlength{\tabcolsep}{10pt}
        \begin{tabular}{c ccc}
          $o$ & $1$ & $2$ & $3$ \\[0.25em]
          \hline
          \rule{0pt}{1.25em} 
         $a_1\,\,$ & $\,1\,$ & $(1-i)/2$ & $0.1057 - 0.3943 i$  \\
         $a_2\,\,$ &  & $(1+i)/2$ & $0.3943 + 0.1057 i$ \\
         $a_3\,\,$ &  &  & $0.3943 - 0.1057 i$ \\
         $a_4\,\,$ &  &  & $0.1057 + 0.3943 i$ \\[0.5em]
         $\alpha_1\,\,$ & $\,1\,$ & $(1-i)/2$ & $0.1057 - 0.3943 i$ \\
         $\alpha_2\,\,$ &  & $(1+i)/2$ & $0.3943 + 0.1057 i$\\
         $\alpha_3\,\,$ &  &  & $0.3943 - 0.1057 i$ \\
         $\alpha_4\,\,$ &  &  & $0.1057 + 0.3943 i$ \\[0.5em]
         $\phantom{\alpha_c}$\\
         $\phantom{\alpha_c}$\\
         $\phantom{\alpha_c}$
        \end{tabular}
        \caption{Coefficients for the S-LPE schemes of lowest order. 
        We remark that $a_i=\alpha_i$. The sets of coefficients presented for each order are unique up to conjugation. The first order scheme with a single timestep is equivalent to the first order Baker–Campbell–Hausdorff expansion. For S-LPE schemes the coefficient $\beta$ in \cref{eqn:LPE_S} is always vanishing.
        }
        \label{tab:SLPE}
    \end{minipage}\hfill
    \begin{minipage}[t]{0.48\textwidth}
        \centering
        \setlength{\tabcolsep}{6pt}
        \begin{tabular}{c ccc}
          $o$ & $2$ & $3$ & $4$ \\[0.25em]
          \hline
          \rule{0pt}{1.25em} 
         $a_1\,\,$ & $\,\phantom{+}1/2\,$ & $(3 + \sqrt{3} i)/12$ & $(3 - \sqrt{15} i)/24$  \\
         $a_2\,\,$ &  & $(3 - \sqrt{3} i)/12$ & $1/4$ \\
         $a_3\,\,$ &  &  & $(3 +  \sqrt{15} i)/24$ \\[0.5em]
         $b_1\,\,$ & $\,-1/2\,$ & $(-3 - \sqrt{3} i)/12$ & $(-3 + i\sqrt{15})/24$ \\
         $b_2\,\,$ &  & $(-3 + \sqrt{3} i)/12$ & $-1/4$\\
         $b_3\,\,$ &  &  & $(-3 - i\sqrt{15})/24$ \\[0.5em]
         $\alpha_1\,\,$ & $\,\phantom{+}1/2\,$ & $(3 + \sqrt{3} i)/12$ & $(3 - \sqrt{15} i)/24$ \\
         $\alpha_2\,\,$ & $\,\phantom{+}1/2\,$ & 1/2 & $(9 - \sqrt{15} i)/24$\\
         $\alpha_3\,\,$ &  & $(3 - \sqrt{3} i)/12$ & $(9 + \sqrt{15} i)/24$ \\
         $\alpha_4\,\,$ &  &  & $(3 + \sqrt{15} i)/24$ \\[0.5em]
        \end{tabular}
        \caption{Coefficients for the S-PPE schemes of lowest order. The sets of coefficients presented for each order are unique up to conjugation. 
        We note that $\alpha_s=0$ is always followed by $\alpha_{s+1}\neq 0$. This corresponds to a final diagonal operation after the sequence of optimizations.
        }
        \label{tab:SPPE}
    \end{minipage}
\end{table}


\subsection{Linear Product Expansion (LPE)}

Consider the ordinary differential equation of the form in \cref{eqn:generic_evolution}, where the solution $\ket{\psi(t)}$ is discretized over a set of times $\{ t_n\}_{n=1,2,\ldots}$, such that $\ket{\psi_n} = \ket{\psi(t_n)}$ and $t_{n+1} - t_n = \dd t$. The LPE scheme introduced in \cref{sec:LPE} provides the following prescription for approximately updating $\ket{\psi_n}$:

\begin{equation}
\label{eqn:LPE_expanded}
\begin{aligned}
     \ket{\psi_{n+1}} 
     &= \Big(\prod_{j=1}^{s} \hat T_{a_j}\Big) \ket{\psi_n} 
     = \ket{\psi_n} + \dd t\,\sum_{j=1}^s a_j \ket{\kappa_j}
\end{aligned}
\end{equation}
where 
\begin{equation}
    \ket{\kappa_j} = \hat{\Lambda} \qty(\ket{\psi_n} + \dd t\,\sum_{\ell=1}^{j-1} a_\ell \ket{\kappa_\ell}).\\
\end{equation}
This expression is in direct correspondence with the evolution equations of an explicit Runge–Kutta method with update function $f(\ket\psi) = \hat \Lambda \ket\psi$.
Although the LPE scheme can be cast as an explicit Runge-Kutta approximation, its scalability relies on avoiding this direct interpretation. Instead, \cref{eqn:LPE_expanded} is treated as a recursive process defined by
\begin{equation}
\label{eqn:LPE_recursion}
    \ket*{\phi_k}\quad\text{s.t.}\quad \ket*{\phi_k} = \hat T_{a_k} \ket*{\phi_{k-1}} \quad (k=1,\ldots,s)
\end{equation}
where $\ket{\psi_{n+1}} \equiv \ket*{\phi_s}$, and $\ket{\psi_n} \equiv \ket*{\phi_0}$. 
This is analogous to the formulation given in \cref{eqn:optimization_UV} in terms of transformations of the variational parameters of the wave function.
As explained in the main text, each sub-problem in \cref{eqn:LPE_recursion} involves at most linear powers of $\hat \Lambda$ making its application to NQS far more practical than a direct implementation of the Taylor expansion.
The numerical values of the coefficients $(a_1, \ldots, a_s)$ are obtained by  Taylor expanding both sides of \cref{eqn:LPE} and matching the terms order by order. This leads to to the following linear system of equations for the coefficients:

\begin{equation}
    e_k(a_1, \ldots, a_s) = \frac{1}{k!}
\end{equation}
for $1\leq k\leq s$. Here, $e_k(a_1, \ldots, a_s)$ is the elementary symmetric polynomial of degree $k$ in $s$ variables, defined for $k\leq s$ as the sum of the products of all possible combinations of $k$ distinct elements chosen from the set $\{a_j\}_{j=1}^s$ \cite{stanley1999}. We solve these equations in Mathematica for $s\leq 4$ and report the solutions in Table~\ref{tab:LPE}.
Interestingly, all coefficients $a_j$ for which $\Im{a_j}\neq 0$ appear in complex conjugate pairs and $\Re{a_j}>0$ $\forall j$.

\subsection{Padé Product Expansion (PPE)}

The PPE scheme introduced in \cref{sec:PPE} provides the following prescription for approximately updating $\ket{\psi_n}$:
\begin{equation}
\label{eqn:PPE_expanded}
\begin{aligned}
     \ket{\psi_{n+1}} 
     &= \Big(\prod_{j=1}^{s} \hat P_{b_j,a_j}\Big) \ket{\psi_n} =\Big(\prod_{j=1}^{s} \hat T_{b_j}^{-1}T_{a_j}\Big)\ket{\psi_n}.
\end{aligned}
\end{equation}
While a correspondence with explicit Runge–Kutta methods could be established via the Neumann series expansion of each $\hat T_{b_j}^{-1}$ term, the scalability of the method relies on avoiding this expansion. Instead, \cref{eqn:PPE_expanded} is treated as a recursive problem defined by
\begin{equation}
    \ket*{\phi_k}\quad\text{s.t.}\quad \hat T_{b_k} \ket*{\phi_k} = \hat T_{a_k} \ket*{\phi_{k-1}} \quad (k=1,\ldots,s),
\end{equation}
and where $\ket{\psi_{n+1}} \equiv \ket*{\phi_s}$, and $\ket{\psi_n} \equiv \ket*{\phi_0}$. This recursion relation can be alternatively stated as 
\begin{equation}
    \ket{\phi_k} = 
    \ket{\psi_{n}} 
    + \dd t \sum_{j=1}^k b_j (-\hat\Lambda) \ket{\phi_j} 
    + \dd t \sum_{j=1}^k a_j \hat\Lambda\ket{\phi_{j-1}} 
    \quad (k=1,\ldots,s),
\end{equation}
which puts the method in direct correspondence with the evolution equations of an implicit-explicit (IMEX) Runge–Kutta method with implicit and explicit update function $g(\ket\psi) = -\hat \Lambda \ket\psi$ and $f(\ket\psi) = \hat \Lambda \ket\psi$ respectively \cite{Ascher1995, Ascher1997}.


As for the LPE scheme,  the superiority in scalability of the method relies on avoiding this expansion and casting instead the expression onto the nested series of optimizations in \cref{eqn:optimization_UV}. The numerical values of the coefficients $(a_1, \ldots, a_s, b_1, \ldots, b_s)$ are obtained by  Taylor expanding both sides of \cref{eqn:PPE} and matching the terms order by order. This leads to to the following linear system of equations for the coefficients:
\begin{equation}
\label{eqn:PPE_conditions}
    \sum_{j=0}^{k}(-1)^{k-j}\,e_j(a_1, \ldots, a_s) h_{k-j}(b_1, \ldots, b_s) = 1/k!
\end{equation}
for $1\leq k\leq 2s$. Here, $h_{k-j}(b_1, \ldots, b_s)$ is the complete homogeneous symmetric polynomial of degree $k$ in $s$ variables, defined as the sum of all monomials of degree $k$ that can be formed from the set $\{b_j\}_{j=1}^s$ allowing repetition of variables \cite{stanley1999}. We adopt the convention that $e_0(a_1, \ldots, a_s) = h_0(b_1, \ldots, b_s) = 1$, and $e_{j>s}(a_1, \ldots, a_s)=0$.
The values of $\{a_j, b_j\}_{j=1}^s$ for $s\leq 3$ satisfying \cref{eqn:PPE_conditions} are provided in Table~\ref{tab:PPE}. Interestingly, we note that $a_j = -b_j^*$ and that $\Re{a_j}>0$, $\Re{b_j}<0$ $\forall j$. As before, if $a_j$ or $b_j$ have nonvanishing imaginary part they appear in conjugate pairs.

\subsubsection{Unitarity of Padè Product Expansions}
\label{app:unitarity}
We here show that PPE schemes are exactly (to all orders in $\dd t$) unitary. 
The fundamental building blocks of PPEs are of the form 
\begin{equation}
    U_i 
    = P_{b_i,a_i} 
    = \hat T_{b_i}^{-1}\hat T_{a_i} 
    = \qty(\id + b_i \hat \Lambda)^{-1}\qty(\id + a_i \hat\Lambda),
\end{equation}
where the timestep $\dd t$ has been absorbed into $\hat\Lambda = -i\hat H \dd t$. 
Since $\H$ is Hermitian, $\hat\Lambda = -\hat\Lambda^{\dagger}$ is anti-Hermtian.
It follows that
\begin{equation}
\label{eqn:unitary_proof_1}
    \begin{aligned}
        U_i^\dagger U_i 
        &= \qty(\id - a_i^* \hat\Lambda)\qty(\id - b_i^* \hat \Lambda)^{-1}
        \qty(\id + b_i \hat \Lambda)^{-1}\qty(\id + a_i \hat\Lambda) \\
        &= \qty(\id - a_i^* \hat\Lambda) \qty(\id + b_i \hat \Lambda)^{-1}
        \qty(\id - b_i^* \hat \Lambda)^{-1}\qty(\id + a_i \hat\Lambda),
    \end{aligned}
\end{equation}
where in the last equality we used that $[(\id+a\hat A)^{-1}, \id + \hat A]=0$: a trivial consequence of the fact that $[\hat A, f(\hat A)]=0$ for any $\hat A \in \mathcal{H}$ and $f:\mathcal{H}\to \mathcal{H}$. From \cref{tab:PPE} we observe that the expansion coefficients are related by $a_i = -b_i^*$. 
Substituting this in \cref{eqn:unitary_proof_1} leads to 
\begin{equation}
    U_i^\dagger U_i 
    = 
    \qty(\id + b_i \hat\Lambda) \qty(\id + b_i \hat \Lambda)^{-1}
    \qty(\id +a_i \hat \Lambda)^{-1}\qty(\id + a_i \hat\Lambda) = \id.
\end{equation}
PPE expansions of arbitrary order $o=2s$ obtained as 
$\hat U = \prod_{i=1}^{s} U_i$ are therefore unitary, that is $\hat U^\dagger \hat U = \id$.
To be precise, the relation $a_i = -b_i^*$ holds only for a subset of all possible solutions to \cref{eqn:PPE_conditions}. 
In general, however, the relation $a_i = -\Re{b_i} \pm \Im{b_i}$ holds. 
Moreover, the $b_i$s always present in conjugate pairs so that the proof can be carried out analogously.

Analogous calculations can be applied to the S-PPE-2 scheme. 
Here, since all coefficients are real, the diagonal terms in $\hat U^\dagger \hat U$ cancel out, making the algorithm exactly unitary. 
This does not hold for S-PPE schemes of higher order for which the $\alpha_i$s are in general complex.

\section{Control variates for the conditional Monte-Carlo estimator [\cref{eqn:cmc}]}
\label{app:cv_cmc}

\begin{figure}[htb]
\center
\includegraphics[width=0.5\columnwidth]{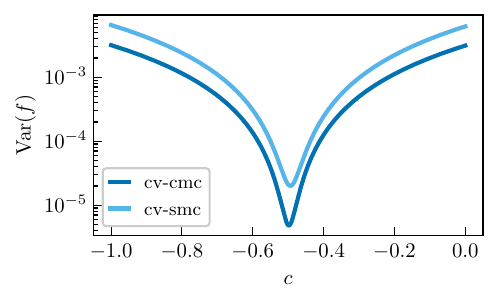}
\caption{Variance comparison of controlled estimators \cref{eqn:smc_cv} (cv-smc) and \cref{eqn:cmc_cv} (cv-cmc) as a function of the control coefficient $c$.}
\label{fig:cv_comparison}
\end{figure}

Control variates are particularly effective when both the control variable and the ideal control coefficient can be derived analytically.
For the simple Monte Carlo estimator of the fidelity in \cref{eqn:smc}, this is done in Ref.~\cite{Sinibaldi2023}, where the identity $\E_{(x,y)\sim\pi}[|A(x,y)|^2]=1$ is employed to reduce the variance of the estimator leading to \cref{eqn:smc_cv}.
To identify an analogous control variable for the cmc estimator we can proceed in one of two ways. 
The first approach recognizes that the cmc estimator arises from conditioning the smc estimator, and therefore the control-variate-enhanced cmc estimator naturally follows from conditioning the cv-smc estimator. 
Specifically, 
\begin{equation}
    \begin{aligned}
        \F(\ket{\psi},\ket{\phi}) 
        = \E_{(x,y)\sim\pi}[F(x,y)] 
        = \E_{x\sim\pi_\psi}[\E_{y\sim\pi_\phi}[F(x,y)|x]],
    \end{aligned}
\end{equation}
corresponds exactly to \cref{eqn:cmc_cv}. 
The second approach observes that $|A(z)|^2$ is correlated with $H_{\rm loc}(x)$ and thus remains a valid control variable for the cmc estimator.
By leveraging the separability of both $\pi(x,y) = \pi_{\psi}(x)\pi_{\phi}(y)$, and $A(x,y) = A_x(x)A_y(y)$, we can use $|A(x,y)|^2$ as control variate for the cmc estimator as well:
\begin{equation}
\label{eqn:cmc_controlled_derivation}
\begin{aligned}
    \F(\ket{\psi},\ket{\phi}) 
    &= \E_{x\sim\pi_\psi}\qty[\Re{H_{\rm loc}} + c \Big( \E_{(x,y)\sim\pi}\bigl[|A(x,y)|^2\bigr] - 1\Big)],\\
    &= \E_{x\sim\pi_\psi}\qty[\Re{H_{\rm loc}} + c \Big(|A_x(x)|^2\E_{y\sim \pi_\phi}\qty[|A_y(y)|^2] - 1\Big)],
\end{aligned}
\end{equation}
which again corresponds to \cref{eqn:cmc_cv}. 
Here, $A_{x}(x) = \phi(x)/\psi(x)$, and $A_{y}(y) = \psi(y)/\phi(y)$.
With a derivation similar to that in Ref.~\cite{Sinibaldi2023}, we find that the optimal control coefficient $c$, which minimizes the variance of the cv-cmc estimator, converges to $c=-1/2$ as $\ket{\psi} \to \ket{\phi}$.

In \cref{fig:cv_comparison} we illustrate the variance reduction achieved by applying control variates to the smc and cmc estimators. 
The data clearly highlight the efficiency of control variates in reducing the variance for both estimators. 
These calculations were performed far from ideal convergence, resulting in slight deviations from the expected optimal value of $c=-1/2$. 
In accordance with the Rao-Blackwell theorem, the variance of the cv-cmc estimator is consistently lower than that of the cv-smc estimator for any choice of $c$.

Note that in both control-variate-enhanced estimators, we retain only the real part of the original estimator, as the fidelity is known to be real. This ensures that $\E[\Im{f}] = 0$ for $f = A, H_{\rm loc}$. Including the imaginary part would amount to introducing an additional control variate with zero mean. However, any variability explained by the correlation between $\Im{f}$ and $\Re{f}$ is already captured through the variate $|A(x, y)|^2$. As a result, this additional control variate does not contribute to further variance reduction and is therefore omitted.

\section{Derivation of the gradient estimators}
\label{app:gradient_derivations}
In this section we derive and discuss the properties of different possible Monte Carlo estimators for the gradient of the fidelity.
To do in an efficient manner we introduce the notation 
\begin{equation}
    z = (x,y),
    \quad
    \pi_\psi(x) = \frac{\abs{\psi(x)}^2}{\braket{\psi}},
    \quad
    \pi(z) = \pi_\psi(x)\pi_\phi(y),
    \quad
    A(z) = \frac{\phi(x)}{\phi(y)}\frac{\psi(y)}{\psi(x)} .
\end{equation}
In the following we consider a complex ansatz $[\psi_\theta(x) \in \mathbb C]$ and real parameters $(\theta \in \mathbb R^{N_p})$.
As we have done throughout the paper, we identify the variational state as $\psi = \psi_\theta$, making implicit the dependence of the variational state $\ket{\psi}$ on the variational parameters. 

\subsection{Derivation of the $\nabla$cmc gradient estimator [\cref{eqn:grad_cmc}]}
We derive the $\nabla$cmc estimator of the gradient in \cref{eqn:grad_cmc} by differentiating the cmc fidelity estimator $\F = \E_{x\sim\pi_\psi}[H_{\rm loc}(x)]$ in \cref{eqn:cmc}. 
We show that this gradient estimator can be written in the form $\grad \F = \bm X^\top \bm\varepsilon$. \\

Applying the chain rule to $\F$ yields two contributions to the gradient 
\begin{equation}
    \grad \mathcal{F} 
    = \grad \sum_x \pi_\psi(x) H_{\rm loc}(x) = \underbrace{\,\,\sum_x  H_{\rm loc}(x) \grad\pi_\psi(x)\,\,}_{\circled{1}} + \underbrace{\,\,\sum_x \pi_\psi(x) \grad  H_{\rm loc}(x)\,\,}_{\circled{2}}.
\end{equation}
First off, we have that 
\begin{equation}
\label{eqn:grad_pix}
    \begin{aligned}
        \grad \pi_\psi(x) 
        = \grad \frac{\braket{\psi}{x}\braket{x}{\psi}}{\braket{\psi}}
        =  \frac{\braket{\grad\psi}{x}\braket{x}{\psi} + \braket{\psi}{x}\braket{x}{\grad\psi}}{\braket{\psi}} 
        - \frac{\braket{\psi}{x}\braket{x}{\psi}}{\braket{\psi}} \frac{\braket{\grad\psi}{\psi} + \braket{\psi}{\grad\psi}}{\braket{\psi}} 
        = 2 \pi_\psi(x) \Delta J^{\rm re}(x),
    \end{aligned}
\end{equation} 
where we denote $A^{\rm re} \equiv \Re{A}$ and $A^{\rm im} \equiv \Im{A}$. It follows that
\begin{equation}
    \begin{aligned}
        \text{\circled{1}}
        &= \sum_x  H_{\rm loc}(x) \grad\pi_\psi(x)
        = \E_{x\sim\pi_\psi}[2  \Delta J^{\rm re}(x)  H_{\rm loc}(x)].
    \end{aligned}
\end{equation}
We next compute
\begin{equation}
    \grad  H_{\rm loc}(x) = \grad  \frac{\mel{x}{\hat H}{\psi}}{\braket{x}{\psi}} 
    = \frac{\mel{x}{\hat H}{\grad \psi}}{\braket{x}{\psi}} - H_{\rm loc}(x)\frac{\braket{x}{\grad\psi}}{\braket{x}{\psi}}
    = \frac{\mel{x}{\hat H}{\grad \psi}}{\braket{x}{\psi}} - H_{\rm loc}(x)J(x),
\end{equation}
and note that
\begin{equation}
    \begin{aligned}
        \E_{\pi_{\psi}}\qty[\frac{\mel{x}{\hat H}{\grad \psi}}{\braket{x}{\psi}}]
    &= \sum_{x} \pi_{\psi}(x) \frac{\mel{x}{\hat H}{\grad \psi}}{\braket{x}{\psi}} 
    = \sum_{x} \pi_{\psi}(x) \frac{\bra{x}\hat H\qty(\sum_{y} \ketbra{y})\ket{\grad\psi}}{\braket{x}{\psi}}
    = \sum_{x,y} \frac{\braket{\psi}{x}\!\!\!\mel{x}{\hat H}{y}}{\braket{\psi}}\braket{y}{\grad \psi} \\
    &= \sum_{y} \frac{\bra{\psi}\Big(\sum_{x}\ketbra{x}\Big)\hat H\ket{y}}{\braket{\psi}}\braket{y}{\grad \psi}  
    = \sum_{x} \pi_{\psi}(x) \frac{\mel{\psi}{\hat H}{x}}{\braket{\psi}{x}} J(x) 
    = \sum_{x} \pi_{\psi}(x) H_{\rm loc}^*(x) J(x) \\
    &= \E_{\pi_{\psi}}[H_{\rm loc}^*(x) J(x)].
    \end{aligned}
\end{equation}
so that, 
\begin{equation}
    \begin{aligned}
        \text{\circled{2}}
        &= \E_{\pi_{\psi}}[\grad H_{\rm loc}(x)] 
        = \E_{\pi_{\psi}}\qty[\frac{\mel{x}{\hat H}{\grad \psi}}{\braket{x}{\psi}}] - \E_{\pi_{\psi}}[H_{\rm loc}(x) J(x)] 
        = \E_{\pi_{\psi}}[J(x)\qty(H_{\rm loc}^*(x) - H_{\rm loc}(x))] \\
        &= -2i \,\E_{\pi_{\psi}}[J(x) H_{\rm loc}^{\rm im}(x)].
    \end{aligned}
\end{equation}
Since $\hat H$ is Hermitian, we know that $\expval{\hat H}{\psi} = \E_{\pi_{\psi}}[H_{\rm loc}(x)] \in \mathbb{R}$ and therefore that $\E_{\pi_{\psi}}[H_{\rm loc}^{\rm im }(x)] = 0$. We can thus use that $H_{\rm loc}^{\rm im } = \Delta H_{\rm loc}^{\rm im }$ to obtain 
\begin{equation}
    \text{\circled{2}} = -2i \E_{\pi_{\psi}}[J(x) \Delta H_{\rm loc}^{\rm im}(x)]
    = -2i \E_{\pi_{\psi}}[\Delta J(x) H_{\rm loc}^{\rm im}(x)].
\end{equation}
Finally,
\begin{equation}
\label{eqn:grad_cmc_long}
\begin{aligned}
    \grad \mathcal{F} 
    &= \text{\circled{1}} + \text{\circled{2}} 
    = \E_{\pi_{\psi}}[2 \Delta J^{\rm re}(x) H_{\rm loc}(x) -2i \Delta J(x) H_{\rm loc}^{\rm im}(x)] \\[0.08cm]
    &= \E_{\pi_{\psi}}[2\Re{\Delta J(x) H_{\rm loc}(x)^{*}}] 
    = 2\E_{\pi_{\psi}}[\Delta J_x^{\rm re}H_{\rm loc}^{\rm re}(x) + \Delta J_x^{\rm im}H_{\rm loc}^{\rm im}(x)] \\[0.08cm]
    &= \E_{\pi_{\psi}}\qty[\left(\begin{array}{c}
            \Delta J^{\rm re}(x) \\[0.2em]
            \Delta J^{\rm im}(x) 
        \end{array}\right)
        \cdot
        \left(\begin{array}{c}
           2\Re{H_{\rm loc}(x)} \\[0.2em]
           2\Im{H_{\rm loc}(x)} 
        \end{array}\right)].
\end{aligned}
\end{equation}
We can now explicit the Monte-Carlo sampling of the expectation value which is in practice evaluated as
\begin{equation}
    \label{eqn:grad_H_mc}
    \begin{aligned}
        \grad \mathcal{F} 
        &\approx \frac{2}{N_s} \sum_{i=1}^{N_s} \Delta J^{\rm re}(x_i) H^{\rm re}_{\rm loc}(x_i) + \Delta J^{\rm im}(x_i)  H^{\rm im}_{\rm loc}(x_i)
        =\frac{2}{N_s^2} \sum_{i,j=1}^{N_s} \Delta J^{\rm re}(x_i) A^{\rm re}(x_i,y_j) + \Delta J^{\rm im}(x_i) A^{\rm im}(x_i,y_j),
    \end{aligned}
\end{equation}
with $N_s$ the number of samples. Note that the second expression follows from the fact that the local estimator itself is evaluated with MC sampling as 
\begin{equation}
    H_{\rm loc}(x) = \frac{\phi(x)}{\psi(x)} \frac{1}{N_s}\sum_{j=1}^{N_s} \frac{\psi(y_j)}{\phi(y_j)}.
\end{equation}

We now want to express the above in a form compatible with NTK and automatic damping strategies. To do so, we define
\begin{equation}
\begin{aligned}
    \bm X = 
    \begin{pmatrix}
    \Re{\bm O} \\[0.5ex]
    \Im{\bm O}
    \end{pmatrix} \in \mathbb{R}^{2N_s\times N_p}, 
    \qquad
    \bm O = \frac{1}{\sqrt{N_s}}
    \begin{pmatrix}
    \Delta J(x_1)^\top \\
    \vdots\\
    \Delta J(x_{N_s})^\top
    \end{pmatrix}\in \mathbb C^{N_s\times N_p}.    
\end{aligned}
\end{equation}
In this way, the Monte Carlo estimate of the Fubini-Study metric tensor reads $\bm S = \bm X^\top\! \bm X$, and the neural tangent kernel $\bm T = \bm X \bm X^\top$. We then define the complex-valued local energy vector as 
\begin{equation}
    \bm f = \frac{2}{\sqrt{N_s}}\Big(H_{\rm loc} (x_1), \ldots, H_{\rm loc} (x_{N_s})\Big) \in \mathbb C^{N_s},
\end{equation}
and 
\begin{equation}
    \bm \varepsilon = 
    \begin{pmatrix}
    \Re{\bm f} \\[0.5ex]
    \Im{\bm f}
    \end{pmatrix} \in \mathbb{R}^{2N_s}.
\end{equation}
It is easy to see that we can express \cref{eqn:grad_H_mc} as $\grad\F = \bm X^\top \bm\varepsilon$.

\subsection{Derivation of the $\nabla$cv-smc gradient estimator [\cref{eqn:grad_cv}]}
We derive the $\nabla$cv-smc estimator of the gradient in \cref{eqn:grad_cv} by differentiating the cv-smc fidelity estimator $\F = \E_{z\sim\pi}[F(z)] = \E_{z\sim\pi}[\Re A(z) + c (\abs{A(z)}^2 - 1)]$ in \cref{eqn:smc_cv}. 
We show that this gradient estimator does not admit the form $\grad \F = \bm X^\top \bm \varepsilon$ and is therefore unsuited for NTK calculations.\\

Once again
\begin{equation}
    \grad \mathcal{F} 
    = \grad \sum_z \pi(z) F(z) = \underbrace{\,\,\sum_z F(z) \grad\pi(z)\,\,}_{\circled{1}} + \underbrace{\,\,\sum_z \pi(z) \grad F(z)\,\,}_{\circled{2}}.
\end{equation}
Since $\grad\pi = \pi_\phi\grad\pi_\psi$, \cref{eqn:grad_pix} gets us $\circled{1} = \E_{z\sim\pi}[2 \Delta F(z) J^{\rm re}(x)]$.
Differentiating $A(z)$ yields $\grad A(z) = A(z) [J(y) - J(x)]$, so that
\begin{equation}
    \grad F(z) 
        = \frac{\grad A(z) + \grad A^*(z)}{2} + c \Big(A(z)\grad A^*(z) + A^*(z)\grad A(z)\Big)
        = \Re{\Big(A(z)+ 2c \abs{A(z)}^2\Big) \Big(J(y) - J(x)\Big)}.
\end{equation}
Inserting this into the expression for $\circled{2}$ leads to 
\begin{equation}
    \begin{aligned}
        \text{\circled{2}}
        =\E_{z\sim\pi}\qty[\qty(A^{\rm re}(z) + 2c\abs{A(z)}^2) \Big(J^{\rm re}(y) - J^{\rm re}(x)\Big) + A^{\rm im}(z) \Big(J^{\rm im}(x) - J^{\rm im}(y)\Big)].
    \end{aligned}
\end{equation}
The gradient is then found to be
\begin{equation}
\label{eqn:grad_cv_estended}
        \grad_{\theta} \mathcal{F} 
        = \text{\circled{1}} + \text{\circled{2}} 
        =
        \E_{z\sim\pi}\qty[\left(\begin{array}{c}
             J^{\rm re}(x) \\[0.2em]
             J^{\rm im}(x) \\[0.2em]
             J^{\rm re}(y) \\[0.2em]
             J^{\rm im}(y)
        \end{array}\right)
        \cdot
        \left(\begin{array}{c}
            2\Delta F(z) -A^{\rm re}(z) - 2c\abs{A(z)}^2 \\[0.2em]
            A^{\rm im}(z) \\[0.2em]
            A^{\rm re}(z) + 2c\abs{A(z)}^2 \\[0.2em]
            -A^{\rm im}(z)
        \end{array}\right)]
        .
    \end{equation}
We remark that this estimator evaluates the Jacobian of $\psi$ not only on the samples of $\psi$ as we would normally expect, but on those of $\phi$ as well. 
Equation \eqref{eqn:grad_cv_estended} makes it manifest that this estimator cannot be expressed in the form $\grad\F = \bm X^\top \bm \varepsilon$.

\subsection{Derivation of the $\nabla$smc gradient estimator [\cref{eqn:grad_smc}]}
\label{app:derivation_mixed_grad}
We derive the $\nabla$smc estimator of the gradient $\F = \E_{(x,y)\sim\pi}[A(x,y)]$  in \cref{eqn:grad_smc} and show that it admits the form $\grad \F = \bm X^\top \bm \varepsilon$.\\

One straightforward approach is to reverse the marginalization yielding the $\nabla$cmc gradient estimator [\cref{eqn:grad_cmc}]
\begin{equation}
\label{eqn:grad_smc_long}
\begin{aligned}
    \grad_{\theta} \mathcal{F} 
    &= \E_{\pi_{\psi}}[2\Re{\Delta J(x)^{*} H_{\rm loc}(x)}] 
    = 2\Re\Big\{\sum_x\pi_\psi(x) \Delta J^{*}(x) H_{\rm loc}(x)\Big\} \\
    &= 2\Re\Big\{\sum_x\pi_\psi(x) \Delta J(x)^{*} \frac{\phi(x)}{\psi(x)} \sum_y \pi_\phi(y) \frac{\psi(y)}{\phi(x)}\Big\} 
    = \E_{z\sim\pi}[2\Re{\Delta J(x) A(z)^{*}}]\\[0.1cm]
    &= \E_{z\sim\pi}\qty[\left(\begin{array}{c}
            \Delta J^{\rm re}(x) \\[0.2em]
            \Delta J^{\rm im}(x) 
        \end{array}\right)
        \cdot
        \left(\begin{array}{c}
           2\Re{A(z)} \\[0.2em]
           2\Im{A(z)} 
        \end{array}\right)].
\end{aligned}
\end{equation}
In practice, the expectation value above is evaluated using MC sampling as
\begin{equation}
    \grad\F \approx \frac{1}{N_s} \sum_{i=1}^{N_s} \Delta J^{\rm re}(x_i) \Re{A(x_i,y_i)} + \Delta J^{\rm im}(x_i)  \Im{A(x_i,y_i)},
\end{equation}
which makes manifest the possibility of expressing the gradient as $\grad\F = \bm X\varepsilon$ with 
\begin{equation}
    \bm f = \frac{2}{\sqrt{N_s}}\Big(A(x_1, y_1), \ldots, A(x_{N_s}, y_{N_s})\Big) \in \mathbb C^{N_s},
\end{equation}
and 
\begin{equation}
    \bm \varepsilon = 
    \begin{pmatrix}
    \Re{\bm f} \\[0.5ex]
    \Im{\bm f}
    \end{pmatrix} \in \mathbb{R}^{2N_s}.
\end{equation}

Although this derivation would suffice, it is still insightful to explore how the same expression can be obtained by manipulating the $\nabla$cv-smc estimator [\cref{eqn:grad_cv}, or equivalently \cref{eqn:grad_cv_estended}]. 
The value in this alternative derivation stems from the properties of $A(z)$ that this approach reveals. We believe these properties could be useful in the future to derive control variables tailored to the gradient. 
For starters, we find that
\begin{equation}
    \E_{z\sim\pi}\qty[\abs{A(z)}^2 f(x)] = \E_{y\sim\pi_\phi}\qty[f(y)].
\end{equation}
Using this identity, one can easily show that
\begin{align}
    &\E_{z\sim\pi}\qty[A^*(z) f(x)] = \E_{z\sim\pi}\qty[A(z) f(y)],\\
    &\E_{z\sim\pi}\qty[A^*(z) f(y)] = \E_{z\sim\pi}\qty[A(z) f(x)],\\
    &\E_{z\sim\pi}\qty[A(z) f(x)] = \E_{z\sim\pi}\qty[A^*(z) f(y)],\\
    &\E_{z\sim\pi}\qty[A(z) f(y)] = \E_{z\sim\pi}\qty[A^*(z) f(x)].
\end{align}
Again, these equations can be combined to show that
\begin{align}
        \E_{z\sim\pi}\qty[f(x) \Re{A(z)}] &= \E_{z\sim\pi}\qty[f(y) \Re{A(z)}],\\
        \E_{z\sim\pi}\qty[f(x) \Im{A(z)}] &= -\E_{z\sim\pi}\qty[f(y) \Im{A(z)}].
\end{align}
Substitution into \cref{eqn:grad_cv_estended} yields \cref{eqn:grad_smc_long}.

\section{Covariance structure, Rao-Blackwellization, and variance reduction}
\label{app:grad_variance}
\begin{figure}[h]
\center
\includegraphics[width=0.5\columnwidth]{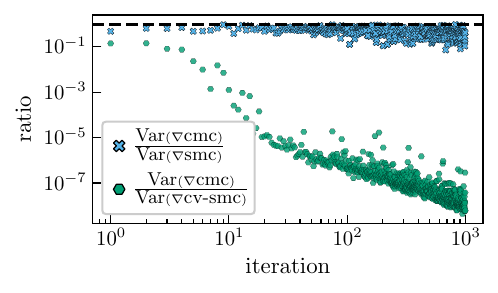}
\caption{Comparison of the variance of the different estimators of the fidelity gradient. As the gradient is a vector in $\mathbb R^{N_p}$, for each estimator we consider the a single direction in parameter space. The results are independent of the chosen directions.
Statistics are computed on the same data of \cref{fig:state_matching}.}
\label{fig:grad_var_ratio}
\end{figure}

We use this appendix to expand on the differences in variance exhibited by the gradient estimators presented in the main text.
First off, we note that while both the $\nabla$cmc and $\nabla$smc estimators take the form of a covariance estimator [c.f.~\cref{eqn:grad_cmc_long,eqn:grad_smc_long}], the $\nabla$cv-smc estimator does not. 
Generally, estimators structured as covariances tend to exhibit lower variance largely due to the inherent variance reduction from centering each component.
Specifically, the sample covariance used to estimate the covariance between two random variables $X$ and $Y$ is given by
\begin{equation} \operatorname{Cov}(X,Y) = \mathbb{E}[(X - \mathbb{E}[X])(Y - \mathbb{E}[Y])] \approx \frac{\gamma}{N_s} \sum_{i=1}^{N_s} \left(X_i - \frac{1}{N_s} \sum_{j=1}^{N_s} X_j\right) \left(Y_i - \frac{1}{N_s} \sum_{j=1}^{N_s} Y_j\right), \end{equation}
where $\gamma = N_s/(N_s - 1)$ is the Bessel correction factor used to remove the bias introduced by centering.
The centering process effectively provides sample-specific control variates that automatically adjust to the non-stationarity of the variables, enhancing estimator stability, and justifying the results presented in the main text and further corroborated by \cref{fig:grad_var_ratio}.

While the poor performance of the $\nabla$cv-smc estimator is attributed to its lack of a covariance structure, the smaller performance difference between the $\nabla$cmc and $\nabla$smc estimators can be understood through the Rao-Blackwell theorem. 
Identical arguments apply for the cmc and smc estimators of the fidelity itself.
In the simplest setting, Rao-Blackwellization replaces a function of two random variables with its conditional expectation resulting in a new estimator whose variance is no greater than that of the original function (Rao-Blackwell theorem).
Consider two random variables, $X$ and $Y$, and a function $J(X, Y)$. Our goal is to compute its expectation $\mathbb{E}[J(X, Y)]$ with respect to the joint distribution of $X$ and $Y$. To do so, we define \cite{ranganath2014black}
\begin{equation}
\mathbb{E}[J(X, Y) \mid X] = \sum_y J(x,y) \cdot P(Y=y \mid X=x)  \equiv \hat{J}(X) 
\end{equation}
and note that $\mathbb{E}[\hat{J}(X)] = \mathbb{E}[J(X, Y)]$. 
This implies that $\hat{J}(X)$ can be used in place of $J(X, Y)$ in a Monte Carlo approximation of $\mathbb{E}[J(X, Y)]$. 
The variance of $\hat{J}(X)$ can be easily computed to be
\begin{equation}
    \text{Var}(\hat{J}(X)) = \text{Var}(J(X, Y)) - \mathbb{E}[(J(X, Y) - \hat{J}(X))^2],
\end{equation}
from which it follows that $\text{Var}(\hat{J}(X)) \leq \text{Var}(J(X, Y))$.
To compute Rao-Blackwellized estimators, we generally need to calculate conditional expectations. In many cases, this cannot be done exactly; however, under a mean-field assumption, the conditional expectation simplifies due to factorization:
\begin{equation} \hat{J}(X) = \mathbb{E}[J(X, Y) \mid X] = \sum_y J(x, y) \cdot P(Y = y \mid X = x) = \sum_y J(x, y) \cdot P(Y = y) = \mathbb{E}_Y[J(X, Y)], \end{equation}
where we used that $P(X,Y) = P(X)P(Y)$ and thus $P(Y \mid X) = P(Y)$.
Therefore, to construct a lower variance estimator when the joint distribution factorizes, all we need to do is integrate out some variables and use this marginal as the estimator.

\section{Lowering sampling cost by reweighting}
\label{app:reweighting}

Direct evaluation of the fidelity (and of its gradient) between two transformed states $\ket*{\tilde\psi} = \hat V \ket*{\psi}$ and $\ket*{\tilde\phi} = \hat U \ket*{\phi}$ requires, in principle, sampling from their Born distributions $\pi_{\tilde\psi}(x)$ and $\pi_{\tilde\phi}(y)$, respectively. 
This process introduces a computational overhead that scales with $N_c$, the number of connected elements in the transformations.
For local spin Hamiltonians, this introduces a sampling overhead proportional to the system size $N$, which often becomes the dominant cost (in the simulations presented in \cref{sec:results}, for instance, around 90\% of the computational time is spent sampling).
The computational burden grows substantially for other Hamiltonians, such as for those arising in natural-orbital chemistry or for the kinetic term in first-quantisation formulations \cite{Li2024}. 

To address this overhead, one can resort to self-normalized importance sampling \cite{Rubinstein2016}, adapting the estimators discussed in \cref{sec:fidelity} to sample directly from the bare distributions.
In line with the procedure outlined in Ref.~\cite{Sinibaldi2023}, we avoid direct sampling from the target state. While previous works were concerned with unitary transformations applied to the target state alone, here we extend the approach to arbitrary transformations that act on both the target and variational states.

As all estimators discussed in \cref{sec:fidelity} are of this form, it is convenient to write down the reweighting procedure just once for the generic estimator $\E_{\sigma\sim\chi}[f(\sigma)]$. In this case, we have \cite{Rubinstein2016}
\begin{align}
\label{eqn:smc_rw}
    \E_{\sigma\sim\chi}[f(\sigma)] 
    = \frac{\E_{\sigma\sim\eta}[w(\sigma)f(\sigma)]}{\E_{\sigma\sim\eta}[w(\sigma)]},
\end{align}
where $w = \eta/\chi$. Specializing to the case where we sample from a joint distribution, e.g.~in~\cref{eqn:smc,eqn:smc_cv,eqn:grad_smc,eqn:grad_cv}, yields
\begin{align}
    \E_{(x,y)\sim\tilde\pi}[f(x,y)]
    = \frac{\sum_{x,y} \pi(x,y) \frac{\tilde \pi(x,y)}{\pi(x,y)} f(x,y)}{\sum_{x,y} \pi(x,y) \frac{\tilde \pi(x,y)}{\pi(x,y)} },
\end{align}
where $\tilde\pi(x,y) = \pi_{\tilde\psi}(x)\pi_{\tilde\phi}(y)$. 
Applying Rao-Blackwellization to this estimator is still possible and results in 
\begin{align}
\label{eqn:cmc_rw}
    \E_{(x,y)\sim\tilde\pi}[f(x,y)]  
    = \E_{x\sim\tilde\pi_\psi}[f_x(x)] \E_{y\sim\tilde\pi_\phi}[f_y(y)] 
    =
    \frac{\sum_{x} \pi_\psi(x) \frac{\tilde \pi_\psi(x)}{\pi_\psi(x)} f_x(x)}{\sum_{x} \pi_\psi(x) \frac{\tilde \pi_\psi(x)}{\pi_\psi(x)} }
    \frac{\sum_{y} \pi_\phi(y) \frac{\tilde \pi_\phi(y)}{\pi_\phi(y)} f_y(y)}{\sum_{y} \pi_\phi(y) \frac{\tilde \pi_\phi(y)}{\pi_\phi(y)} },
\end{align}
where we made use of the separability of the Born distributions, and of the integrator $f(x,y)=f_x(x)f_y(y)$ which we have observed in \cref{sec:fidelity}.

\section{Machine precision on small systems, the limitation of Monte Carlo sampling, and the importance of curvature}
\label{app:numerically_exact}
We now demonstrate the theoretical possibility of solving infidelity-based optimizations to machine precision. Specifically, we revisit the quench dynamics studied in \cref{sec:results_big}, involving a quench from $h=\infty$ to $h=h_c/10$, but now on a smaller $4\times 4$ lattice—small enough to allow exact summation over the entire Hilbert space.

\begin{figure}[htb]
\center
\includegraphics[width=0.98\columnwidth]{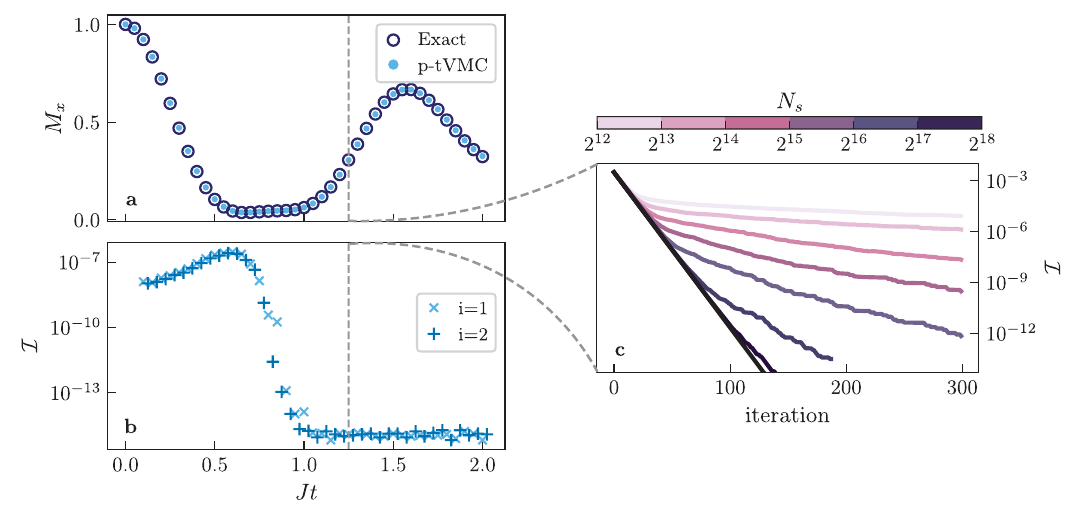}
\caption{Quenched dynamics ($h=\infty\to h_c/10$) on a $4\times 4$ lattice. (a,b) Average magnetization and optimization infidelity as a function of time. 
The variational results (full dots) are obtained in full summation using the S-LPE-2 scheme with $\dd t=0.05$. We  compare this to the exact calculation (open circles).
(c) Optimization profiles for the update $Jt=1.25 \to 1.30$ for different sample sizes $N_s$. The convergence is negatively impacted by the lack a number of samples sufficient to properly reconstruct the curvature matrix and resolve the ideal descent direction. Stochastic optimization achieves performance compatible with full summation (black line) only for very large sample sizes, comparable or superior to the size of the Hilbert space. 
The regularization coefficient $\lambda$ is fixed to $\lambda=10^{-8}$ for $N_s=\infty,2^{18},2^{17},2^{16}$. Smaller sample sizes require stronger regularization. We use $\lambda=10^{-7}$ for $N_s=2^{15},2^{14}$ and $\lambda=10^{-6}$ for $N_s=2^{13},2^{12}$.
Parameters: $\alpha=0.05$, $\bm\Theta_{\rm{CNN}} = (10,10,10,10;3)$.
}
\label{fig:numerical_precision}
\end{figure}

In \cref{fig:numerical_precision}(a-b), we show the variational p-tVMC dynamics computed by evaluating all statistical averages relevant to the optimization exactly. These results exhibit perfect agreement with the exact solution, with optimizations converging to the target state within machine precision.

By contrast, performing the same simulations using Monte Carlo sampling—i.e., evaluating statistical averages over a finite number of samples—yields results similar to those shown in \cref{results_small}. While convergence remains qualitatively good, it no longer reaches machine precision. As discussed in \cref{sec:auto_damping}, this discrepancy is likely due to poor estimation of the gradient and curvature matrix when using limited sample sizes, which leads to unreliable updates. Indeed, increasing the number of samples steadily improves the results, bringing them closer to the exact, noiseless solution.
Achieving results comparable to full-summation, however, appears to require a number of samples on the order of the Hilbert space dimension itself. This is illustrated in \cref{fig:numerical_precision}(c), where we display optimization trajectories for the following infidelity minimization problem:
\begin{equation}
\theta^* = \argmin{\theta}\,\, \mathcal{I}(\ket{\psi_{t}}, \ket{\psi_{\theta}}).
\end{equation}
The initial parameter configuration corresponds to the state at time $Jt=1.25$, obtained from the noiseless p-tVMC simulation shown in \cref{fig:numerical_precision}(a-b). The target state $\ket{\psi_{t}}$ at time $Jt=1.3$, by contrast, is taken from the exact evolution of the TFIM dynamics and encoded exactly, using one complex-valued parameter per entry of the state vector.

\begin{figure}[htb]
\center
\includegraphics[width=0.5\columnwidth]{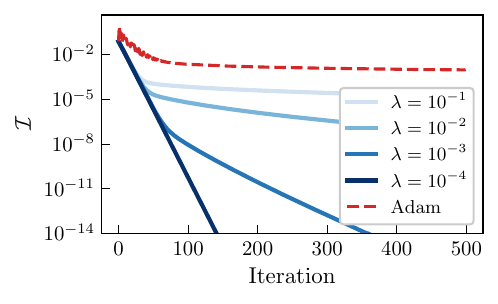}
\caption{
Performance of NGD with varying damping parameters $\lambda$, and of Adam on the same state-compression problem investigate in \cref{fig:numerical_precision}(c). All gradients and curvature matrices are computed exactly. 
Parameters: $\alpha_{\mathrm{ngd}}=0.05$, $\bm\Theta_{\rm{CNN}} = (10,10,10,10;3)$, $\alpha_{\mathrm{adam}}=10^{-4}$. Note that larger values of the learning rate for Adam are unstable and result in divergences in the objective. Different warm up strategies progressively raising and then lowering the learning rate yield no visible improvement over the presented results.
}
\label{fig:ngdvadam}
\end{figure}

We now use this simple toy problem to provide a clear illustration of the importance of incorporating curvature information into the optimization strategy to ensure reliable convergence. To isolate the effect of the optimization method from other sources of error—such as Monte Carlo noise and model expressivity—we consider an idealized setting in which all statistical averages required to compute the objective's gradient and curvature matrix are evaluated exactly by summing over the entire Hilbert space at each optimization step. Moreover, as shown in \cref{fig:numerical_precision}, the chosen CNN model with 
$\bm\Theta_{\mathrm{CNN}} = (10, 10, 10, 10; 3)$ is able to represent the target state at time $Jt=1.3$ with machine precision accuracy.
In \cref{fig:ngdvadam}, we compare the performance of NGD and Adam on the same state-compression task studied in \cref{fig:numerical_precision}(c). The results show that Adam converges significantly more slowly and often becomes trapped in local minima, failing to accurately compress the target state. By contrast, NGD achieves rapid and reliable convergence.
To further explore the role of curvature information, we modulate the damping parameter $\lambda$, which interpolates between NGD and standard gradient descent. For large values of $\lambda$, the curvature information is heavily suppressed by regularization, and the behavior approaches that of vanilla gradient descent or Adam. As $\lambda$ is decreased, curvature plays a more prominent role in shaping the optimization path, culminating in exponential convergence for $\lambda=10^{-4}$.
Overall, this toy problem highlights the critical role of NGD in learning large, global transformations of the quantum state, especially in regimes where precise convergence is required.

\section{Exact application of diagonal operators}
\label{app:diag_ops}
Let $\kpsiv = \sum_x \psiv(x) \ket x $ be a variational state and $\hat A = \sum_{x,y} A_{xy} \ketbra{x}{y}$ a generic operator acting on it. We want to find a way to reduce the application of $\hat A$ on $\kpsiv$ to a change of parameters $\theta \to \theta'$. In other words, we want to find $\theta'$ such that 
\begin{equation}
    \label{eqn:}
    \ket{\psi_{\theta'}} = \hat A \ket{\psi_{\theta}}. 
\end{equation}
If successful, this procedure would allow us to apply $\hat A$ on $\ket{\psi_{\theta}}$ exactly and at no computational expense. 
Unfortunately, for a generic parametrization of the state, or a generic operator, this is not possible. 
The problem greatly simplifies if we restrict to diagonal operators of the form $\hat A = \sum_x A_x \ketbra{x}$, whose application on a generic variational state reads  $\hat A \ket{\psi_{\theta}} = \sum_x A_x \psiv(x) \ket{x}$. Even in this simple case, the transformation $\theta\to\theta'$ satisfying $\ket{\psi_{\theta'}} = \hat A \ket{\psi_{\theta}}$ is not guaranteed to exist. Consider, however, the improved ansatz $\ket{\psi_{\theta,\phi}} = \sum_x \psi_{\theta,\phi}(x) \ket x$ where 
\begin{equation}
    \psi_{\theta,\phi}(x) = (A_x)^{\phi}\psiv(x) \quad\text{with}\quad \phi\in\mathbb C.
\end{equation}
The application of $\hat A$ on this state reads
\begin{equation}
    \hat A \ket{\psi_{\theta,\phi}} = \sum_x A_x \psi_{\theta,\phi}(x) \ket x = \sum_x (A_x)^{\phi+1}\psiv(x) \ket{x} = \ket{\psi_{\theta,\phi+1}}
\end{equation}
The action of $\hat A$ on $\ket{\psi_{\theta,\phi}}$ can thus be \emph{exactly} reduced to the parameter transformation $(\theta,\phi)\to(\theta,\phi+1)$ which can be computed at virtually no computational expense. Note that while the additional multiplicative layer has a structure determined by $\hat A$, the network $\psiv(x)$ to which this layer is added is completely arbitrary.

This procedure can be easily generalized to multiple diagonal operations which, of course, commute with each other. Given $\hat A = \sum_x A_x \ketbra{x}$ and $\hat B = \sum_x B_x \ketbra{x}$, we define the improved ansatz as 
\begin{equation}
\label{eqn:two_matrix_ansatz}
    \psi_{\theta,\phi_A, \phi_B}(x) = (B_x)^{\phi_B}(A_x)^{\phi_A}\psiv(x).
\end{equation}
The application of $\hat A$ without $\hat B$ is equivalent to $(\theta, \phi_A, \phi_B)\to(\theta, \phi_A+1, \phi_B)$. The application of $\hat B$ without $\hat A$ is equivalent to $(\theta, \phi_A, \phi_B)\to(\theta, \phi_A, \phi_B+1)$. The simultaneous application of $\hat A$ and $\hat B$ is equivalent to the transformation $(\theta, \phi_A, \phi_B)\to(\theta, \phi_A+1, \phi_B+1)$. Note that we never act on the network itself ($\theta$ is never changed).

\subsection{$ZZ$-operations}
Let $\ket{\psi_{\theta}}$ be an arbitrary ansatz for the state of a system of $N$ spin-$1/2$ particles, and
\begin{equation}
    \hat A 
    = \exp{\alpha \sz_\mu\sz_\nu} 
    = \sum_x e^{\,\alpha \,x_\mu x_\nu} \ketbra x
    \equiv \sum_x A_x \ketbra x
    \quad
    \text{with}
    \quad 
    \mu,\nu\in[1,\ldots,N].
\end{equation}
the $ZZ$-operation acting on spins $\mu$ and $\nu$. To encode the action of $\hat A$ as a change of parameters we define the improved ansatz 
\begin{equation}
\label{eqn:zzasnatz_1}
    \psi_{\theta,\phi}(x) = (A_x)^{\phi}\psiv(x) = e^{\,\alpha\phi \,x_\mu x_\nu}\psiv(x) \equiv e^{\,\phi \,x_\mu x_\nu}\psiv(x),
\end{equation}
where, in the last equality, we absorb $\alpha$ in the parameter $\phi$ without loss of generality. As expected, $\ket{\psi_{\theta,\phi}} \to \hat A \ket{\psi_{\theta,\phi}} \equiv (\theta,\phi) \to (\theta, \phi+1)$.
This ansatz accounts for the application of $ZZ$-operations on a fixed pair of  spins, namely spin $\mu$ and $\nu$. For this reason the additional parameter $\phi$ is a scalar. In general, however, we want to reserve the right to apply the operation between any pair of spins and/or on multiple pairs simultaneously. 

Consider now the application of $ZZ$-rotations on two pairs of spins: ($\mu,\nu$) and ($\mu',\nu'$). Said differently, we want to find an ansatz to incorporate the action of $\hat A = \exp{\alpha_{\mu\nu} \sz_\mu\sz_\nu}$, of $\hat B = \exp{\alpha_{\mu'\nu'} \sz_{\mu'}\sz_{\nu'}}$, and of their product $AB$, as a simple change of parameters \footnote{Here $\alpha_{\mu\nu}\in\mathbb C$ is the phase of the gate operation acting on the pair ($\mu,\nu$).}.
To do so we can incorporate the single-operator ansatz from \cref{eqn:zzasnatz_1} 
into the two-operator structure in \cref{eqn:two_matrix_ansatz} as
\begin{equation}
        \psi_{\theta,\bm\phi}(x) = (B_x)^{\phi_{\mu'\nu'}}(A_x)^{\phi_{\mu\nu}}\psiv(x)
        = \exp{x_\mu \phi_{\mu\nu} x_\nu} \exp{x_{\mu'} \phi_{\mu'\nu'} x_{\nu'}}\psiv(x)
        = \exp{x_\mu \phi_{\mu\nu} x_\nu + x_{\mu'} \phi_{\mu'\nu'} x_{\nu'}}\psiv(x).
\end{equation}
Note that now $\bm\phi = (\phi_{\mu\nu}, \phi_{\mu'\nu'})$ is a two-dimensional vector. This can be further generalized to account for $ZZ$-operations between any two spins via the ansatz 
\begin{equation}
    \psi_{\theta,\bm \phi}(x) = \operatorname{exp}\Big\{\sum_{ij} x_i \phi_{ij} x_j\Big\}\psiv(x) = \exp{\bm x\bm\phi\bm x^\top}\psiv(x).
\end{equation}
Note that the multiplicative layer added to the network is exactly a two-body Jastrow ansatz where $\bm\phi=[\phi_{\mu\nu}]$ is an $N\times N$ matrix. Application of the operator $\hat A_{\mu\nu} = \exp{\alpha_{\mu\nu} \sz_\mu\sz_\nu}$ is equivalent to the parameter transformation $\phi_{\mu\nu} \to \phi_{\mu\nu} + \alpha_{\mu\nu}$. 
In general we parameterize the log-amplitude of the wave function $\log \psi_{\theta,\bm\phi}(x) = \bm x\bm\phi\bm x^\top + \log\psiv(x)$, so that the multiplicative layer actually becomes an additive one.

\section{Neural network architectures}
\label{app:architecture}
In this section we review the two architectures used in this work. 

\subsection{Convolutional neural networks}
\label{app:CNN}
Convolutional Neural Networks (CNNs) are particularly well-suited for processing and analyzing grid-like data, such as images or quantum systems on a lattice. 
The architecture of a CNN consists of multiple layers indexed by $\ell \in [1, N_L]$, typically structured as alternating non-linear and affine transformations. 
For a system defined on a square lattice of linear length $L$ and number of particles $N = L^2$, the input layer ($\ell = 1$) receives the configuration vector $x = (s_1, \ldots, s_{N})$, which is reshaped into an $L \times L$ matrix as $X = \operatorname{vec}^{-1}(x)$, where $\operatorname{vec}$ represents the vectorization operation \cite{Minganti2019}.

The building block of CNNs is the convolutional layer, where filters (or kernels) are applied to local regions of the input, learning spatial hierarchies of features.
This is analogous to performing convolution operations over a lattice, capturing local correlations across the system.
Let the output of the $\ell$-th layer be
\begin{equation}
    X^{(\ell)} = [X^{(\ell)}]^{\alpha}_{i,j} \in\mathbb{C}^{C_\ell}\otimes\mathbb{C}^{H_\ell\times W_\ell}
    \qq{with}
    \begin{cases}
    i\,\in[0, H_\ell-1],\\
    j \,\in [0,W_\ell-1],\\
    \alpha\in[0,C_\ell-1]
    \end{cases}
\end{equation}
where $(H_\ell, W_\ell)$ are the height and width of the processed data at layer step $\ell$, and $C_\ell$ is the number of channels.
The convolution operation yielding the data structure of the $(\ell+1)$-th layer is 
\begin{equation}
\label{eqn:CNN}
\qty(X^{(\ell+1)})^{\beta}_{m,n}
= 
\sigma_\ell\qty( \bigl[X^{(\ell)} \circledast F^{(\ell)}\bigr]^{\beta}_{m,n})
= 
\sigma_\ell\qty(\sum_{\alpha} \sum_{i,j} \bigl[F^{(\ell)}\bigr]_{i,j}^{\alpha\beta}\,\bigl[X^{(\ell)}\bigr]_{m+i,\, n+j}^{\alpha} )
\qq{with}
\begin{cases}
i \,\in[0, h_\ell-1],\\
j \,\in [0,w_\ell-1],\\
\beta\in[0,c_\ell-1]
\end{cases}
\end{equation}
where $\sigma_\ell$ is the activation function, $(h_\ell, w_\ell)$ is the size of the convolutional kernel $F^{(\ell)}$, and $c_\ell$ is its output dimension.

For the activation functions, we follow the approach in Ref.~\cite{Schmitt_2020}. The activation function in the first layer conists of the first three non-vanishing terms of the series expansion of $\operatorname{logcosh}(z)$, ensuring the incorporation of the system's $\mathbb{Z}_2$ symmetry in the absence of bias in the first layer.
It is defined as
\begin{equation}
    \sigma_1(z) = \frac{z^2}{2} - \frac{z^4}{12} + \frac{z^6}{45}  .
\end{equation}
In subsequent layers, its derivative is used
\begin{equation}
    \sigma_{\ell>1}(z) = \frac{z}{2} - \frac{z^3}{3} + \frac{2}{15}z^5.
\end{equation}
We use circular padding to respect periodic boundary conditions, ensuring moreover that the spatial dimensions remain constant across layers, $H_{\ell} = W_{\ell} = L$. Both dilation and stride are set to one across all layers, and a fixed kernel size $h_{\ell} = w_{\ell} = k$ is used. A final dense layer is used to pool the output of the CNN into a scalar output.
After the convolutional layers, a fully connected layer reduces the output of the convolutional sequence to a scalar.

\begin{figure}[htb]
    \begin{minipage}[b]{0.48\columnwidth}
        \centering
        \includegraphics[width=\linewidth]{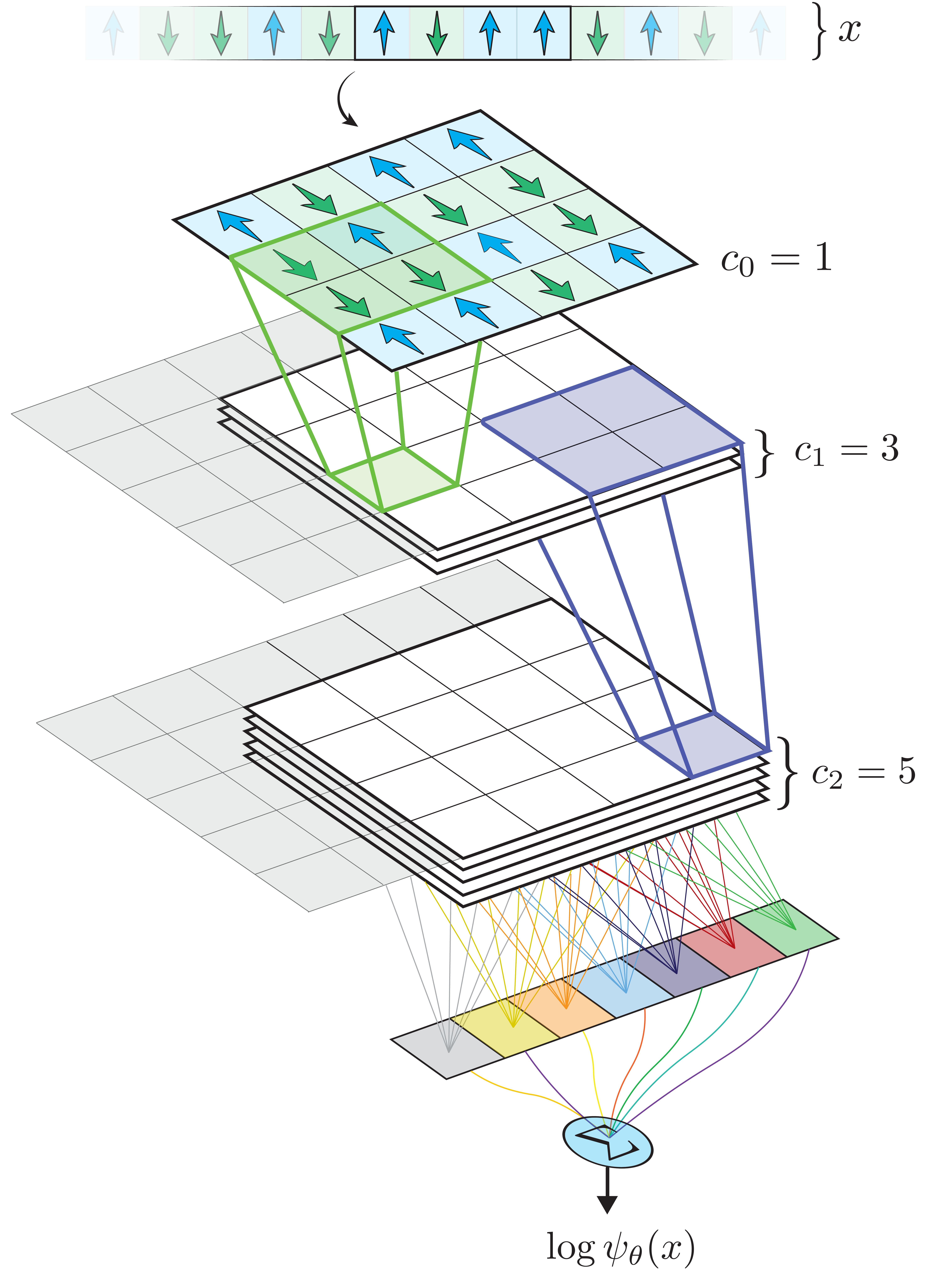}
        \caption{
        Illustrative representation of the CNN architecture described in \cref{app:CNN}.\\
        \ }
        \label{fig:NN_sketch_CNN}
    \end{minipage}
    \hfill
    \begin{minipage}[b]{0.48\columnwidth}
        \centering
        \includegraphics[width=\linewidth]{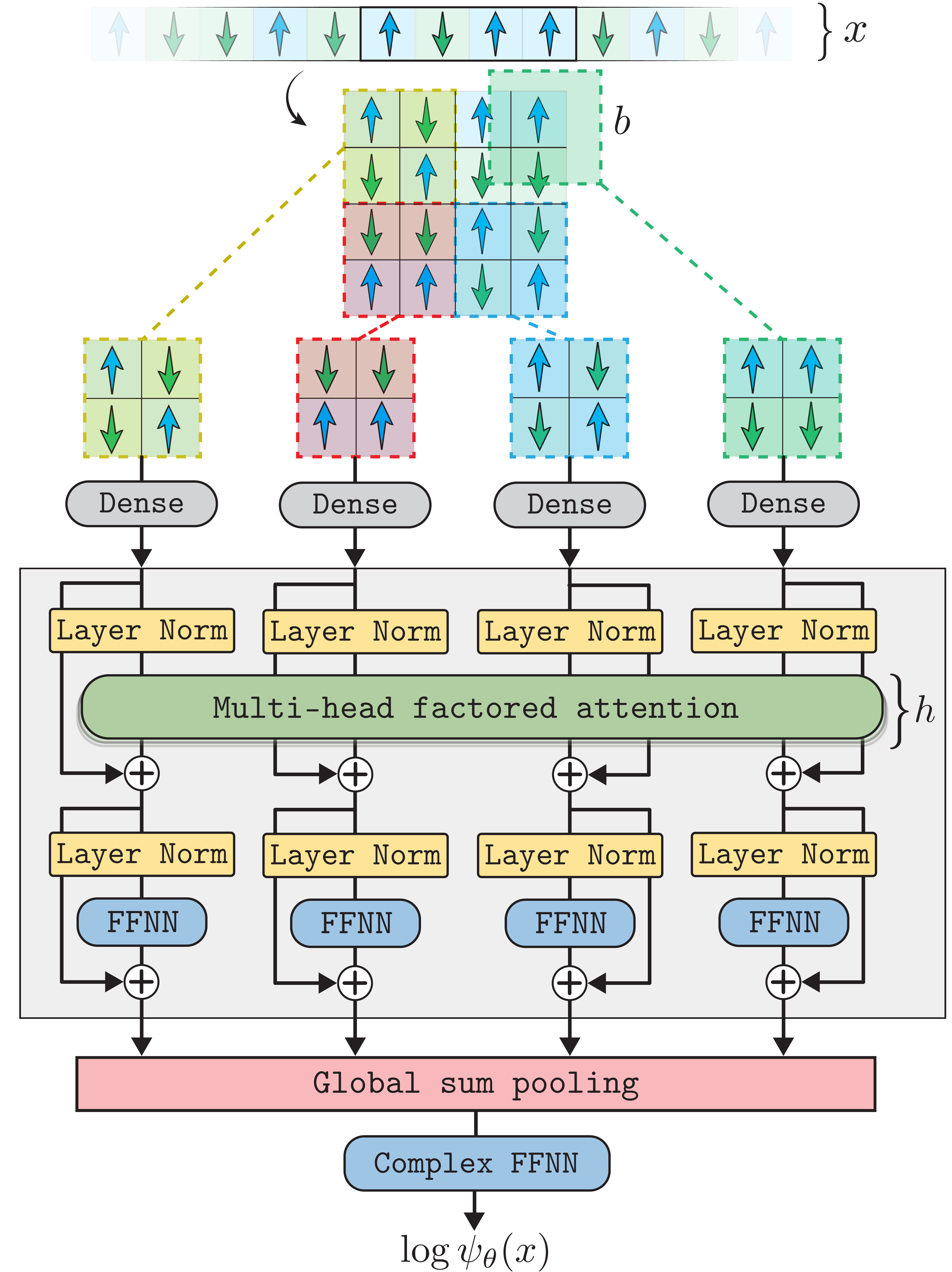}
        \caption{Illustrative representation of the ViT architecture described in \cref{app:vit}.
        We refer to Refs.~\cite{Viteritti2023,Viteritti2023a} for further details.
        }
        \label{fig:NN_sketch_ViT}
    \end{minipage}
\end{figure}

The CNN structure can is summarized by a tuple $\bm\Theta_{\rm CNN} = (c_1, \ldots, c_{N_L}; k)$, where $c_\ell$ represents the number of channels in each layer, and $k$ denotes the kernel size.  A sketch of the architecture is shown in \cref{fig:NN_sketch_CNN}.

\subsection{Vision Transformer}
\label{app:vit}
The Vision Transformer (ViT) is a state-of-the-art architecture in machine learning. While originally developed for image classification and segmentation, it has recently been adopted as a powerful ansatz for quantum many-body wavefunctions \cite{Viteritti2023a,Viteritti2023}. Below, we describe the key components and parameters of this architecture as applied to NQS.
For a spin-$1/2$ system defined on a square lattice of linear length $L$, the input consists of a configuration vector ${\bm x} = (s_1, \ldots, s_{N}) $ with $ s_i \in \{\pm 1\} $. This vector is reshaped into an $ L \times L $ matrix $\bm X = \operatorname{vec}^{-1}({\bm x}) $, with $\operatorname{vec}$ representing the vectorization operation \cite{Minganti2019}. This matrix corresponds to a specific configuration of spins on the lattice.
\begin{description}
    \item[Patch Extraction and Embedding.] 
    The matrix is divided into $n \in \mathbb N$ non-overlapping patches of size $ b \times b $, namely $\bm X_1, \ldots, \bm X_n$ with $\bm X_i \in \{\pm 1\}^{b\times b}$. Of course, for this to be possible, the total number of sites must be an integer multiple of $b$.
    Each patch is flattened, and linearly projected into a high-dimensional embedding space of dimension $d$.
    This transforms the spin values into a vector representation that is processed by the encoder blocks: 
    \begin{equation}
        \bigl\{\bm X_i \in \{\pm 1\}^{b\times b}\bigr\}_{i=1}^n 
        \,\,\to\,\, 
        \bigl\{\operatorname{vec}(\bm X_i)\in \{\pm 1\}^{b^2} \bigl\}_{i=1}^n 
        \,\,\to\,\,
        \bigl\{{\bm x}_i\in \mathbb R^{d} \bigl\}_{i=1}^n.
    \end{equation}
    \item[Encoder Blocks.] 
    The core of the ViT architecture consists of $ N_{L} $ encoder blocks. 
    The input to each block is a set of $n$ vectors $\bigl\{{\bm x}_i^{(\ell)}\in \mathbb R^{d} \bigl\}_{i=1}^n$ with $\ell = 1,\ldots, N_L$ and ${\bm x}_i^{(1)} = \bm x_i$. 
    Each encoder block correlates the input vectors thereby capturing long-range correlations and intricate interactions in the quantum system. Each block consists of the following components:
    \begin{description}
        \item[Multi-Head Factored Attention Mechanism.] 
        The generic \textit{self-attention} mechanism is defined by three rectangular matrices of parameters $\bm Q, \bm K, \bm V$ known as Query, Key, and Value. 
        For each of the $n$ input vectors $\bm x_i$, three new vectors are computed: $\bm q_i = \bm Q{\bm x}_i$, $\bm k_i = \bm K{\bm x}_i$, $\bm v_i = \bm V{\bm x}_i$. These vectors are then combined with one another to create $n$ attention vectors $\bm A_i = \sum_{j=1}^n\alpha(\bm q_i, \bm k_j) \, \bm v_j$ where the attention weights $\alpha(\bm q_i, \bm k_j)$ modulate the contribution of each value vector $\bm v_j$ to the globally mixed vector $\bm A_i$. 
        To improve the performance of the model, a \textit{multi-head attention mechanism} is often preferred to this self-attention. In this case, a set of $h \in \mathbb N$ matrices $\{\bm Q^\mu, \bm K^\mu , \bm V^\mu \}_{\mu = 1}^h$ are used to compute in parallel a set of $h$ sets of $n$ self-attention vectors $\{\bm A_1^\mu, \ldots, \bm A_n^\mu\}_{\mu = 1}^h$ with $\bm A_i^\mu \in \mathbb R^{r}$ and $r = d/h$. While in general we would have 
        \begin{equation}
            \bm A_i^\mu = \sum_{j=1}^n\alpha(\bm Q{\bm x}_i^\mu, \bm K {\bm x}_j^\mu) \, \bm V {\bm x}_j,
        \end{equation}
        the simplification proposed in \cite{Viteritti2023, Viteritti2023a} takes 
        \begin{equation}
            \bm A_i^\mu = \sum_{j=1}^n\alpha_{ij}^\mu \, \bm V^\mu {\bm x}_j,
        \end{equation}
        where $\bm V^\mu$ is an $r \times d$ matrix and $\alpha_{ij}^\mu$ is a tensor of parameters depending on the positions of groups of spins (patches $i,j$) and independent instead of the actual spin configuration at those positions (no dependence on ${\bm x}_i$ or $\bm x_j$). To summarize:
        \begin{equation}
            \bigl\{{\bm x}_i \in \mathbb{R}^{d}\bigr\}_{i=1, \ldots, n} 
            \,\,\to\,\,
            \{\bm A_i^\mu \in \mathbb{R}^r\}_{\substack{i=1,\ldots, n\\\mu=1,\ldots, h}}.
        \end{equation}    
        \item[Feed-Forward Neural Network.] 
        The $h$ sets of vectors from the attention layer $\bm A_i^\mu$ are first concatenated into a set of $n$ d-dimensional vectors $\{\bm y_i\}_{i=1}^n$ with 
        \begin{equation}
            \bm y_i = \operatorname{Concat}(\bm A^1_i, \ldots, \bm A^h_i) \in \mathbb{R}^d.
        \end{equation}
        A feed-forward network (FFN) is finally used to process these vectors independently. The hidden layer in this FFN has a dimensionality of $ n_{\mathrm up}\,d $ with $ n_{\rm up} $ the upscaling factor. A GeLU (Gaussian Error Linear Unit) activation function is used between the layers of the FFN, introducing a non-linearity that helps the model capture more complex features. The output of the encoder block is a set of $n$ vectors 
        \begin{equation}
            {\bm x}^{(\ell + 1)}_i = \operatorname{FFN}(\bm y_i) \in \mathbb{R}^{n_{\mathrm up} \, d}.
        \end{equation}
        This output is then used as the input to the subsequent block. 
        \item[Skip Connections and Layer Normalization.] Skip connections are applied across the attention and FFN layers. These connections help alleviate the vanishing gradient problem, allowing the model to train deeper architectures. Layer normalization is applied before both the attention and FFN layers to stabilize the training process and improve convergence.
    \end{description}
    \item[Output and Wave Function Representation.]
    After the spin configurations pass through the encoder blocks, the output vectors corresponding to each patch are summed to create a final hidden representation vector 
    \begin{equation}
        \bm z = \sum_{i=1}^n \bm x_i^{(N_L+1)}.
    \end{equation}
    This vector represents the configuration in a high-dimensional space and is passed through a final complex-valued fully connected neural network yielding the log-amplitude of our variational wave function:
    \begin{equation}
        \log \psi_{\theta}({\bm x}) = \sum_{\alpha=1}^K \operatorname{logcosh}(b_\alpha + \bm w_\alpha\cdot \bm z),
    \end{equation}
    where $\bm W = \{b_\alpha, \bm w_\alpha\}_{\alpha=1}^K$ are complex-valued parameters so that $\{\Re(\bm W), \Im(\bm W)\} \subset \theta$. 
\end{description}
We summarize the ViT configuration with the tuple $\Theta_{\mathrm{ViT}} = (b, h, r, n_{\rm up}; N_L)$. A schematic representation is shown in \cref{fig:NN_sketch_ViT}.

\end{document}